\documentclass[a4paper,11pt]{article}
\pdfoutput=1
\usepackage{jheppub}
\usepackage[utf8]{inputenc}

\usepackage[vcentermath]{youngtab}
\usepackage{pdflscape}
\usepackage{tikz}
\usetikzlibrary{positioning}
\usetikzlibrary{arrows}
\usetikzlibrary{arrows.meta}

\usepackage{romannum}
\usepackage{adjustbox}
\usepackage[colorinlistoftodos]{todonotes}
\tikzset{gauge1/.style={draw=none,minimum size=0.6cm,fill=white,circle, draw}}
\tikzset{none/.style={draw=none}}
\tikzset{redgauge/.style={draw=none,minimum size=0.4cm,fill=red,circle, draw}}
\tikzset{bluegauge/.style={draw=none,minimum size=0.4cm,fill=blue,circle, draw}}
\tikzset{orangegauge/.style={draw=none,minimum size=0.4cm,fill=orange,circle, draw}}
\tikzset{magentagauge/.style={draw=none,minimum size=0.4cm,fill=magenta,circle, draw}}
\tikzset{cyangauge/.style={draw=none,minimum size=0.4cm,fill=cyan,circle, draw}}
\tikzset{olivegauge/.style={draw=none,minimum size=0.4cm,fill=olive,circle, draw}}
\tikzset{gaugeX/.style={draw=none,minimum size=0.4cm,fill=white,circle, draw}}
\tikzset{smallgauge/.style={draw=none,minimum size=0.1cm,fill=black,circle, draw}}
\tikzset{gauge2/.style={draw=none,minimum size=0.35mm,fill=red,circle, draw}}
\tikzset{flavor1/.style={draw=none,minimum size=0.35mm,fill=blue, regular polygon,regular polygon sides=4,draw}}
\tikzset{smalldot/.style={draw=none,minimum size=0.1mm,fill=black, circle,draw}}
\tikzset{verysmalldot/.style={draw=none,minimum size=0.05mm,fill=black, circle,draw}}
\tikzset{doubleguys/.style={double, double distance = 3pt}}
\tikzset{tripleguys/.style={triple}}
\tikzset{new edge style 1/.style={dashed}}
\tikzset{thickline/.style={line width=0.06cm}}
\tikzset{cyane/.style={line width=0.3mm,cyan}}
\tikzset{bluee/.style={line width=0.3mm,blue}}
\tikzset{orangee/.style={line width=0.3mm,orange}}
\tikzset{magentae/.style={line width=0.3mm,magenta}}
\tikzset{rede/.style={line width=0.3mm,red}}
\tikzset{olivee/.style={line width=0.3mm,olive}}
\tikzset{darke/.style={line width=0.3mm,black}}
\tikzset{brace/.style={decorate,decoration={brace,amplitude=5pt}}}
\pgfdeclarelayer{edgelayer}
\pgfdeclarelayer{nodelayer}
\usetikzlibrary{decorations.pathreplacing}
\pgfsetlayers{edgelayer,nodelayer,main}

\usetikzlibrary{decorations.pathreplacing}
\usepackage{subcaption}
\usepackage{xcolor,colortbl}
\usepackage{mathtools}
\usepackage{ragged2e}
\newlength\ubwidth
\newcommand\parunderbrace[2]{\settowidth\ubwidth{$#1$}\underbrace{#1}_{\parbox{\ubwidth}{\scriptsize\RaggedRight#2}}}
\usepackage{amsthm}
\newtheorem*{DefHiggsRing}{Higgs Ring}
\newtheorem*{DefHiggsVar}{Higgs Variety}
\newtheorem*{DefHiggsScheme}{Higgs Scheme}
\usepackage{afterpage}

\title{\boldmath Brane Webs and Magnetic Quivers for SQCD}

\author{Antoine Bourget, }
\author{Santiago Cabrera, }
\author{Julius F. Grimminger, }
\author{Amihay Hanany}
\author{and Zhenghao Zhong}
\affiliation{Theoretical Physics Group, The Blackett Laboratory, Imperial College London, Prince Consort Road
London, SW7 2AZ, UK}
\emailAdd{a.bourget@imperial.ac.uk}
\emailAdd{santiago.cabrera13@ic.ac.uk}
\emailAdd{julius.grimminger17@imperial.ac.uk}
\emailAdd{a.hanany@imperial.ac.uk}
\emailAdd{zhenghao.zhong14@imperial.ac.uk}

\abstract{It is widely considered that the classical Higgs branch of 4d $\mathcal{N}=2$ SQCD is a well understood object. However there is no satisfactory understanding of its structure. There are two complications: (1) the Higgs branch chiral ring contains nilpotent elements, as can easily be checked in the case of $\mathrm{SU}(N)$ with 1 flavour. (2) the Higgs branch as a geometric space can in general be decomposed into two cones with nontrivial intersection, the baryonic and mesonic branches. To study the second point in detail we use the recently developed tool of magnetic quivers for five-brane webs, using the fact that the classical Higgs branch for theories with 8 supercharges does not change through dimensional reduction. We compare this approach with the computation of the hyper-Kähler quotient using Hilbert series techniques, finding perfect agreement if nilpotent operators are eliminated by the computation of a so called \emph{radical}. We study the nature of the nilpotent operators and give conjectures for the Hilbert series of the full Higgs branch, giving new insights into the vacuum structure of 4d $\mathcal{N}=2$ SQCD. In addition we demonstrate the power of the magnetic quiver technique, as it allows us to identify the decomposition into cones, and provides us with the global symmetries of the theory, as a simple alternative to the techniques that were used to date.}

\begin{document}

\maketitle

\section{Introduction and Summary}

In the last few decades, it has become clear that the study of string theory and brane dynamics can provide great insights and help solving quantum field theory problems in various space-time dimensions. In this paper, we offer yet another demonstration of this paradigm on a concrete example, although the techniques we use are applicable in a much wider context. 

Specifically, we consider supersymmetric gauge theories with 8 supercharges where the gauge group is $\mathrm{SU}(N_c)$ and the matter content consists of $N_f$ hypermultiplets in the fundamental 
representation of the gauge group. For definiteness, let us focus on four spacetime dimensions, in which this theory is the well-known $\mathcal{N}=2$ SQCD theory. We set all masses and Fayet-Iliopoulos terms to zero. The moduli space of classical vacua is parametrized by the vacuum expectation values (VEVs) of the various scalar fields in the theory, and it can be organized in several branches, according to the residual gauge group which is left unbroken by these VEVs. On the Coulomb branch, parametrized by the VEVs of scalars in the vector multiplet, the gauge group is generically broken to the Abelian group with the same rank, $\mathrm{U}(1)^{N_c-1}$. On the Higgs branch, parametrized by the VEVs of scalars $Q$ and $\tilde{Q}$ in the hypermultiplets, the gauge group is generically maximally broken (not necessarily completely broken). Between these two extremes, there is a variety of mixed branches. In this paper, we focus on the Higgs branch \cite{Seiberg:1994aj,Argyres:1996eh,Gaiotto:2008nz,Xie:2014pua}. 

We recall that the classical Higgs branch in a 4d $\mathcal{N}=2$ theory does not receive any quantum correction \cite{Seiberg:1994aj}, at least as long as there is no phase transition. It can therefore be computed classically using the  hyper-K\"ahler quotient construction \cite{Hitchin:1986ea}. This purely algebraic construction does not depend on the dimension of the theory with 8 supercharges under consideration, and as a consequence, the Higgs branches studied in this paper can be either thought of as emanating from 3d $\mathcal{N}=4$, 4d $\mathcal{N}=2$, or 5d $\mathcal{N}=1$ theories. In the present discussion we use the most well-known four-dimensional vocabulary (superpotential, F-terms, D-terms). More specifically, we are interested in the complex structure of the Higgs branch, and therefore only need to solve the complex moment map equations (which correspond to the F-terms), and then quotient by the complexified gauge group \cite{Billo:1994yj,Antoniadis:1996ra}. The gauge invariant fields which can be constructed from the hypermultiplets are mesons and baryons, so these parametrize the Higgs branch. The reason why we only care about the complex structure is that we use the Hilbert series \cite{stanley1978hilbert,Benvenuti:2006qr,Benvenuti:2010pq} of the Higgs branch as our main characterization tool: the Higgs branch holomorphic coordinate ring has a natural grading by the nonnegative integers, which corresponds to the highest weight under the $\mathrm{SU}(2)_R$ acting on the Higgs branch. Physically, this is the scaling dimension of the operators, mesons having dimension 2 and baryons having dimension $N_c$.

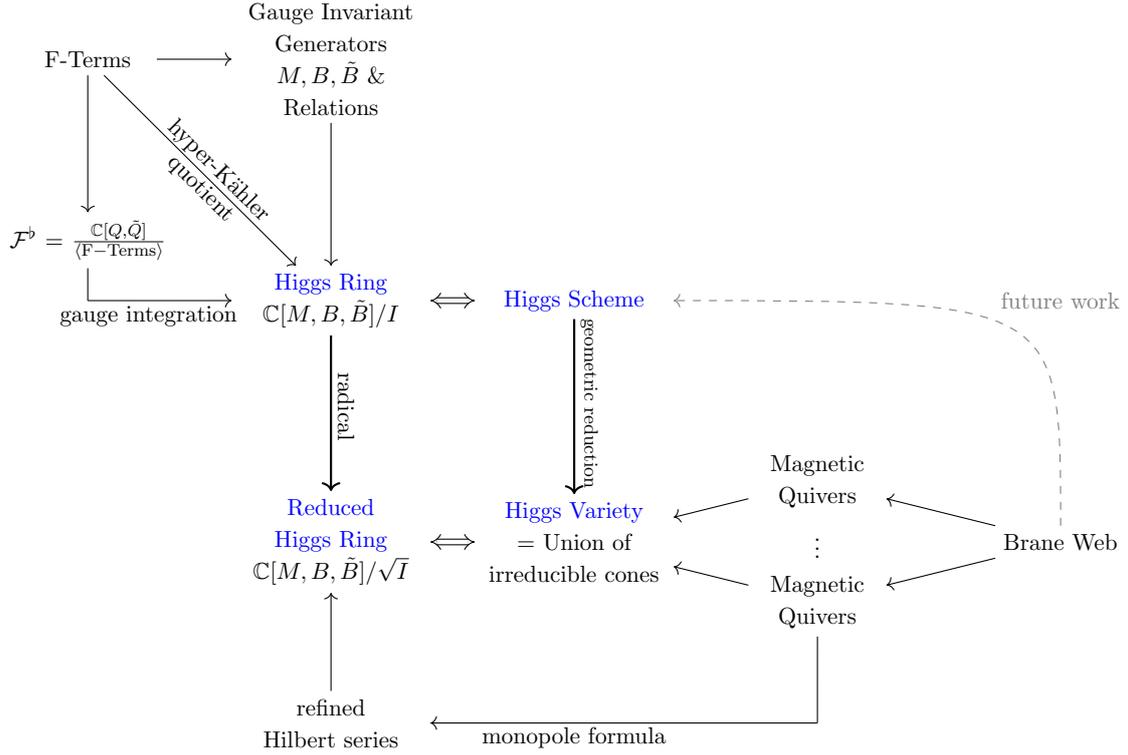
\begin{figure}[t]
    \centering
   \begin{tikzpicture}[scale=0.8, every node/.style={scale=0.8}]
\node[text width=3cm,align=center](1) at (0,4) {\hyperlink{HiggsRing}{Higgs Ring} $\mathbb{C}[M,B,\tilde{B}]/I$};
\node[text width=3cm,align=center](2) at (4,4) {\hyperlink{HiggsScheme}{Higgs Scheme}};
\node[text width=3cm,align=center](3) at (0,0) {\hyperlink{HiggsVar}{Reduced Higgs Ring} $\mathbb{C}[M,B,\tilde{B}]/\sqrt{I}$};
\node[text width=3cm,align=center](4) at (4,0) {\hyperlink{HiggsVar}{Higgs Variety} = Union of irreducible cones};
\node[text width=2cm,align=center](5) at (8,1) {Magnetic Quivers};
\node at (8,0) {$\vdots$};
\node[text width=2cm,align=center](6) at (8,-1) {Magnetic Quivers};
\node[text width=2cm,align=center](7) at (12,0) {Brane Web};
\draw[implies-implies,double equal sign distance] (1) -- (2);
\draw[implies-implies,double equal sign distance] (3) -- (4);
\draw[<-] (4) -- (5);
\draw[<-] (4) -- (6);
\draw[<-] (5) -- (7);
\draw[<-] (6) -- (7);
\draw[thick,->] (1) -- (3);
\node[rotate=-90] at (0.25,2.25) {radical};
\draw[thick,->] (2) -- (4);
\node[rotate=-90] at (4.25,2.3) {\footnotesize{geometric reduction}};
\node[text width=3cm,align=center](8) at (0,-3) {refined Hilbert series};
\draw[->](8)--(3);
\draw (6)--(8,-3);
\draw[<-] (8)--(8,-3);
\node at (4,-3.25) {monopole formula};
\node[text width=2cm,align=center](9) at (-4,8) {F-Terms};
\node[text width=3cm,align=center](10) at (-4,5) {$\mathcal{F}^{\flat} = \frac{\mathbb{C}[Q,\tilde{Q}]}{\langle \mathrm{F-Terms}\rangle}$};
\node[text width=3cm,align=center](11) at (0,8) {Gauge Invariant Generators $M,B,\tilde{B}$ \& Relations};
\draw[->] (9)--(1);
\node[rotate=-45] at (-1.85,6.15) {hyper-Kähler};
\node[rotate=-45] at (-2.15,5.85) {quotient};
\draw[->] (9)--(11);
\draw[->] (11)--(1);
\draw[->] (9)--(10);
\draw (10)--(-4,4);
\draw[->] (-4,4)--(1);
\node at (-3,3.75) {gauge integration};
\draw[gray,dashed,->] (7) .. controls (12,4) .. (2);
\node[gray] at (12,4) {future work};
\end{tikzpicture} 
    \caption{Summary of the various incarnations of the Higgs branch. The second column contains the coordinate rings, the third column contains the geometric objects. In the second row, the ring contains nilpotent operators, it can be computed from the F-term relations through the hyper-Kähler quotient. Upon taking the radical (see appendix \ref{AppendixAlgebra}), the third row contains no nilpotent operators. The associated Higgs branch is an algebraic variety, which is a symplectic singularity or union thereof. As such, it is described by magnetic quivers that can be read from the brane web.}
    \label{figSummary}
\end{figure}

It is important to stress that we actually use this correspondence in the other direction: we \emph{define}:
\hypertarget{HiggsRing}{\begin{DefHiggsRing}
    \emph{The polynomial ring in the variables $Q$ and $\tilde{Q}$ subjected to the F-term equations (called the $\mathcal{F}^{\flat}$ ring) and projected on gauge invariant polynomials. See the upper-left corner of Figure \ref{figSummary}. The} \hyperlink{HiggsScheme}{Higgs scheme} \emph{is then defined as the geometrical object whose coordinate ring is the Higgs ring.}\footnote{It is a scheme in the sense of algebraic geometry, as reviewed briefly in Appendix \ref{AppendixAlgebra}.}
\end{DefHiggsRing}}
 The Hilbert series of the \hyperlink{HiggsScheme}{Higgs scheme}/\hyperlink{HiggsRing}{Higgs ring} \footnote{The Hilbert series of a geometric space is defined as the Hilbert series of its coordinate ring. We use both interchangeably.} is then evaluated as 
\begin{equation}
\label{generalFormulaHBHS}
    H^{G}_{N_f}(t) = \int_G \mathrm{d} \mu_{G^{\mathbb{C}}} \mathrm{HS} \left( \frac{\mathbb{C}[Q,\tilde{Q}]}{\langle\textrm{F-terms}\rangle} ,t\right) \,.
\end{equation}
This is a two step process, in which one first needs to compute the Hilbert series of the quotient polynomial ring defined by the F-terms, and then perform an integration over the complexified gauge group $G^{\mathbb{C}}$, with Haar measure $\mathrm{d} \mu_{G^{\mathbb{C}}}$. The details are reviewed in Section \ref{sectionHBSQCD}. As is shown there, evaluating (\ref{generalFormulaHBHS}) explicitly for $\mathrm{SU}(N_c)$ SQCD involves formidable computational challenges, and the explicit computation can be done only in a handful of low-rank cases. Similar computations were performed for $\mathcal{N}=1$ SQCD with vanishing superpotential in \cite{Gray:2008yu}.

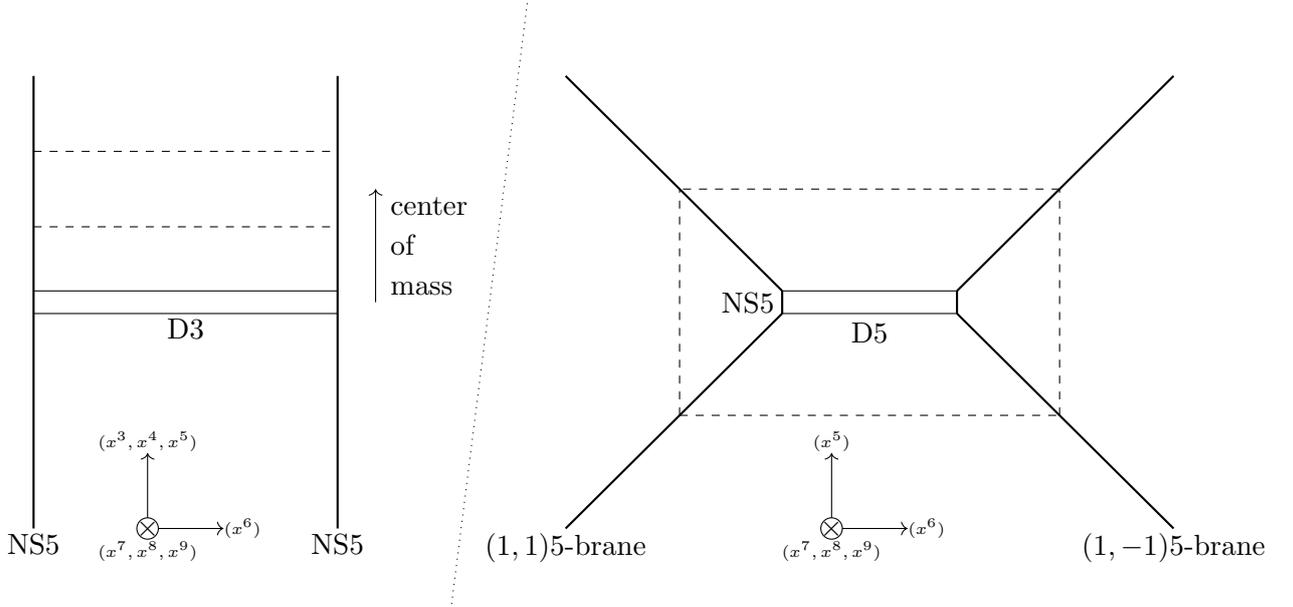
\begin{figure}[t]
    \centering
    \begin{tikzpicture}
    \tikzstyle{gauge1} = [inner sep=0.8mm,draw=none,fill=white,minimum size=0.35mm,circle, draw];
    \draw[thick] (0,0)--(0,6) (4,0)--(4,6);
    \node at (0,-0.2) {NS5};
    \node at (4,-0.2) {NS5};
    \draw[] (0,3.15)--(4,3.15) (0,2.85)--(4,2.85);
    \node at (2,2.65) {D3};
    \draw[dashed] (0,4)--(4,4) (0,5)--(4,5);
    \draw[->] (4.5,3)--(4.5,4.5);
    \node[text width=3] at (4.75,3.75) {center of mass};
    \draw[->] (1.5,0)--(2.5,0);
    \node at (2.75,0) {\tiny $(x^6)$};
    \draw[->] (1.5,0)--(1.5,1);
    \node at (1.5,1.15) {\tiny $(x^3,x^4,x^5)$};
    \node[inner sep=1mm, fill=white, circle, draw] at (1.5,0) {};
    \node at (1.5,0) {$\times$};
    \node at (1.5,-0.3) {\tiny $(x^7,x^8,x^9)$};
    
    \draw[thick] (7,0)--(9.85,2.85) (7,6)--(9.85,3.15) (9.85,2.85)--(9.85,3.15);
    \draw (9.85,2.85)--(12.15,2.85) (9.85,3.15)--(12.15,3.15);
    \draw[thick] (12.15,3.15)--(12.15,2.85) (12.15,3.15)--(15,6) (12.15,2.85)--(15,0);
    \draw[dashed] (8.5,1.5)--(8.5,4.5) (8.5,1.5)--(13.5,1.5) (13.5,1.5)--(13.5,4.5) (13.5,4.5)--(8.5,4.5);
    \node at (9.4,3) {NS5};
    \node at (11,2.6) {D5};
    \node at (7,-0.25) {$(1,1)5$-brane};
    \node at (15,-0.25) {$(1,-1)5$-brane};
    
    \draw[->] (10.5,0)--(11.5,0);
    \node at (11.75,0) {\tiny $(x^6)$};
    \draw[->] (10.5,0)--(10.5,1);
    \node at (10.5,1.15) {\tiny $(x^5)$};
    \node[inner sep=1mm, fill=white, circle, draw] at (10.5,0) {};
    \node at (10.5,0) {$\times$};
    \node at (10.5,-0.3) {\tiny $(x^7,x^8,x^9)$};
    
    \draw[dotted] (5.5,-1)--(6.5,7);
    \end{tikzpicture}
    \caption{Depicted are Type IIB brane configurations with supersymmetric gauge theories living on the lightest branes. A different point in their moduli space is depicted with a dashed line. The 3d $\mathcal{N}=4$ gauge theory living on the two D3 branes has 2 (quaternionic) Coulomb branch moduli, while the 5d $\mathcal{N}=1$ gauge theory living on the D5 branes has only one (real) Coulomb branch modulus, as the center of mass is fixed. Hence the 3d theory has a $\mathrm{U}(2)$ gauge group at the origin of the moduli space, while the 5d theory has a $\mathrm{SU}(2)$ gauge group. Both have the Weyl group $S_2=\mathbb{Z}_2$. Decoupling the center of mass of the D3 branes leads to a centerless $\mathrm{U}(2)/\mathrm{U}(1)=\mathrm{SU}(2)/\mathbb{Z}_2$ gauge group. A similar argument holds for 4d $\mathcal{N}=2$ gauge theories living on D4 branes suspended between NS5 branes in Type IIA, which is a T-dual configuration of the above. Because of the logarithmic bending of the NS5 branes the center of mass is fixed and the gauge group is $\mathrm{SU}(2)$. Moving the center of mass in the 4d or 5d case corresponds to changing the asymptotic behaviour \cite{Karch:1998yv}. The moduli space of the gauge theory corresponds to the brane motions which keep the asymptotic form of the heavy branes intact \cite{Karch:1998uy}. }
    \label{SUvsU}
\end{figure}

Now that we have introduced the problem that we address, let us introduce the tools that we use, namely webs of five-branes in Type IIB string theory \cite{Aharony:1997ju,Aharony:1997bh,DeWolfe:1999hj,Benini:2009gi,Bergman:2013aca,Cabrera:2018jxt}. $(p,q)5$-branes are bound states of D5 and NS5 branes which can be obtained from the D5 branes of Type IIB string theory by suitable action of the $\mathrm{SL}(2,\mathbb{Z})$ duality group. The brane webs are two-dimensional slices of supersymmetry preserving configurations of various $(p,q)$5-branes, whose common world-volume is 5-dimensional, thus allowing to construct $\mathcal{N}=1$ supersymmetric gauge theories in 5d. Crucially, the gauge theory on the world-volume of $N_c$ parallel D5 branes suspended between the appropriate $(p,q)$ branes has gauge group $\mathrm{SU}(N_c)$, and not $\mathrm{U}(N_c)$, due to the bending of the branes \cite{Witten:1997sc}, see Figure \ref{SUvsU}. This is to be contrasted with 3d setups, in which there is no asymptotic bending, allowing the $\mathrm{U}(1)$ factor in the gauge group to contribute to the low-energy dynamics.

As is apparent in this paper, the Higgs branch of SQCD seen as an algebraic variety appears naturally as a union of hyper-K\"ahler cones, or symplectic singularities \cite{beauville2000symplectic,fu2006survey}. In order to clearly distinguish it from the \hyperlink{HiggsScheme}{Higgs scheme}, of which it is the reduced part, we call it the \emph{\hyperlink{HiggsVar}{Higgs variety}} (see Figure \ref{figSummary}; we explain why this distinction is important in the next paragraph). In the last few years, a rich family of symplectic singularities have been described using combinatorial data encoded in quivers. This was initially confined to the realm of 3d $\mathcal{N}=4$ quiver theories, in which case both the Higgs and the (quantum corrected) Coulomb branches are hyper-K\"ahler cones. In particular, the Hilbert series of the Coulomb branch of such theories can be evaluated using the \emph{monopole formula} \cite{Cremonesi:2013lqa}, which counts three-dimensional dressed monopole operators. However, it should be emphasised that the monopole formula can be used in theories with spacetime dimension $d \geq 3$ to count dressed codimension-3 BPS operators, as was recently demonstrated in  \cite{DelZotto:2014kka,Ferlito:2017xdq,Cabrera:2018jxt,Hanany:2018vph,Hanany:2018uhm,Cabrera:2019izd}. In particular, the appearance of a symplectic singularity corresponding to the space of such operators does not necessarily imply a relation to 3d physics. As a consequence, we use the concept of \emph{magnetic quivers}, already introduced in \cite{Hanany:1996ie} and used in \cite{DelZotto:2014kka,Ferlito:2017xdq,Cabrera:2018jxt,Hanany:2018vph,Hanany:2018uhm,Cabrera:2019izd} as an abstract way of describing symplectic singularities. The connection between the quiver and the associated symplectic singularities (which can be seen as the Coulomb branch of the corresponding 3d $\mathcal{N}=4$ theory) have been studied from the mathematical point of view in a series of papers \cite{Nakajima:2015txa,Braverman:2016wma,Nakajima:2017bdt}, and it turns out that these singularities are always irreducible and normal. On the other hand, the Higgs branches that we consider are in general made up of several cones \cite{Seiberg:1994aj,Gaiotto:2008ak,Ferlito:2016grh}. As a consequence, they cannot be described by a unique quiver, but instead to each irreducible component, and to each intersection thereof, is associated a quiver characterising the components and their intersections. The magnetic quivers are schematically computed as follows \cite{Cabrera:2018jxt}: 
\begin{itemize}
    \item First one finds all the maximal decompositions of the brane webs into consistent subwebs. 
    \item Then to each decomposition one associates a magnetic quiver in which the ranks of the gauge nodes is given by the number of coinciding identical subwebs, and the number of links between nodes is computed as the stable intersection number of the corresponding tropical curves.  
    \item The process is repeated for non-maximal decompositions of brane webs to obtain the magnetic quivers associated to intersections of several cones. 
\end{itemize}
See appendix \ref{Appendixmag} for more details. Once the magnetic quivers are known, not only can the Hilbert series be evaluated using the monopole formula, but since the magnetic quivers involve only unitary nodes, it can be refined through fugacities for the (maximal torus of the) global symmetry group. Upon this refinement, after a suitable fugacity map, the coefficients of the Hilbert series become characters of the global symmetry algebra, which can be compactly encapsulated into a \emph{highest-weight generating function} (HWG) \cite{Hanany:2014dia,Hanany:2016djz}. This formalism is quickly reviewed below in Section \ref{sectionUgroup}. 

We now come to an important subtle point. The Higgs ring as defined above contains non-trivial nilpotent elements. This is most easily seen in the case where the number of flavors is $N_f=1$: there the meson is a scalar $M$ and the F-term equations imply that $M^2=0$. The Hilbert series computation (\ref{generalFormulaHBHS}) is sensitive to such nilpotent operators, but in the \hyperlink{HiggsVar}{Higgs variety}, this equation implies $M=0$. More generally, the geometric locus of points defined by a set of algebraic equations has a coordinate ring in which there are no nilpotent operators (this is the classical Nullstellensatz of commutative algebra, see Appendix \ref{AppendixAlgebra}). This leads to a crucial distinction between two concepts:
\hypertarget{HiggsVar}{\begin{DefHiggsVar}
    \emph{The Higgs branch seen as an algebraic variety, i.e. the zero locus of the relations defining the Higgs ring. We call its coordinate ring the} reduced Higgs ring\emph{. It is defined by the} radical \emph{of the mesonic and baryonic structure equations and contains no nilpotent operators.}
\end{DefHiggsVar}}
\hypertarget{HiggsScheme}{\begin{DefHiggsScheme}
    \emph{The Higgs branch seen as the object whose coordinate ring is the full \hyperlink{HiggsRing}{Higgs ring}, with Hilbert series given by (\ref{generalFormulaHBHS}).}
\end{DefHiggsScheme}}
The brane web and magnetic quiver description, being intrinsically geometric, is only sensitive to the algebraic variety structure of the Higgs branch. This is summarised in Figure \ref{figSummary}. 

\begin{table}[t]
	\begin{adjustbox}{center}
	\begin{tabular}{|c|c|c|c|}
		\hline
	Range of $N_f$ & Baryonic Branch & Intersection & Mesonic Branch\\ \hline
	$N_f>2N_c+2$ & Fig.\ref{complast} & - & - \\ \hline
$N_f = 2N_c+2$  & Fig.\ref{Nf2Ncpl2} &- &-\\ \hline	
$N_f = 2N_c+1$  &Fig.\ref{Nf2Ncpl1} &- &-\\ \hline
$N_f=2N_c$ &Fig.\ref{Nf2Nc} &- &-\\ \hline
$N_f=2N_c-1$ &Fig.\ref{Nf2Ncminus1} &-&-\\ \hline
$N_c < N_f < 2N_c-1$&Fig.\ref{NfinbetweenBary} & Fig.\ref{NfinbetweenIntersection}&\begin{tabular}{c} Fig.\ref{NfinbetweenME} (even)\\ Fig.\ref{NfinbetweenO} (odd) \end{tabular}\\ \hline
$N_f=N_c$ & Fig.\ref{nfequalncB}  & Fig.\ref{intersectionNfeqNc} & \begin{tabular}{c}
Fig.\ref{compfirst} (even) \\ Fig.\ref{NfsmallerNcOdd} (odd) 
\end{tabular}\\ \hline
$N_f < N_c$ &- & - &\begin{tabular}{c}
Fig.\ref{compfirst} (even) \\ Fig.\ref{NfsmallerNcOdd} (odd) 
\end{tabular}\\ \hline
	\end{tabular}
	\end{adjustbox}
	\caption{This Table gives for each value of $N_f$ and $N_c$ the figure in which one can find the grid diagrams, brane webs and magnetic quivers for each component of the Higgs variety. The empty squares $-$ mean that the branch is empty ($N_f < N_c$) or is included in another branch ($N_f \geq 2N_c-1$). See also Table \ref{finitebranches} for a schematic representation of the magnetic quivers. }
	\label{tableSummary}
	\end{table}
	
We now turn to a description of the computational complexity of the various steps that appear in Figure \ref{figSummary}. In the upper-left part, most of the steps involve a Gr\"obner basis computation: this is required to evaluate the Hilbert series of the $\mathcal{F}^{\flat}$ moduli space, to compute the gauge invariant operators and their relations,\footnote{In the case of SQCD, this is known explicitly, but the generic problem is hard. } and finally to extract the radical of the \hyperlink{HiggsRing}{Higgs ring} (or equivalently to perform the geometric reduction of the \hyperlink{HiggsScheme}{Higgs scheme}). The computational complexity of such algorithms is very high. To give an order of magnitude, on a standard computer, computations for SQCD with $N_f,N_c > 5$ are unfeasible. On the other hand, the steps that belong to the lower-right part of Figure \ref{figSummary} are easily implemented for generic range of parameters, as demonstrated in this paper on the SQCD example with generic $N_f$ and $N_c$. The computation of the magnetic quivers from the brane webs can be done by hand. The evaluation of the corresponding Hilbert series using the monopole formula is more time consuming. However, using HWG techniques, full explicit results can be obtained for any $N_f$ and $N_c$. This tremendous advantage of magnetic quivers and HWG over standard algebraic algorithms in terms of computational complexity constitutes one of the main motivations for this work. 
 
The paper is organized as follows. In Section \ref{sectionHBSQCD}, we review the construction of the \hyperlink{HiggsRing}{Higgs ring} as a hyper-K\"ahler quotient, beginning with the simple case where the gauge group is $\mathrm{U}(N_c)$, before moving to the more interesting case where the gauge group is $\mathrm{SU}(N_c)$. We also present the solution in terms of mesons and baryons, and compute Hilbert series in a collection of low rank cases. In Section \ref{section3}, we introduce the brane web method and apply it systematically to compute all the relevant magnetic quivers for $\mathrm{SU}(N_c)$ SQCD with $N_f$ flavors at finite coupling. The results of this case by case analysis is presented in a collection of figures as summarized in Table \ref{tableSummary}. This can be considered as an extension of the computations of \cite{Cabrera:2018jxt}, which were performed in the infinite gauge coupling regime. In Section \ref{sectionRadical}, we show that the brane web method indeed describes correctly the \hyperlink{HiggsVar}{Higgs variety}, using explicit computations of radicals. Finally in Section \ref{sectionDiscreteFactors} we show how the nilpotent elements can be accounted for in certain examples, and speculate about a general understanding of this point in the framework of the brane web method.

\section{\texorpdfstring{The Higgs Branch of 4d $\mathcal{N}=2$ SQCD}{The Higgs Branch of 4d N=2 SQCD}}
\label{sectionHBSQCD}

Our aim in this section is to study the Higgs branch of the four-dimensional $\mathcal{N}=2$ theory with gauge group $\mathrm{SU}(N_c)$ and $N_f$ hypermultiplets in the fundamental representation of $\mathrm{SU}(N_c)$. Before we do so, we look in subsection \ref{sectionUgroup} at the theory where the gauge group is $\mathrm{U}(N_c)$ instead of $\mathrm{SU}(N_c)$, which is easier to analyze. We then move to special unitary gauge group in section \ref{sectionSUgroup}.

\subsection{Unitary gauge group}
\label{sectionUgroup}

As a warm-up, we first consider the four-dimensional $\mathcal{N}=2$ theory with gauge group $\mathrm{U}(N_c)$ and $N_f$ hypermultiplets in the fundamental representation of $\mathrm{U}(N_c)$. The scalar components of these multiplets are denoted $Q$ and $\tilde{Q}$. Here $Q$ is a $N_c \times N_f$ matrix with components $Q^i_a$ ($a=1,\dots , N_c$ and $i=1,\dots,N_f$) and $\tilde{Q}$ is a $N_f \times N_c$ matrix with components $\tilde{Q}^a_i$, that transform in the obvious way. The superpotential\footnote{Note that in the superpotential the same notation is used for the scalar components and for the corresponding chiral superfields.} can be written $\mathcal{W} = \mathrm{Tr} \, \tilde{Q} \phi Q$, and the F-terms equations are 
\begin{equation}
    Q \tilde{Q}  = 0 \, . 
\end{equation}

The best way to characterize the Higgs branch is to describe it as an algebraic variety, i.e. giving some algebraic equations that define it. A coarser observable would be to give its Hilbert series, which is nothing but its graded dimension. The Hilbert series can be refined so as to encode the action of the global symmetry. Here it turns out that the full description as an algebraic variety is available. We introduce the meson matrix 
\begin{equation}
    M = \tilde{Q} Q \, ,  
\end{equation}
which is the only gauge invariant that can be constructed out of $Q$ and $\tilde{Q}$. 
Then the Higgs branch can be described by the following system of equations: 
\begin{equation}
\label{idealSU}
    M^2 = 0 \, , \qquad \mathrm{Tr}(M) = 0 \, ,  \qquad \mathrm{rank}(M) \leq N_c \, .
\end{equation}
When $N_f \leq 2N_c+1$, the inequality on the rank can be dropped. 
The Higgs branch Hilbert series can be found in all cases by combining the two following observations:  
\begin{itemize}
    \item When $N_f \geq 2N_c$, this is the theory $T_{\rho}[\mathrm{SU}(N_f)]$, where $\rho$ is the partition $(N_f-N_c,N_c)$. The Higgs branch of this theory is the closure of the nilpotent orbit\footnote{For a general introduction of nilpotent orbits in physics, see \cite{Cabrera:2016vvv}. The first appearance of these objects in the context of string theory is \cite{Bachas:2000dx}, to the best of our knowledge. } of $\mathfrak{sl}(N_f , \mathbb{C})$ associated to the partition $\rho^T = (2^{N_c},1^{N_f-2N_c})$. We have gathered in Appendix \ref{appendixNO} some useful definitions and properties of nilpotent orbits in relation with integer partitions. The Higgs branch has quaternionic dimension $N_c(N_f-N_c)$, and its Highest Weight Generating function (HWG) is given by \cite{Ferlito:2016grh}
    \begin{equation}
    \label{hwgSimpleU}
    PE \left[ \sum\limits_{i=1}^{N_c} \mu_i \mu_{N_f-i} t^{2i} \right] \, . 
\end{equation}
We recall that the HWG is a convenient way of packaging a Hilbert series which we have refined according to a given global symmetry group, here $\mathrm{SU}(N_f)$. The fugacities $\mu_1 , \dots , \mu_{N_f -1}$ are highest-weight fugacities: a term $\prod_{i=1}^{N_f-1} \mu_i^{r_i}$ in the HWG stands for the character of $\mathfrak{su}(N_f)$ with Dynkin labels $[r_1 , \dots , r_{N_f -1}]$ in the Hilbert series. 
    \item The Higgs branch depends only on the values of $\mathrm{min}(N_c,[N_f/2])$ and $N_f$. This means that for $N_f<2N_c$, we have 
    \begin{equation}
        H^U_{N_c,N_f} = H^U_{[N_f/2],N_f} \, . 
    \end{equation}
\end{itemize}
It is worth pointing out that the Higgs branch is in all cases the closure of a nilpotent orbit. In addition, it can not be written as a union of several cones.\footnote{There are closures of nilpotent orbits that are unions of cones, for instance the very even orbits of $D$-type groups $O(2n)$.} 
This property is lost when we consider theories with special unitary gauge groups (see section \ref{sectionSUgroup}).

In order to compute the Higgs branch Hilbert series, one can use the technique known as \emph{hyper-K\"ahler quotient}: one first computes the Hilbert series for the ring $\mathbb{C}[Q,\tilde{Q}]/\langle  \textrm{F-terms} \rangle$, weighted by fugacities of the gauge group, and then performs a gauge integration to project on to the gauge invariant operators. The first step presents no difficulty if the number of flavors is big enough: if $N_f \geq 2N_c-1$ then the ring $\mathbb{C}[Q,\tilde{Q}]/\langle  \textrm{F-terms} \rangle$ is a complete intersection
, which means that its unrefined Hilbert series can be written as $PE[2 N_c N_f t - N_c^2 t^2]$. The refined Hilbert series is obtained by replacing in this expression $2 N_c N_f$ by the character of the bifundamental representation of $\mathrm{U}(N_c) \times \mathrm{U}(N_f)$ plus its conjugate, and $N_c^2$ by the character of the adjoint representation of $\mathrm{U}(N_c)$. It is then straightforward to integrate over the $N_c$ fugacities of $\mathrm{U}(N_c)$. 

\begin{table}[t]
	\begin{adjustbox}{center}
	\begin{tabular}{|c|c|}
    \hline
    $N_f$ & Hilbert series $H^U_{1,N_f}$  \\ \hline
 1 & \begin{tabular}{l}1\end{tabular}   \\ \hline
 2 & \begin{tabular}{l}\noindent\(\dfrac{1+t^2}{(1-t)^2 (1+t)^2}\)\end{tabular}  \\ \hline
 3 & \begin{tabular}{l}\noindent\(\dfrac{1+4 t^2+t^4}{(1-t)^4 (1+t)^4}\)\end{tabular} \\ \hline
 4 & \begin{tabular}{l}\noindent\(\dfrac{\left(1+t^2\right) \left(1+8 t^2+t^4\right)}{(1-t)^6 (1+t)^6}\)\end{tabular}  \\ \hline
 5 & \begin{tabular}{l}\noindent\(\dfrac{1+16 t^2+36 t^4+16 t^6+t^8}{(1-t)^8 (1+t)^8}\)\end{tabular} \\ \hline
 6 & \begin{tabular}{l}\noindent\(\dfrac{\left(1+t^2\right) \left(1+24 t^2+76 t^4+24 t^6+t^8\right)}{(1-t)^{10}(1+t)^{10}}\)\end{tabular}\\ \hline
 7 & \begin{tabular}{l}\noindent\(\dfrac{1+36 t^2+225 t^4+400 t^6+225 t^8+36 t^{10}+t^{12}}{(1-t)^{12} (1+t)^{12}}\)\end{tabular}  \\ \hline 
    \end{tabular}
    \end{adjustbox}
    
        \vspace{0.5cm}  
        \begin{adjustbox}{center}
    \begin{tabular}{|c|c|}
    \hline
    $N_f$ &  Hilbert series $H^U_{2,N_f}$  \\ \hline
 1 & \begin{tabular}{l}1\end{tabular}  \\ \hline
 2 & \begin{tabular}{l}\noindent\(\dfrac{1+t^2}{(1-t)^2 (1+t)^2}\)\end{tabular} \\ \hline
 3 & \begin{tabular}{l}\noindent\(\dfrac{1+4 t^2+t^4}{(1-t)^4 (1+t)^4}\)\end{tabular} \\ \hline
 4 & \begin{tabular}{l}\noindent\(\dfrac{\left(1+t^2\right)^2 \left(1+5 t^2+t^4\right)}{(1-t)^8 (1+t)^8}\)\end{tabular} \\ \hline
 5 & \begin{tabular}{l}\noindent\(\dfrac{1+12 t^2+53 t^4+88 t^6+53 t^8+12 t^{10}+t^{12}}{(1-t)^{12} (1+t)^{12}}\)\end{tabular} \\ \hline
 6 & \begin{tabular}{l}\noindent\(\dfrac{\left(1+t^2\right)^2 \left(1+17 t^2+119 t^4+251 t^6+119 t^8+17 t^{10}+t^{12}\right)}{(1-t)^{16} (1+t)^{16}}\)\end{tabular} \\ \hline
 7 & \begin{tabular}{l}\noindent\(\dfrac{1+28 t^2+357 t^4+1952 t^6+5222 t^8+7224 t^{10}+5222 t^{12}+1952 t^{14}+357 t^{16}+28 t^{18}+t^{20}}{(1-t)^{20} (1+t)^{20}}\)\end{tabular}  \\\hline 
    \end{tabular}
    \end{adjustbox}
    \caption{Hilbert series for the $\mathrm{U}(1)$ and $\mathrm{U}(2)$ theories with $n$ flavors.}
    \label{tabUSQCD1}
\end{table}
\begin{table}[t]
    \begin{adjustbox}{center}
    \begin{tabular}{|c|c|}
    \hline
    $N_f$ &  Hilbert series $H^U_{3,N_f}$  \\ \hline
 1 & 1   \\ \hline
 2 & \begin{tabular}{l}\noindent\(\dfrac{1+t^2}{(1-t)^2 (1+t)^2}\)\end{tabular}\\ \hline
 3 & \begin{tabular}{l}\noindent\(\dfrac{1+4 t^2+t^4}{(1-t)^4 (1+t)^4}\)\end{tabular} \\ \hline
 4 & \begin{tabular}{l}\noindent\(\dfrac{\left(1+t^2\right)^2 \left(1+5 t^2+t^4\right)}{(1-t)^8 (1+t)^8}\)\end{tabular} \\ \hline
 5 & \begin{tabular}{l}\noindent\(\dfrac{1+12 t^2+53 t^4+88 t^6+53 t^8+12 t^{10}+t^{12}}{(1-t)^{12} (1+t)^{12}}\)\end{tabular}  \\ \hline
 6 & \begin{tabular}{l}\noindent\(\dfrac{\left(1+t^2\right)^3 \left(1+14 t^2+72 t^4+133 t^6+72 t^8+14 t^{10}+t^{12}\right)}{(1-t)^{18} (1+t)^{18}}\)\end{tabular}\\ \hline
 7 & \begin{tabular}{l}\noindent\(\dfrac{\left( \begin{array}{c}1+24 t^2+251 t^4+1472 t^6+5129 t^8+10808 t^{10}+13854 t^{12}\\
 +10808 t^{14} +5129 t^{16}+1472 t^{18}+251 t^{20}+24 t^{22}+t^{24}\end{array}\right)}{(1-t)^{24} (1+t)^{24}}\)\end{tabular} \\ \hline 
    \end{tabular}
    \end{adjustbox}
    \caption{Hilbert series for $\mathrm{U}(3)$ with $n$ flavors. For the $\mathrm{U}(N_c)$ theories with $N_c \geq 4$, the results for $H^U_{N_c,N_f}$ coincide with those obtained for $N_c=3$. }
    \label{tabUSQCD2}
\end{table}

On the other hand, if the number of flavors is small, the computation of the Hilbert series of $\mathbb{C}[Q,\tilde{Q}]/\langle  \textrm{F-terms} \rangle$ is more difficult. As an example, for $N_c=3$ and $N_f=4$, one finds for the unrefined Hilbert series
\begin{equation}
    \frac{1-9 t^2+36 t^4-64 t^6-120 t^7+393 t^8-208 t^9-273
   t^{10}+384 t^{11}-126 t^{12}-48 t^{13}+42 t^{14}-8
   t^{15}}{(1-t)^{24}} \, . 
\end{equation}
The refined expression is not very illuminating, so we do not reproduce it here, but it is necessary to compute it in order to integrate out the gauge fugacities. 

We gather in Tables \ref{tabUSQCD1} and \ref{tabUSQCD2} the unrefined Higgs branch Hilbert series for small values of $N_f$ and $N_c$. These results can be partially compared with the results presented in Tables 4 and 5 of \cite{Hanany:2016gbz}. Whenever comparison is possible, there is agreement.

\subsection{Special Unitary gauge group}
\label{sectionSUgroup}

We now turn to the four-dimensional $\mathcal{N}=2$ theory with gauge group $\mathrm{SU}(N_c)$ and $N_f$ hypermultiplets in the fundamental representation of $\mathrm{SU}(N_c)$. The superpotential has the same expression as in the previous subsection, but the matrix $\phi$ is now traceless, so that the F-term equations become 
\begin{equation}
    Q \tilde{Q} - \frac{1}{N_c}( \mathrm{Tr} \, Q \tilde{Q} ) \mathbf{1}_{N_c}  = 0 \, . 
\end{equation}
The gauge invariant operators are built from 
\begin{itemize}
    \item The $N_f^2$ mesons $M = \tilde{Q}  Q$ (in components, $M_j^i = \tilde{Q}^a_j Q^i_a$). We also introduce the related matrix (which is not traceless in general!)
    \begin{equation}
    \label{DefMprime}
        M' = M - \frac{1}{N_c} \mathrm{Tr}(M) \mathbf{1}_{N_f} \, . 
    \end{equation}
    The mesons have weight 2. They transform in the representation with highest weight $1+\mu_1 \mu_{N_f-1}$. Crucially, in addition to the adjoint representation, we also have the trace $\mathrm{Tr}(M)$ which is non-vanishing, as opposed to when the gauge group is $\mathrm{U}(N_c)$. 
    \item The $2 \binom{N_f}{N_c}$ baryons $B_I$ and $\tilde{B}^I$ for $I \subset \{1 , \dots , N_f\}$ a subset with $N_c$ elements. More precisely, for $I = \{i_1 , \dots , i_{N_c}\}$ with $i_1<\dots < i_{N_c}$, we have 
    \begin{equation}
        B^I = B^{i_1 \dots i_{N_c}} = \epsilon^{a_1  , \dots , a_{N_c}} Q^{i_1}_{a_1} \cdots Q^{i_{N_c}}_{a_{N_c}}
    \end{equation}
  and similarly for $\tilde{B}$, 
  \begin{equation}
       \tilde{B}_{I} = \tilde{B}_{i_1 \dots  i_{N_c}} = \tilde{Q}_{i_1}^{a_1} \cdots \tilde{Q}_{i_{N_c}}^{a_{N_c}} \epsilon_{a_1 \dots a_{N_c}} \, . 
  \end{equation}
The baryons and anti-baryons have weight $N_c$. They transform in the representations with highest weight $\mu_{N_c}$ and $\mu_{N_f-N_c}$ of $\mathrm{SU}(N_f)$, and with respective charges $1$ and $-1$ under the baryonic $\mathrm{U}(1) \subset \mathrm{U}(N_f)$. We also need the Hodge duals,\footnote{Note that $\epsilon_{i_1 \dots i_{N_f}}$ is not an invariant tensor of the symmetry group $\mathrm{U}(N_f)$. However we use it as a convenient notation, keeping in mind that in all cases the quantities involving $\epsilon_{i_1 \dots i_{N_f}}$ are set equal to $0$. } 
\begin{eqnarray}
(\star B)_{i_{N_c+1} \dots i_{N_f}} &=& \epsilon_{i_1 \dots i_{N_f}} B^{i_1 \dots i_{N_c}} \\
(\star \tilde{B})^{i_{N_c+1} \dots i_{N_f}} &=& \epsilon^{i_1 \dots i_{N_f}} \tilde{B}_{i_1 \dots i_{N_c}} 
\end{eqnarray}
\end{itemize}
The mesons and baryons satisfy the following \emph{complete} set of relations \cite{Argyres:1996eh,Gaiotto:2008nz,HiggsFuture}\footnote{The completeness of this set of relations will be discussed at length and in more general situations in a companion paper \cite{HiggsFuture}. }
\begin{enumerate}
\item \begin{equation}
(\star B) \tilde{B} = \star (M^{N_c}) \quad \textnormal{for }N_f \geq N_c
\label{rel1}
\end{equation}
\newline 
This is at weight $2N_c$ and transforms at baryon charge $0$. In components, the relation reads 
\begin{equation}
    \tilde{B}_{i_1 \dots  i_{N_c}} (\star B)_{i_{N_c+1} \dots i_{N_f}} = N_c! M^{j_1}_{i_1} \cdots M^{j_{N_c}}_{i_{N_c}} \epsilon_{j_1 \dots j_{N_c} i_{N_c+1} \dots i_{N_f}} \, . 
\end{equation}
\item \begin{equation}
    M \cdot (\star B)   = 0\textnormal{ and }(\star \tilde{B}) \cdot M = 0\textnormal{ for }N_f \geq N_c+1\,.
    \label{rel2}
    \end{equation} 
\newline
In components, 
\begin{equation}
    M_{j_1}^i (\star B)_{i j_{2} \dots j_{N_f-N_c}} = 0 \qquad \textrm{and} \qquad M_{i}^{j_1} (\star \tilde{B})^{i j_{2} \dots j_{N_f-N_c}} = 0 \, . 
\end{equation}
\item \begin{equation}
    M' \cdot B = \tilde{B} \cdot M' = 0\textnormal{ for }N_f \geq N_c\,
    \label{rel3}
    \end{equation}
\item \begin{equation}
    M \cdot M' = 0\,.
    \label{rel4}
\end{equation}
\item \begin{equation}
    B^{[i_1 i_2\dots  i_{N_c}} B^{j_1] j_2 \dots  j_{N_c}} = 0\textnormal{ and } \tilde{B}_{[i_1 i_2\dots  i_{N_c}}  \tilde{B}_{j_1] j_2 \dots  j_{N_c}} = 0\,.
    \label{rel5}
\end{equation}
\end{enumerate}

We want to compute the Higgs branch Hilbert series. Here again, the two equivalent methods presented in the previous section can be used: 
\begin{itemize}
    \item The evaluation of the Higgs branch Hilbert series of the variety $\mathbb{C}[Q,\tilde{Q}]/\langle \textrm{F-terms} \rangle$, followed by an integration over the gauge group to project on the gauge invariant sector ; 
    \item The evaluation of the Hilbert series of $\mathbb{C}[M,B,\tilde{B}]/\langle\textrm{Relations 1. 2. 3. 4. 5.}\rangle$
\end{itemize}
We have checked that both methods give the same results in all the explicit computations we could perform. This provides a check that the set of relations between mesons and baryons is indeed complete; a deeper analysis of this question will be presented in \cite{HiggsFuture}. The results of the computations are presented in Table \ref{tabSUSQCD}. Note that the Hilbert series can be refined, introducing fugacities for the global symmetry $\mathrm{U}(N_f)$. However, it is difficult to write them in a simple and compact form as in (\ref{hwgSimpleU}). This fact can be deduced from the observation that the numerator of the Hilbert series of Table \ref{tabSUSQCD} are not palindromic in general. Geometrically, this corresponds to the Higgs branch being made up of the union of several cones, which intersect along cones of lower dimension. In the next subsection, we explore how this geometric fact is reflected in the structure of the equations.

\begin{table}[t]
	\begin{adjustbox}{center}
	\begin{tabular}{|c|c|}
		\hline
	$N_f$    & Hilbert series $H^{SU}_{2,N_f}$\\ \hline
1  &\begin{tabular}{l}
$1+t^2$
	 \end{tabular} \\ \hline
2 &\begin{tabular}{l}
\noindent\(\dfrac{1+4t^2-t^4}{(1-t)^2(1+t)^2}\)	 
	 \end{tabular} \\ \hline
3 &\begin{tabular}{l}
\noindent\(\dfrac{\left(1+t^2\right) \left(1+8t^2+t^4\right)}{(1-t)^6 (1+t)^6}\)	 
	 \end{tabular} \\ \hline
4 &\begin{tabular}{l}
\noindent\(\dfrac{\left(1+t^2\right) \left(1 + 17 t^2 + 48 t^4 + 17 t^6 + t^8\right)}{(1-t)^{10} (1+t)^{10}}\)	 
	 \end{tabular} \\ \hline
	\end{tabular}
	\end{adjustbox}
 \vspace{0.5cm}	

	\begin{adjustbox}{center}
	\begin{tabular}{|c|c|}
		\hline
	$N_f$    & Hilbert series $H^{SU}_{3,N_f}$\\ \hline
1  &\begin{tabular}{l}
$1+t^2$
	 \end{tabular} \\ \hline
2 &\begin{tabular}{l}
\noindent\(\dfrac{\left(1+t^2\right) \left(1+t^2-2t^4+t^6\right)}{(1-t)^2 (1+t)^2}\)	 
	 \end{tabular} \\ \hline
3 &\begin{tabular}{l}
\noindent\(\dfrac{1 + t + 6 t^2 + 7 t^3 + 13 t^4 + 2 t^5 + 3 t^8 + 3 t^9 + 
 2 t^{10} - t^{11} - t^{12}}{(1-t)^4(1+t)^4(1+t+t^2)}\)	 
	 \end{tabular} \\ \hline
4 &\begin{tabular}{l}
\noindent\(\dfrac{\left( \begin{array}{c} 1 + 4 t + 18 t^2 + 56 t^3 + 151 t^4 + 320 t^5 + 581 t^6 + 856 t^7 + 
 1044 t^8 + 1012 t^9 \\+ 790 t^{10} + 460 t^{11} + 177 t^{12} + 4 t^{13} - 
 46 t^{14} - 36 t^{15} - 15 t^{16} - 4 t^{17} - t^{18}\end{array} \right)}{(1-t)^8 (1+t)^8 \left(1+t+t^2\right)^4}\)	 
	 \end{tabular} \\ \hline
	\end{tabular}
	\end{adjustbox}
	
 \vspace{0.5cm}

	\begin{adjustbox}{center}
	\begin{tabular}{|c|c|}
		\hline
	$N_f$    & Hilbert series $H^{SU}_{4,N_f}$\\ \hline
1  &\begin{tabular}{l}
$1+t^2$
	 \end{tabular} \\ \hline
2 &\begin{tabular}{l}
\noindent\(\dfrac{\left(1+t^2\right) \left(1+t^2-2t^4+t^6\right)}{(1-t)^2 (1+t)^2}\)	 
	 \end{tabular} \\ \hline
3 &\begin{tabular}{l}
\noindent\(\dfrac{\left(1+t^2\right) \left(1 + 4 t^2 + 2 t^4 - 4 t^6 + 6 t^8 - 4 t^{10} + t^{12}\right)}{(1-t)^4 (1+t)^4}\)	 
	 \end{tabular} \\ \hline
4 &\begin{tabular}{l}
\noindent\(\dfrac{\left( \begin{array}{c}1 + 9 t^2 + 30 t^4 + 15 t^6 + 15 t^8 - 42 t^{10} + 23 t^{12}\\ - 3 t^{14} +17 t^{16} - 14 t^{18} + 6 t^{20} - t^{22} \end{array}\right) }{(1-t)^8
   (1+t)^8 \left(1+t^2\right)}\)	 
	 \end{tabular} \\ \hline
	\end{tabular}
	\end{adjustbox}
	
 \vspace{0.5cm}

	\begin{adjustbox}{center}
	\begin{tabular}{|c|c|}
		\hline
	$N_f$    & Hilbert series $H^{SU}_{5,N_f}$\\ \hline
1  &\begin{tabular}{l}
$1+t^2$
	 \end{tabular} \\ \hline
2 &\begin{tabular}{l}
\noindent\(\dfrac{\left(1+t^2\right) \left(1+t^2-2t^4+t^6\right)}{(1-t)^2 (1+t)^2}\)	 
	 \end{tabular} \\ \hline
3 &\begin{tabular}{l}
\noindent\(\dfrac{\left(1+t^2\right) \left(1 + 4 t^2 + 2 t^4 - 4 t^6 + 6 t^8 - 4 t^{10} + t^{12}\right)}{(1-t)^4 (1+t)^4}\)	 
	 \end{tabular} \\ \hline
4 &\begin{tabular}{l}
\noindent \( \dfrac{ \left( \begin{array}{c}(1+t^2) (1 + 7 t^2 + 13 t^4 - 6 t^6 - 16 t^8 + 35 t^{10}\\ - 55 t^{12} + 70 t^{14} -  56 t^{16} + 28 t^{18} - 8 t^{20} + t^{22})\end{array}\right) 
}{(1-t)^8(1+t)^8}  	 \)
	 \end{tabular} \\ \hline
	\end{tabular}
	\end{adjustbox}
	
		\caption{Hilbert series $H^{SU}_{N_c,N_f}$ for $\mathrm{SU}(N_c)$ SQCD with $N_f$ flavors with $2 \leq N_c \leq 5$ and $1 \leq N_f \leq 4$, computed using the hyper-K\"ahler quotient. }
		\label{tabSUSQCD}
\end{table}

\subsection{Decomposition in cones}
\label{sectionDecompositionPrimary}

Using the classical dictionary between algebra (represented here by the ideal generated by the equations of section \ref{sectionSUgroup}) and geometry (the Higgs branch), we are led to use the concept of primary decomposition. Roughly speaking, if an algebraic variety is made of a union of a collection of algebraic varieties, then the associated ideal is the intersection of the corresponding ideals. So one way to see algebraically the various cones of the Higgs branch is to disentangle the equations of section \ref{sectionSUgroup} and to decompose the ideal as an intersection of ``simpler" ideals, namely \emph{primary} ideals. Such a decomposition is known to exist as per the Lasker-Noether theorem. We refer the reader to Appendix \ref{AppendixAlgebra} for details and references on these topics from commutative algebra. Suffice it to say here that this primary decomposition can be obtained algebraically, though the algorithm is extremely time consuming.

Let us consider as an example the case $N_c=2$, $N_f=2$. In this case, the gauge invariant operators are 
\begin{equation}
    M := \begin{pmatrix} a & b \\ c & d \end{pmatrix} \, , \qquad B \, , \qquad \tilde{B} \, . 
\end{equation}
The primary decomposition tells us that the ideal generated by the equations of section \ref{sectionSUgroup} is the intersection of two ideals: 
\begin{itemize}
    \item The first one corresponds to $B=\tilde{B}=0$, $\mathrm{Tr}(M)=0$ and $M^2=0$. This is a purely mesonic branch, and the geometry is given by $a^2+bc=0$. This defines the hyper-Kähler cone $\mathbb{C}^2/\mathbb{Z}_2$, with Hilbert series equal to $\mathrm{PE}[3t^2-t^4]$. Equivalently, this is the closure of the non-trivial nilpotent orbit of $\mathfrak{sl}(2,\mathbb{C})$. 
    \item The second one corresponds to $b=c=a-d=0$ and $B \tilde{B}= a^2$. In other words, the meson matrix is a pure trace, so it can be treated as a scalar. So we have three scalars, $\mathrm{Tr}(M)$, $B$ and $\tilde{B}$ satisfying again the $\mathbb{C}^2/\mathbb{Z}_2$ equation. The Hilbert series for this cone is again $\mathrm{PE}[3t^2-t^4]$. This branch is baryonic, as it involves the baryons.\footnote{The general definition of what we call the baryonic branch is given in the next section. } 
\end{itemize}
It is clear that the intersection of the two cones is trivial. Therefore, we obtain the decomposition 
\begin{equation}
    H^{SU}_{2,2} =\frac{1+t^2}{(1-t^2)^2} + \frac{1+t^2}{(1-t^2)^2}  -1 \,. 
\end{equation}

In the following sections, we show how such a decomposition can be obtained with much less effort from brane webs.

\section{Brane Web Method}
\label{section3}

In this chapter we are using the tropical brane web method from \cite{Cabrera:2018jxt}. As this method uses the decomposition of brane webs, we recapitulate some basic results \cite{Aharony:1997bh,DeWolfe:1999hj,Benini:2009gi,Bergman:2013aca}. We consider a set up of 5-branes and 7-branes in Type \Romannum{2}B string theory on Minkowski space (Table \ref{setup}).

\begin{figure}[t]
\centering
\scalebox{0.7}{\begin{tikzpicture}
	\begin{pgfonlayer}{nodelayer}
		\node [style=gauge1] (0) at (-2.975, 2) {};
		\node [style=gauge1] (1) at (3.025, -2) {};
		\node [style=gauge1] (2) at (1.025, 2) {};
		\node [style=gauge1] (3) at (-0.975, -2) {};
		\node [style=none] (5) at (3.025, 0) {};
		\node [style=none] (6) at (1.025, 0.25) {};
		\node [style=none] (9) at (-0.975, -0.25) {};
		\node [style=gauge1] (10) at (3.025, 0) {};
		\node [style=gauge1] (11) at (-2.975, 0) {};
		\node [style=none] (12) at (3.025, -0.25) {};
		\node [style=none] (13) at (-2.975, 0.25) {};
		\node [style=gauge1] (14) at (5.025, 0) {};
		\node [style=gauge1] (15) at (-4.975, 0) {};
		\node [style=none] (16) at (1, 2.5) {[0,1]};
		\node [style=none] (17) at (4, 0.25) {D5};
		\node [style=none] (18) at (5.75, 0) {[1,0]};
		\node [style=none] (19) at (3.5, -2.5) {[1,1]};
		\node [style=none] (20) at (-4, -2) {};
		\node [style=none] (21) at (-4, -4) {};
		\node [style=none] (22) at (-2, -4) {};
		\node [style=gauge1] (23) at (-4, -4) {};
		\node [style=none] (24) at (-4, -1.575) {$x^5$};
		\node [style=none] (25) at (-1.575, -4) {$x^6$};
		\node [style=none] (26) at (-4, -4.6) {$x^7,x^8,x^9$};
		\node [style=none] (27) at (1.5, 1) {NS5};
		\node [style=none] (28) at (-4, -4) {{\Huge $\times$}};
	\end{pgfonlayer}
	\begin{pgfonlayer}{edgelayer}
		\draw (0) to (9.center);
		\draw (9.center) to (3);
		\draw (9.center) to (12.center);
		\draw (2) to (6.center);
		\draw (13.center) to (6.center);
		\draw (6.center) to (1);
		\draw (11) to (10);
		\draw (10) to (14);
		\draw (15) to (11);
		\draw [style=->] (21.center) to (20.center);
		\draw [style=->] (21.center) to (22.center);
	\end{pgfonlayer}
\end{tikzpicture}

}
\caption{The brane web realisation of $\mathrm{SU}(3)$ SQCD at finite coupling with $N_f=4$ massless flavours at the origin of its moduli space. Horizontal lines correspond to D5 branes, vertical lines correspond to NS5 branes, lines at an angle $\tan(\alpha)=\frac{p}{q}$ from the $x^5$ axis correspond to $(p,q)5$-branes and circles correspond to $[p,q]7$-branes. Note that $(p,q)5$-branes end on $[p,q]7$-branes. The parallel D5 branes are supposed to coincide in space and are drawn slightly appart for clarity.}
\label{SU(3)4sing}
\end{figure}
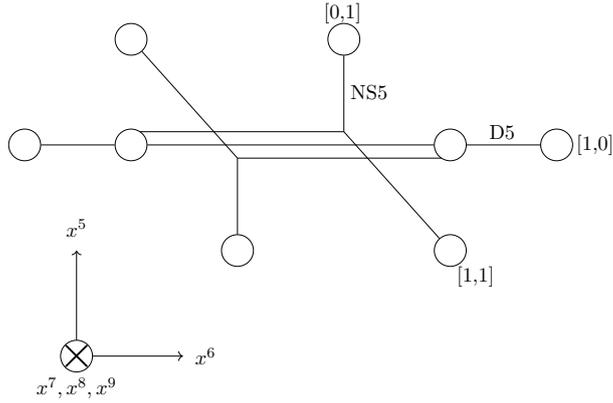

\begin{table}[h]
\begin{center}
\begin{tabular}{c|c|c|c|c|c|c|c|c|c|c}
& $x^0$ & $x^1$ & $x^2$ & $x^3$ & $x^4$ & $x^5$ & $x^6$ & $x^7$ & $x^8$ & $x^9$\\
\hline
NS5 & $-$ & $-$ & $-$ & $-$ & $-$ & $-$ & & & &  \\
\hline
D5 & $-$ & $-$ & $-$ & $-$ & $-$ & & $-$ & & & \\
\hline
$(p,q)5$-brane & $-$ & $-$ & $-$ & $-$ & $-$ & \multicolumn{2}{c|}{angle $\alpha$} & & & \\
\hline
$[p,q]7$-Brane & $-$ & $-$ & $-$ & $-$ & $-$ & & & $-$ & $-$ & $-$ \\
\end{tabular}
\caption{Brane webs setup, where $-$ mark the spacetime directions spanned by the various branes and $\tan(\alpha)=q \tau_2/(p+q\tau_1)$ where the axiodilaton is $\tau = \tau_1 + i \tau_2$. In this paper all the brane webs are drawn for the value $\tau = i$, so that $\tan(\alpha)=q/p$. }
\label{setup}
\end{center}
\end{table}
We can realise 5d $\mathcal{N}$=1 gauge theories as effective field theories on the 5-branes. The moduli of the gauge theory correspond to positions of 5-branes. The Chern-Simons (CS) level of the 5d ${\cal N}=1$ field theory on the D5 branes can be constructed by changing the slopes of the connected branes. For our computations it is irrelevant, as it has no impact on the classical Higgs branch which is equal to the classical Higgs branch of 4d ${\cal N}=2$, where the notion of a CS level does not exist. Therefore in the following we pick the most convenient CS level to us. We give an example of $\mathrm{SU}(3)$ SQCD with $N_f=4$ massless flavours and CS level $0$, whose brane web realisation at finite gauge coupling is depicted in Figure \ref{SU(3)4sing}.

\begin{figure}[t]
  \centering
\begin{tikzpicture}
	\begin{pgfonlayer}{nodelayer}
		\node [style=none] (0) at (-4, 5) {};
		\node [style=none] (1) at (0, 0) {};
		\node [style=none] (2) at (2, 5) {};
		\node [style=none] (3) at (3.75, 3.75) {};
		\node [style=none] (4) at (3.75, 3.75) {};
		\node [style=none] (5) at (5.75, 1.5) {};
		\node [style=none] (6) at (3.25, 1.75) {$\mathcal{M}_\textrm{M}$};
		\node [style=none] (7) at (1.75, 2.75) {$\mathcal{M}_\textrm{B}$};
		\node [style=none] (8) at (-4, 5.5) {$\mathcal{M}_C$};
		\node [style=none] (9) at (2.75, 6.25) {};
		\node [style=none] (10) at (7.25, 2) {};
		\node [style=none] (11) at (5.75, 4.5) {$\mathcal{M}_H$};
	\end{pgfonlayer}
	\begin{pgfonlayer}{edgelayer}
		\draw (0.center) to (1.center);
		\draw (1.center) to (2.center);
		\draw [bend left=105, looseness=1.25] (2.center) to (3.center);
		\draw [bend right=75, looseness=0.75] (2.center) to (3.center);
		\draw [style=black] (1.center) to (5.center);
		\draw [style=black, in=30, out=60, looseness=1.25] (4.center) to (5.center);
		\draw [style=black, bend right=60] (4.center) to (5.center);
		\draw [style=brace] (9.center) to (10.center);
		\draw [style=thickline] (4.center) to (1.center);
	\end{pgfonlayer}
\end{tikzpicture}

\caption{Schematic picture of the Coulomb Branch, $\mathcal{M}_C$, and Higgs Branch, $\mathcal{M}_H$, of SCQD with 8 supercharges at finite coupling. The Higgs Branch is a union of two cones, the \textit{mesonic} cone, $\mathcal{M}_\textrm{M}$, and the \textit{baryonic} cone, $\mathcal{M}_\textrm{B}$, with non-trivial intersection (line in bold).}
\label{branches}
\end{figure}
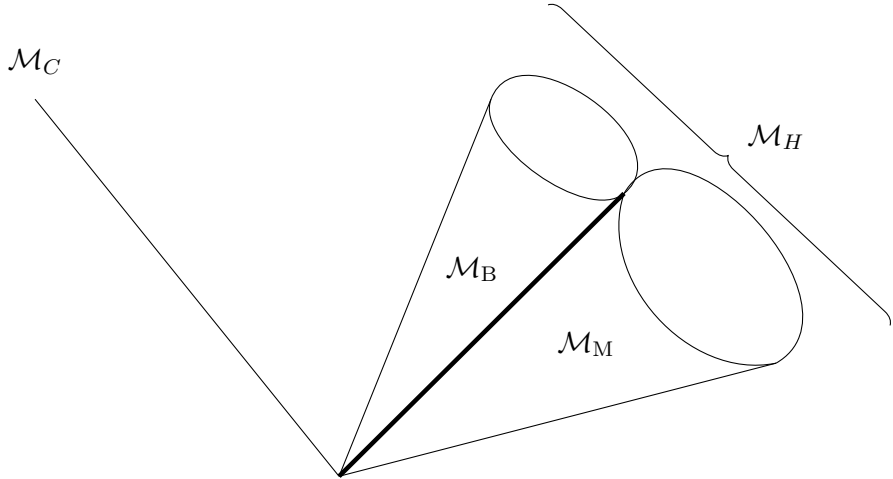

The moduli space of 5d $\mathcal{N}=1$ SQCD contains a Coulomb Branch, $\mathcal{M}_C$, and a Higgs Branch, $\mathcal{M}_H$. The Higgs Branch can be a union of two cones\footnote{For the case of SQCD with $\mathrm{SU}(N)$ gauge group at finite coupling.}, the \textit{mesonic} cone, $\mathcal{M}_\textrm{M}$, and the \textit{baryonic} cone, $\mathcal{M}_\textrm{B}$, with non-trivial intersection (Figure \ref{branches}). We define the baryonic branch as the irreducible component of the Higgs branch which contains a $\mathrm{U}(1)_\textrm{B}$ factor in its global symmetry. 

\begin{figure}[t]
    \begin{subfigure}[b]{0.45\textwidth}
        \centering
   \scalebox{0.6}{   \begin{tikzpicture}
	\begin{pgfonlayer}{nodelayer}
		\node [style=gauge1] (0) at (-4, 2) {};
		\node [style=gauge1] (1) at (2, -2) {};
		\node [style=gauge1] (2) at (0, 2) {};
		\node [style=gauge1] (3) at (-2, -2) {};
		\node [style=none] (5) at (2, 0) {};
		\node [style=none] (6) at (0, 0.25) {};
		\node [style=none] (9) at (-2, -0.25) {};
		\node [style=gauge1] (10) at (2, 0) {};
		\node [style=gauge1] (11) at (-4, 0) {};
		\node [style=none] (12) at (2, -0.25) {};
		\node [style=none] (13) at (-4, 0.25) {};
		\node [style=gauge1] (14) at (4, 0) {};
		\node [style=gauge1] (15) at (-6, 0) {};
	\end{pgfonlayer}
	\begin{pgfonlayer}{edgelayer}
		\draw [style=cyane] (0) to (9.center);
		\draw [style=cyane] (9.center) to (3);
		\draw [style=cyane] (9.center) to (12.center);
		\draw [style=rede] (2) to (6.center);
		\draw [style=rede] (13.center) to (6.center);
		\draw [style=rede] (6.center) to (1);
		\draw [style=olivee] (11) to (10);
		\draw [style=bluee] (10) to (14);
		\draw [style=bluee] (15) to (11);
	\end{pgfonlayer}
\end{tikzpicture}}\\
        \centering
  \scalebox{0.6}{    \begin{tikzpicture}
	\begin{pgfonlayer}{nodelayer}
		\node [style=none] (0) at (-4, 5) {};
		\node [style=none] (1) at (0, 0) {};
		\node [style=none] (2) at (2, 5) {};
		\node [style=none] (3) at (3.75, 3.75) {};
		\node [style=none] (4) at (3.75, 3.75) {};
		\node [style=none] (5) at (5.75, 1.5) {};
		\node [style=none] (6) at (3.25, 1.75) {\Large{$\mathcal{M}_\textrm{M}$}};
		\node [style=none] (7) at (1.75, 2.75) {\Large{$\mathcal{M}_\textrm{B}$}};
		\node [style=none] (8) at (-4, 5.5) {\Large{$\mathcal{M}_C$}};
		\node [style=none] (9) at (2.75, 6.25) {};
		\node [style=none] (10) at (7.25, 2) {};
		\node [style=none] (11) at (5.75, 4.5) {\Large{$\mathcal{M}_H$}};
		\node [style=smalldot] (12) at (2, 3.5) {};
		\node [style=none] (13) at (0, 5.25) {};
		\node [style=none] (14) at (0, 5.45) {\Large{We are here}};
	\end{pgfonlayer}
	\begin{pgfonlayer}{edgelayer}
		\draw (0.center) to (1.center);
		\draw (1.center) to (2.center);
		\draw [bend left=105, looseness=1.25] (2.center) to (3.center);
		\draw [bend right=75, looseness=0.75] (2.center) to (3.center);
		\draw [style=black] (1.center) to (5.center);
		\draw [style=black, in=30, out=60, looseness=1.25] (4.center) to (5.center);
		\draw [style=black, bend right=60] (4.center) to (5.center);
		\draw [style=brace] (9.center) to (10.center);
		\draw [style=thickline] (4.center) to (1.center);
		\draw [style=->, bend right] (13.center) to (12);
	\end{pgfonlayer}
\end{tikzpicture}

}
		\caption{The baryonic branch is indicated in the brane web. Coloured branes are assumed to be on different $(x^7,x^8,x^9)$ positions in space.}
		\label{SU(3)4b}
    \end{subfigure}
    \hfill
    \begin{subfigure}[b]{0.45\textwidth}
        \centering
        \scalebox{0.6}{
\begin{tikzpicture}
	\begin{pgfonlayer}{nodelayer}
		\node [style=gauge1] (0) at (-3, 2) {};
		\node [style=gauge1] (1) at (3, -2) {};
		\node [style=gauge1] (2) at (1, 2) {};
		\node [style=gauge1] (3) at (-1, -2) {};
		\node [style=none] (4) at (3, 0) {};
		\node [style=none] (5) at (3, 0.25) {};
		\node [style=gauge1] (7) at (3, 0) {};
		\node [style=gauge1] (8) at (-3, 0) {};
		\node [style=none] (9) at (3, -0.25) {};
		\node [style=none] (10) at (-3, 0.25) {};
		\node [style=gauge1] (11) at (5, 0) {};
		\node [style=gauge1] (12) at (-5, 0) {};
		\node [style=none] (13) at (-3, -0.25) {};
		\node [style=none] (14) at (-1, 0) {};
		\node [style=none] (15) at (1, 0) {};
	\end{pgfonlayer}
	\begin{pgfonlayer}{edgelayer}
		\draw [style=olivee] (10.center) to (5.center);
		\draw [style=olivee] (9.center) to (13.center);
		\draw [style=magentae] (0) to (14.center);
		\draw [style=magentae] (14.center) to (3);
		\draw [style=magentae] (14.center) to (15.center);
		\draw [style=magentae] (15.center) to (2);
		\draw [style=magentae] (15.center) to (1);
		\draw [style=bluee] (12) to (8);
		\draw [style=bluee] (7) to (11);
	\end{pgfonlayer}
\end{tikzpicture}
}\\
\centering
 \scalebox{0.6}{  \begin{tikzpicture}
	\begin{pgfonlayer}{nodelayer}
		\node [style=none] (0) at (-4, 5) {};
		\node [style=none] (1) at (0, 0) {};
		\node [style=none] (2) at (2, 5) {};
		\node [style=none] (3) at (3.75, 3.75) {};
		\node [style=none] (4) at (3.75, 3.75) {};
		\node [style=none] (5) at (5.75, 1.5) {};
		\node [style=none] (6) at (2.75, 1.25) {\Large{$\mathcal{M}_\textrm{M}$}};
		\node [style=none] (7) at (1.75, 2.75) {\Large{$\mathcal{M}_\textrm{B}$}};
		\node [style=none] (8) at (-4, 5.5) {\Large{$\mathcal{M}_C$}};
		\node [style=none] (9) at (2.75, 6.25) {};
		\node [style=none] (10) at (7.25, 2) {};
		\node [style=none] (11) at (5.75, 4.5) {\Large{$\mathcal{M}_H$}};
		\node [style=smalldot] (12) at (3.5, 1.75) {};
		\node [style=none] (13) at (6.75, 0.5) {};
		\node [style=none] (14) at (6.75, 0.8) {\Large{We are here}};
	\end{pgfonlayer}
	\begin{pgfonlayer}{edgelayer}
		\draw (0.center) to (1.center);
		\draw (1.center) to (2.center);
		\draw [bend left=105, looseness=1.25] (2.center) to (3.center);
		\draw [bend right=75, looseness=0.75] (2.center) to (3.center);
		\draw [style=black] (1.center) to (5.center);
		\draw [style=black, in=30, out=60, looseness=1.25] (4.center) to (5.center);
		\draw [style=black, bend right=60] (4.center) to (5.center);
		\draw [style=brace] (9.center) to (10.center);
		\draw [style=thickline] (4.center) to (1.center);
		\draw [style=->, bend right=315, looseness=0.75] (13.center) to (12);
	\end{pgfonlayer}
\end{tikzpicture}}
		\caption{The mesonic branch is indicated in the brane web. Coloured branes are assumed to be on different $(x^7,x^8,x^9)$ positions in space.}
		\label{SU(3)4m}
    \end{subfigure}
    \begin{subfigure}[b]{0.45\textwidth}
        \centering
    \scalebox{0.6}{
\begin{tikzpicture}
	\begin{pgfonlayer}{nodelayer}
		\node [style=gauge1] (0) at (-3, 2) {};
		\node [style=gauge1] (1) at (3, -2) {};
		\node [style=gauge1] (2) at (1, 2) {};
		\node [style=gauge1] (3) at (-1, -2) {};
		\node [style=none] (4) at (3, 0) {};
		\node [style=none] (5) at (3, 0.25) {};
		\node [style=gauge1] (6) at (3, 0) {};
		\node [style=gauge1] (7) at (-3, 0) {};
		\node [style=none] (8) at (3, -0.25) {};
		\node [style=none] (9) at (-3, 0.25) {};
		\node [style=gauge1] (10) at (5, 0) {};
		\node [style=gauge1] (11) at (-5, 0) {};
		\node [style=none] (12) at (-3, -0.25) {};
		\node [style=none] (13) at (-1, 0) {};
		\node [style=none] (14) at (1, 0.25) {};
	\end{pgfonlayer}
	\begin{pgfonlayer}{edgelayer}
		\draw [style=olivee] (8.center) to (12.center);
		\draw [style=bluee] (11) to (7);
		\draw [style=bluee] (6) to (10);
		\draw [style=magentae] (0) to (13.center);
		\draw [style=magentae] (13.center) to (6);
		\draw [style=magentae] (2) to (14.center);
		\draw [style=magentae] (14.center) to (9.center);
		\draw [style=magentae] (13.center) to (3);
		\draw [style=magentae] (14.center) to (1);
	\end{pgfonlayer}
\end{tikzpicture}}\\
\centering
\scalebox{0.6}{\begin{tikzpicture}
	\begin{pgfonlayer}{nodelayer}
		\node [style=none] (0) at (-4, 5) {};
		\node [style=none] (1) at (0, 0) {};
		\node [style=none] (2) at (2, 5) {};
		\node [style=none] (3) at (3.75, 3.75) {};
		\node [style=none] (4) at (3.75, 3.75) {};
		\node [style=none] (5) at (5.75, 1.5) {};
		\node [style=none] (6) at (3.5, 1.75) {$\mathcal{M}_\textrm{M}$};
		\node [style=none] (7) at (1.75, 2.75) {$\mathcal{M}_\textrm{B}$};
		\node [style=none] (8) at (-4, 5.5) {$\mathcal{M}_C$};
		\node [style=none] (9) at (2.75, 6.25) {};
		\node [style=none] (10) at (7.25, 2) {};
		\node [style=none] (11) at (5.75, 4.5) {$\mathcal{M}_H$};
		\node [style=smalldot] (12) at (2.25, 2.25) {};
		\node [style=none] (13) at (3.25, 0) {};
		\node [style=none] (14) at (3.25, -0.5) {we are here};
	\end{pgfonlayer}
	\begin{pgfonlayer}{edgelayer}
		\draw (0.center) to (1.center);
		\draw (1.center) to (2.center);
		\draw [bend left=105, looseness=1.25] (2.center) to (3.center);
		\draw [bend right=75, looseness=0.75] (2.center) to (3.center);
		\draw [style=black] (1.center) to (5.center);
		\draw [style=black, in=30, out=60, looseness=1.25] (4.center) to (5.center);
		\draw [style=black, bend right=60] (4.center) to (5.center);
		\draw [style=brace] (9.center) to (10.center);
		\draw [style=thickline] (4.center) to (1.center);
		\draw [style=->, bend right=315, looseness=0.75] (13.center) to (12);
	\end{pgfonlayer}
\end{tikzpicture}}
\caption{The intersection of the baryonic branch and the mesonic branch is indicated in the brane web. Coloured branes are assumed to be on different $(x^7,x^8,x^9)$ positions in space.}
\label{SU(3)4i}
    \end{subfigure}
    \hfill
    \begin{subfigure}[b]{0.45\textwidth}
        \centering
\scalebox{0.6}{
\begin{tikzpicture}
	\begin{pgfonlayer}{nodelayer}
		\node [style=gauge1] (0) at (-2, 4) {};
		\node [style=none] (1) at (-1, 3) {};
		\node [style=none] (2) at (1, 3) {};
		\node [style=gauge1] (3) at (1, 4) {};
		\node [style=none] (4) at (2, 2) {};
		\node [style=none] (5) at (-1, 2) {};
		\node [style=none] (7) at (-3.5, 1) {};
		\node [style=none] (8) at (-3.5, 0.925) {};
		\node [style=none] (10) at (-1, 0) {};
		\node [style=none] (11) at (4, 1) {};
		\node [style=none] (12) at (4, 0) {};
		\node [style=none] (13) at (5.5, 1.075) {};
		\node [style=none] (14) at (5.5, 0.925) {};
		\node [style=gauge1] (18) at (-5, 1) {};
		\node [style=gauge1] (19) at (-3.5, 1) {};
		\node [style=gauge1] (20) at (-1, -1) {};
		\node [style=gauge1] (21) at (5, -1) {};
		\node [style=gauge1] (22) at (5.5, 1) {};
		\node [style=gauge1] (23) at (7.5, 1) {};
		\node [style=none] (24) at (3.825, 1.075) {};
		\node [style=none] (25) at (4, 0.925) {};
		\node [style=none] (26) at (-2, 1) {};
		\node [style=none] (27) at (-2, 1) {};
		\node [style=none] (28) at (-1.925, 1.075) {};
		\node [style=none] (30) at (-1.95, 0.925) {};
		\node [style=none] (31) at (-3.5, 1.075) {};
	\end{pgfonlayer}
	\begin{pgfonlayer}{edgelayer}
		\draw [style=bluee](0) to (1.center);
		\draw [style=bluee](1.center) to (2.center);
		\draw [style=bluee](2.center) to (3);
		\draw [style=bluee](2.center) to (4.center);
		\draw [style=bluee](1.center) to (5.center);
		\draw [style=bluee](5.center) to (4.center);
		\draw [style=bluee](4.center) to (11.center);
		\draw [style=bluee](18) to (19);
		\draw [style=bluee](10.center) to (20);
		\draw [style=bluee](10.center) to (12.center);
		\draw [style=bluee](11.center) to (12.center);
		\draw [style=bluee](12.center) to (21);
		\draw [style=bluee](24.center) to (13.center);
		\draw [style=bluee](25.center) to (14.center);
		\draw [style=bluee](22) to (23);
		\draw [style=bluee](5.center) to (27.center);
		\draw [style=bluee](27.center) to (10.center);
		\draw [style=bluee](28.center) to (31.center);
		\draw [style=bluee](8.center) to (30.center);
	\end{pgfonlayer}
\end{tikzpicture}
}\\
\centering
\scalebox{0.6}{\begin{tikzpicture}
	\begin{pgfonlayer}{nodelayer}
		\node [style=none] (0) at (-4, 5) {};
		\node [style=none] (1) at (0, 0) {};
		\node [style=none] (2) at (2, 5) {};
		\node [style=none] (3) at (3.75, 3.75) {};
		\node [style=none] (4) at (3.75, 3.75) {};
		\node [style=none] (5) at (5.75, 1.5) {};
		\node [style=none] (6) at (3.5, 1.75) {$\mathcal{M}_\textrm{M}$};
		\node [style=none] (7) at (1.75, 2.75) {$\mathcal{M}_\textrm{B}$};
		\node [style=none] (8) at (-4, 5.5) {$\mathcal{M}_C$};
		\node [style=none] (9) at (2.75, 6.25) {};
		\node [style=none] (10) at (7.25, 2) {};
		\node [style=none] (11) at (5.75, 4.5) {$\mathcal{M}_H$};
		\node [style=smalldot] (12) at (-2.5, 3) {};
		\node [style=none] (13) at (-0.5, 5.25) {};
		\node [style=none] (14) at (-0.25, 5.75) {We are here};
	\end{pgfonlayer}
	\begin{pgfonlayer}{edgelayer}
		\draw (0.center) to (1.center);
		\draw (1.center) to (2.center);
		\draw [bend left=105, looseness=1.25] (2.center) to (3.center);
		\draw [bend right=75, looseness=0.75] (2.center) to (3.center);
		\draw [style=black] (1.center) to (5.center);
		\draw [style=black, in=30, out=60, looseness=1.25] (4.center) to (5.center);
		\draw [style=black, bend right=60] (4.center) to (5.center);
		\draw [style=brace] (9.center) to (10.center);
		\draw [style=thickline] (4.center) to (1.center);
		\draw [style=->, bend right=45] (13.center) to (12);
	\end{pgfonlayer}
\end{tikzpicture}
}
\caption{The Coulomb branch is indicated in the brane web.}
\label{SU(3)4C}
    \end{subfigure}
    \caption{Depiction of different phases in the brane web corresponding to different points in the moduli space of the gauge theory that lives on the web.}
\end{figure}
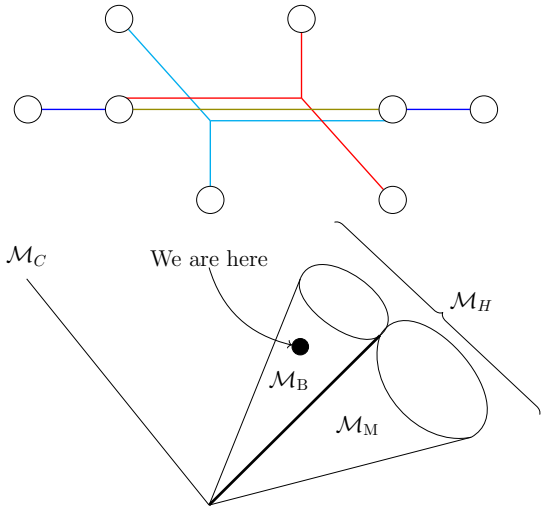
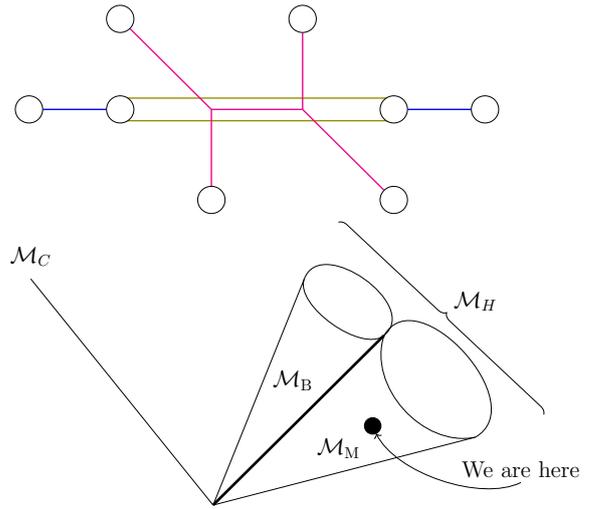
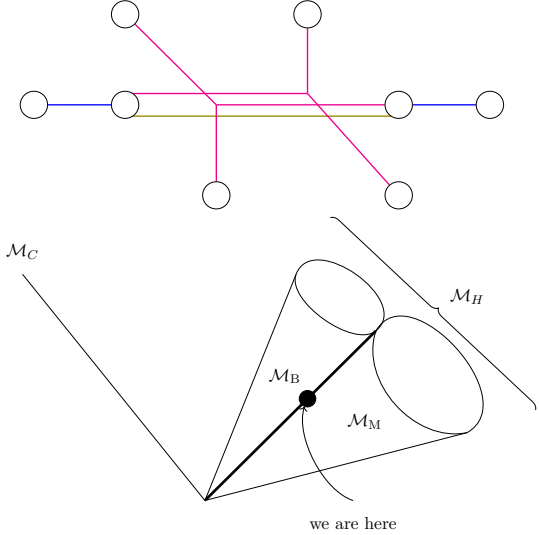
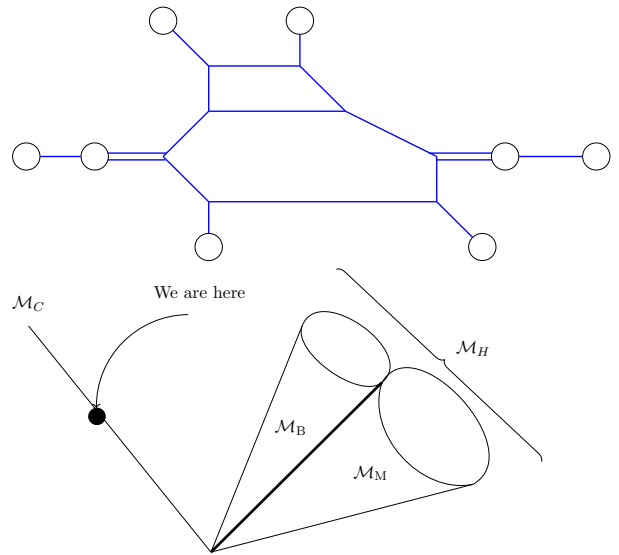

\begin{figure}[t]
    \centering
 \scalebox{0.6}{
\begin{tikzpicture}
	\begin{pgfonlayer}{nodelayer}
		\node [style=gauge1] (0) at (-2.5, 2) {};
		\node [style=gauge1] (1) at (-2.5, 0) {};
		\node [style=gauge1] (2) at (-4.5, 0) {};
		\node [style=gauge1] (3) at (0.5, 2) {};
		\node [style=none] (4) at (-1.5, 1) {};
		\node [style=none] (5) at (0.5, 1) {};
		\node [style=none] (6) at (1.5, 0.125) {};
		\node [style=none] (7) at (-1.5, 0.125) {};
		\node [style=none] (8) at (-0.5, -1) {};
		\node [style=none] (9) at (1.5, -1) {};
		\node [style=gauge1] (10) at (2.5, 0) {};
		\node [style=gauge1] (11) at (2.5, -2) {};
		\node [style=gauge1] (12) at (4.5, 0) {};
		\node [style=none] (13) at (-2.5, -0.125) {};
		\node [style=none] (14) at (2.5, -0.125) {};
		\node [style=gauge1] (15) at (-0.5, -2) {};
		\node [style=none] (16) at (-2.5, 0.125) {};
		\node [style=none] (17) at (2.5, 0.125) {};
	\end{pgfonlayer}
	\begin{pgfonlayer}{edgelayer}
		\draw [style=darke](0) to (4.center);
		\draw [style=darke](4.center) to (7.center);
		\draw [style=darke](4.center) to (5.center);
		\draw [style=darke](5.center) to (3);
		\draw [style=darke](7.center) to (8.center);
		\draw [style=darke](5.center) to (6.center);
		\draw [style=olivee] (13.center) to (14.center);
		\draw [style=darke] (6.center) to (9.center);
		\draw [style=darke] (9.center) to (11);
		\draw [style=darke] (8.center) to (15);
		\draw [style=darke] (8.center) to (9.center);
		\draw [style=bluee] (12) to (10);
		\draw [style=bluee] (2) to (1);
		\draw [style=darke](16.center) to (7.center);
		\draw [style=darke](6.center) to (17.center);
	\end{pgfonlayer}
\end{tikzpicture}
}\\
\centering
\scalebox{0.6}{\begin{tikzpicture}
	\begin{pgfonlayer}{nodelayer}
		\node [style=none] (0) at (-4, 5) {};
		\node [style=none] (1) at (0, 0) {};
		\node [style=none] (2) at (2, 5) {};
		\node [style=none] (3) at (3.75, 3.75) {};
		\node [style=none] (4) at (3.75, 3.75) {};
		\node [style=none] (5) at (5.75, 1.5) {};
		\node [style=none] (6) at (3.5, 1.75) {$\mathcal{M}_\textrm{M}$};
		\node [style=none] (7) at (1.75, 2.75) {$\mathcal{M}_\textrm{B}$};
		\node [style=none] (8) at (-4, 5.5) {$\mathcal{M}_C$};
		\node [style=none] (9) at (2.75, 6.25) {};
		\node [style=none] (10) at (7.25, 2) {};
		\node [style=none] (11) at (5.75, 4.5) {$\mathcal{M}_H$};
		\node [style=smalldot] (12) at (-0.5, 3.25) {};
		\node [style=none] (13) at (0, 5) {};
		\node [style=none] (14) at (0, 5.5) {We are here};
	\end{pgfonlayer}
	\begin{pgfonlayer}{edgelayer}
		\draw (0.center) to (1.center);
		\draw (1.center) to (2.center);
		\draw [bend left=105, looseness=1.25] (2.center) to (3.center);
		\draw [bend right=75, looseness=0.75] (2.center) to (3.center);
		\draw [style=black] (1.center) to (5.center);
		\draw [style=black, in=30, out=60, looseness=1.25] (4.center) to (5.center);
		\draw [style=black, bend right=60] (4.center) to (5.center);
		\draw [style=brace] (9.center) to (10.center);
		\draw [style=thickline] (4.center) to (1.center);
		\draw [style=->, bend right, looseness=0.75] (13.center) to (12);
	\end{pgfonlayer}
\end{tikzpicture}
}
    \caption{The mixed branch is indicated in the brane web. Coloured branes are assumed to be on different $(x^7,x^8,x^9)$ positions in space. }
    \label{SU(3)4mixed}
\end{figure}
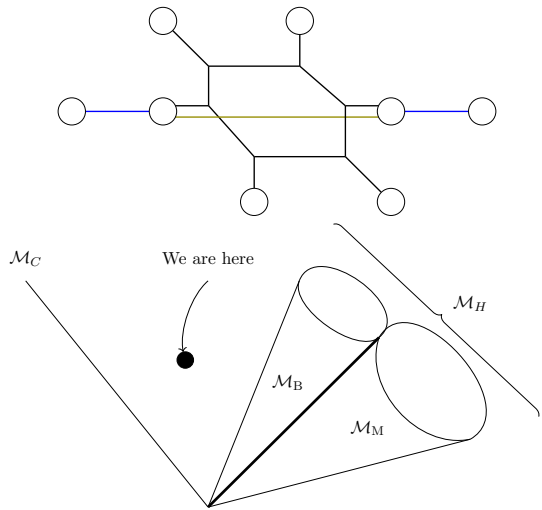

We can now depict various points in our moduli space by moving the 5-branes in the brane web, without changing the position of the 7-branes. A point on the baryonic branch is depicted in Figure \ref{SU(3)4b}, a point on the mesonic branch is depicted in Figure \ref{SU(3)4m}, a point on the intersection of the baryonic branch and the mesonic branch is depicted in Figure \ref{SU(3)4i}. A point on the Coulomb branch is depicted in Figure \ref{SU(3)4C} and a point on the mixed branch is depicted in Figure \ref{SU(3)4mixed}.

While directions along the Coulomb branch correspond to faces changing size and shape, the directions along the Higgs branch correspond to $(p,q)5$-branes moving along the $(x^7,x^8,x^9)$ directions.

\clearpage

According to \cite{Cabrera:2018jxt} the Hilbert series of the individual cones of the Higgs branch can be computed by obtaining a \textit{magnetic} Quiver and using the monopole formula\footnote{It should be noted that this computation is formally equivalent to computing the Coulomb Branch of a 3d ${\cal N}=4$ quiver. However we view the magnetic quiver as a fundamental property of the theory rather than some type of duality that relates to a Coulomb branch of a different theory.} introduced in \cite{Cremonesi:2013lqa}. Below, we review the algorithm presented in \cite{Cabrera:2018jxt}:

\begin{enumerate}
\item Draw the brane web.
\item Find all inequivalent maximal decompositions.
\item Read the magnetic quiver for each such decomposition to obtain the individual cones. (See appendix \ref{Appendixmag}.)
\item Find the common part of each two maximal decompositions into subwebs.
\item Read the magnetic Quiver for each such decomposition to obtain the intersection of two cones.
\item Continue to find intersections of three cones and so on.
\end{enumerate}
In our case of SCQD there are only two cones, a mesonic and a baryonic cone. The baryonic cone corresponds to the phase where the non-flavour part of the web is split into two subwebs \cite{Aharony:1997bh}.\\

We present the examples of $\mathrm{SU}(5)$, $\mathrm{SU}(4)$ and $\mathrm{SU}(3)$ with $N_f=4$ massless flavours. The former consists of only a mesonic branch, the latter two of a mesonic and a baryonic branch with non-trivial intersections.

\subsection{\texorpdfstring{$\mathrm{SU}(5)$ with $N_f=4$ -- Single Mesonic Cone}{SU(5) with Nf=4 flavours}}

The toric diagram for $\mathrm{SU}(5)$ with $N_f=4$ massless flavours at finite coupling and CS level $\pm2$ is given by
\begin{equation}
 \scalebox{0.5}{
\begin{tikzpicture}
	\begin{pgfonlayer}{nodelayer}
		\node [style=smallgauge] (0) at (0, 4) {};
		\node [style=smallgauge] (1) at (0, 3) {};
		\node [style=smallgauge] (2) at (0, 2) {};
		\node [style=smallgauge] (3) at (0, 1) {};
		\node [style=smallgauge] (4) at (0, 0) {};
		\node [style=smallgauge] (5) at (0, -1) {};
		\node [style=smallgauge] (6) at (-1, 4) {};
		\node [style=smallgauge] (7) at (-1, 3) {};
		\node [style=smallgauge] (8) at (-1, 2) {};
		\node [style=smallgauge] (9) at (-1, 1) {};
		\node [style=smallgauge] (10) at (-1, 0) {};
		\node [style=smallgauge] (11) at (1, -1) {};
	\end{pgfonlayer}
	\begin{pgfonlayer}{edgelayer}
		\draw (6) to (7);
		\draw (7) to (8);
		\draw (8) to (9);
		\draw (9) to (10);
		\draw (10) to (5);
		\draw (5) to (11);
		\draw (0) to (11);
		\draw (6) to (0);
		\draw (0) to (1);
		\draw (1) to (2);
		\draw (2) to (3);
		\draw (3) to (4);
		\draw (4) to (5);
	\end{pgfonlayer}
\end{tikzpicture}
}\quad,
\end{equation}
the corresponding brane web is
\begin{equation}
\centering
\scalebox{0.8}{
\begin{tikzpicture}
	\begin{pgfonlayer}{nodelayer}
		\node [style=gaugeX] (0) at (-1, 0) {};
		\node [style=gaugeX] (1) at (-2, 0) {};
		\node [style=gaugeX] (2) at (-3, 0) {};
		\node [style=gaugeX] (3) at (-4, 0) {};
		\node [style=gaugeX] (4) at (0, 1) {};
		\node [style=gaugeX] (5) at (-1, -1) {};
		\node [style=gaugeX] (6) at (2.025, -1) {};
		\node [style=none] (7) at (0, -0.2) {};
		\node [style=none] (8) at (2.025, -0.2) {};
		\node [style=gaugeX] (9) at (5, 1) {};
		\node [style=none] (13) at (2.025, 0.1) {};
		\node [style=none] (14) at (2.025, 0.025) {};
		\node [style=none] (15) at (2.025, -0.05) {};
		\node [style=none] (16) at (-1, 0.1) {};
		\node [style=none] (17) at (-1, 0.025) {};
		\node [style=none] (18) at (-1, -0.05) {};
		\node [style=none] (19) at (-1, -0.125) {};
		\node [style=none] (21) at (2.025, -0.125) {};
		\node [style=none] (25) at (-0.5, 0.25) {4};
		\node [style=none] (26) at (1, 0.25) {5};
		\node [style=none] (27) at (0, 1.5) {[0,1]};
		\node [style=none] (28) at (-1, -1.5) {[1,1]};
		\node [style=none] (29) at (2.025, -1.5) {[0,1]};
		\node [style=none] (30) at (5, 1.5) {[5,1]};
		\node [style=none] (32) at (-1, 0.075) {};
		\node [style=none] (33) at (-1, 0) {};
		\node [style=none] (34) at (-1, -0.075) {};
		\node [style=none] (35) at (-2, 0.075) {};
		\node [style=none] (36) at (-2, 0) {};
		\node [style=none] (37) at (-2, -0.075) {};
		\node [style=none] (38) at (-3, 0.025) {};
		\node [style=none] (39) at (-3, -0.05) {};
		\node [style=none] (40) at (-2, 0.025) {};
		\node [style=none] (41) at (-2, -0.05) {};
		\node [style=none] (42) at (-3, 0) {};
		\node [style=none] (43) at (-1.5, 0.25) {3};
		\node [style=none] (44) at (-2.5, 0.25) {2};
		\node [style=none] (45) at (-3.5, 0.25) {1};
		\node [style=none] (46) at (2.025, -0.25) {};
	\end{pgfonlayer}
	\begin{pgfonlayer}{edgelayer}
		\draw (16.center) to (13.center);
		\draw (14.center) to (17.center);
		\draw (18.center) to (15.center);
		\draw (19.center) to (21.center);
		\draw (4) to (7.center);
		\draw (7.center) to (5);
		\draw (7.center) to (8.center);
		\draw (8.center) to (6);
		\draw (13.center) to (9);
		\draw (35.center) to (32.center);
		\draw (36.center) to (33.center);
		\draw (37.center) to (34.center);
		\draw (38.center) to (40.center);
		\draw (41.center) to (39.center);
		\draw (3) to (42.center);
		\draw (46.center) to (13.center);
	\end{pgfonlayer}
\end{tikzpicture}}
\end{equation}
and there is only one maximal decomposition
\begin{equation}
    \centering
 \scalebox{0.8}{
\begin{tikzpicture}
	\begin{pgfonlayer}{nodelayer}
		\node [style=gaugeX] (0) at (-2, 0) {};
		\node [style=gaugeX] (1) at (-3, 0) {};
		\node [style=gaugeX] (2) at (-4, 0) {};
		\node [style=gaugeX] (3) at (-5, 0) {};
		\node [style=gaugeX] (4) at (-1, 1) {};
		\node [style=gaugeX] (5) at (-2, -1) {};
		\node [style=gaugeX] (6) at (1.025, -1) {};
		\node [style=none] (7) at (-1, -0.2) {};
		\node [style=none] (8) at (1.025, -0.2) {};
		\node [style=gaugeX] (9) at (4, 1) {};
		\node [style=none] (10) at (1.025, 0.1) {};
		\node [style=none] (11) at (1.025, 0.025) {};
		\node [style=none] (12) at (1.025, -0.05) {};
		\node [style=none] (13) at (-2, 0.1) {};
		\node [style=none] (14) at (-2, 0.025) {};
		\node [style=none] (15) at (-2, -0.05) {};
		\node [style=none] (16) at (-2, -0.125) {};
		\node [style=none] (17) at (1.025, -0.125) {};
		\node [style=none] (18) at (-1.5, 0.3) {\color{magenta}{4}};
		\node [style=none] (19) at (0, 0.3) {\color{magenta}{5}};
		\node [style=none] (20) at (-1, 1.5) {[0,1]};
		\node [style=none] (21) at (-2, -1.5) {[1,1]};
		\node [style=none] (22) at (1.025, -1.5) {[0,1]};
		\node [style=none] (23) at (4, 1.5) {[5,1]};
		\node [style=none] (24) at (-2, 0.075) {};
		\node [style=none] (25) at (-2, 0) {};
		\node [style=none] (26) at (-2, -0.075) {};
		\node [style=none] (27) at (-3, 0.075) {};
		\node [style=none] (28) at (-3, 0) {};
		\node [style=none] (29) at (-3, -0.075) {};
		\node [style=none] (30) at (-4, 0.025) {};
		\node [style=none] (31) at (-4, -0.05) {};
		\node [style=none] (32) at (-3, 0.025) {};
		\node [style=none] (33) at (-3, -0.05) {};
		\node [style=none] (34) at (-4, 0) {};
		\node [style=none] (35) at (-2.5, 0.3) {\color{magenta}{2}};
		\node [style=none] (36) at (-3.5, -0.3) {\color{olive}{2}};
		\node [style=none] (37) at (-4.5, -0.3) {\color{blue}{1}};
		\node [style=none] (38) at (-2.5, -0.3) {\color{blue}{1}};
		\node [style=none] (46) at (1.025, -0.25) {};
	\end{pgfonlayer}
	\begin{pgfonlayer}{edgelayer}
		\draw [style=magentae] (4) to (7.center);
		\draw [style=magentae] (7.center) to (5);
		\draw [style=magentae] (7.center) to (8.center);
		\draw [style=magentae] (16.center) to (17.center);
		\draw [style=magentae] (15.center) to (12.center);
		\draw [style=magentae] (14.center) to (11.center);
		\draw [style=magentae] (13.center) to (10.center);
		\draw [style=magentae] (24.center) to (27.center);
		\draw [style=magentae] (25.center) to (28.center);
		\draw [style=olivee] (31.center) to (33.center);
		\draw [style=olivee] (30.center) to (32.center);
		\draw [style=magentae] (8.center) to (6);
		\draw [style=magentae] (9) to (10.center);
		\draw [style=bluee] (29.center) to (26.center);
		\draw [style=bluee] (3) to (34.center);
		\draw [style=magentae] (46.center) to (10.center);
	\end{pgfonlayer}
\end{tikzpicture}}\quad,
\end{equation}
which leads to the magnetic quiver
\begin{equation}
    \centering
    \scalebox{0.8}{
\begin{tikzpicture}
	\begin{pgfonlayer}{nodelayer}
		\node [style=none] (4) at (-0.1, 1.25) {};
		\node [style=none] (5) at (0.1, 1.25) {};
		\node [style=none] (6) at (-0.1, 0) {};
		\node [style=none] (7) at (0.1, 0) {};
		\node [style=none] (8) at (0, 1.75) {1};
		\node [style=none] (9) at (1.5, -0.5) {1};
		\node [style=none] (10) at (0, -0.5) {2};
		\node [style=none] (11) at (-1.5, -0.5) {1};
		\node [style=none] (12) at (0.5, 0.5) {2};
		\node [style=olivegauge] (13) at (0, 0) {};
		\node [style=bluegauge] (14) at (1.5, 0) {};
		\node [style=bluegauge] (15) at (-1.5, 0) {};
		\node [style=magentagauge] (16) at (0, 1.25) {};
	\end{pgfonlayer}
	\begin{pgfonlayer}{edgelayer}
		\draw (5.center) to (7.center);
		\draw [in=90, out=-90] (4.center) to (6.center);
		\draw (13) to (14);
		\draw (13) to (15);
	\end{pgfonlayer}
\end{tikzpicture}}\quad.
\end{equation}
The Higgs Branch is therefore conjectured to be the closure of the next-to-minimal nilpotent orbit of $\mathfrak{sl}(4,\mathbb{C})$, with the corresponding HWG
\begin{equation}
\label{HWG45}
\mathrm{HWG}_{\textrm{Brane Web}}=\mathrm{PE}[\mu_1\mu_3t^2+\mu_2^2t^4]\, .
\end{equation}
The global symmetry of the Higgs Branch is $\mathrm{SU}(4)$. Upon unrefining all the $\mathrm{SU}(4)$ fugacities and expanding the HWG (\ref{HWG45}) into a standard Hilbert series, one obtains 
\begin{equation}
\label{HSBW45}
   \mathrm{HS}_{\textnormal{Brane Web}}=\frac{\left(1+t^2\right)^2 \left(1+5   t^2+t^4\right)}{(1-t^2)^8} \, . 
\end{equation}

\subsection{\texorpdfstring{$\mathrm{SU}(4)$ with $N_f=4$ -- Two Cones with Trivial Intersection}{SU(4) with 4 flavours}}

The toric diagram for $\mathrm{SU}(4)$ with $N_f=4$ massless flavours at finite coupling and CS level $\pm1$ is given by

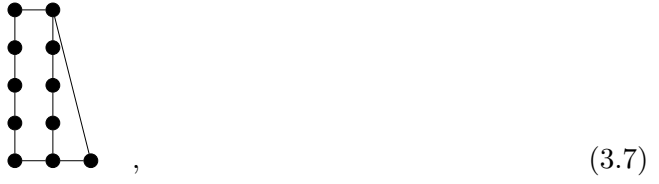
\begin{equation}
     \centering
\scalebox{0.5}{\begin{tikzpicture}
	\begin{pgfonlayer}{nodelayer}
		\node [style=smallgauge] (0) at (0, 4) {};
		\node [style=smallgauge] (1) at (0, 3) {};
		\node [style=smallgauge] (2) at (0, 2) {};
		\node [style=smallgauge] (3) at (0, 1) {};
		\node [style=smallgauge] (4) at (0, 0) {};
		\node [style=smallgauge] (6) at (-1, 4) {};
		\node [style=smallgauge] (7) at (-1, 3) {};
		\node [style=smallgauge] (8) at (-1, 2) {};
		\node [style=smallgauge] (9) at (-1, 1) {};
		\node [style=smallgauge] (10) at (-1, 0) {};
		\node [style=smallgauge] (11) at (1, 0) {};
	\end{pgfonlayer}
	\begin{pgfonlayer}{edgelayer}
		\draw (6) to (7);
		\draw (7) to (8);
		\draw (8) to (9);
		\draw (9) to (10);
		\draw (0) to (11);
		\draw (6) to (0);
		\draw (0) to (1);
		\draw (1) to (2);
		\draw (2) to (3);
		\draw (3) to (4);
		\draw (10) to (4);
		\draw (4) to (11);
	\end{pgfonlayer}
\end{tikzpicture}
}\quad,
\end{equation}
the corresponding brane web is
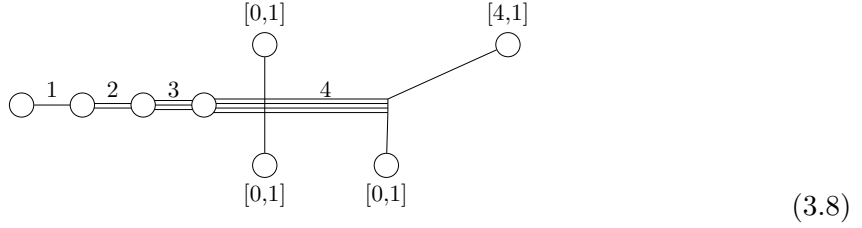
\begin{equation}
    \centering
\scalebox{0.8}{\begin{tikzpicture}
	\begin{pgfonlayer}{nodelayer}
		\node [style=gaugeX] (0) at (-2, 0) {};
		\node [style=gaugeX] (1) at (-3, 0) {};
		\node [style=gaugeX] (2) at (-4, 0) {};
		\node [style=gaugeX] (3) at (-5, 0) {};
		\node [style=gaugeX] (4) at (-1, 1) {};
		\node [style=gaugeX] (5) at (-1, -1) {};
		\node [style=gaugeX] (6) at (1, -1) {};
		\node [style=gaugeX] (9) at (3, 1) {};
		\node [style=none] (10) at (1.025, 0.1) {};
		\node [style=none] (11) at (1.025, 0.025) {};
		\node [style=none] (12) at (1.025, -0.05) {};
		\node [style=none] (13) at (-2, 0.1) {};
		\node [style=none] (14) at (-2, 0.025) {};
		\node [style=none] (15) at (-2, -0.05) {};
		\node [style=none] (16) at (-2, -0.125) {};
		\node [style=none] (17) at (1.025, -0.125) {};
		\node [style=none] (18) at (0, 0.25) {4};
		\node [style=none] (20) at (-1, 1.5) {[0,1]};
		\node [style=none] (21) at (-1, -1.5) {[0,1]};
		\node [style=none] (22) at (1, -1.5) {[0,1]};
		\node [style=none] (23) at (3, 1.5) {[4,1]};
		\node [style=none] (24) at (-2, 0.075) {};
		\node [style=none] (25) at (-2, 0) {};
		\node [style=none] (26) at (-2, -0.075) {};
		\node [style=none] (27) at (-3, 0.075) {};
		\node [style=none] (28) at (-3, 0) {};
		\node [style=none] (29) at (-3, -0.075) {};
		\node [style=none] (30) at (-4, 0.025) {};
		\node [style=none] (31) at (-4, -0.05) {};
		\node [style=none] (32) at (-3, 0.025) {};
		\node [style=none] (33) at (-3, -0.05) {};
		\node [style=none] (34) at (-4, 0) {};
		\node [style=none] (35) at (-2.5, 0.25) {3};
		\node [style=none] (36) at (-3.5, 0.25) {2};
		\node [style=none] (37) at (-4.5, 0.25) {1};
	\end{pgfonlayer}
	\begin{pgfonlayer}{edgelayer}
		\draw (13.center) to (10.center);
		\draw (11.center) to (14.center);
		\draw (15.center) to (12.center);
		\draw (16.center) to (17.center);
		\draw (10.center) to (9);
		\draw (27.center) to (24.center);
		\draw (28.center) to (25.center);
		\draw (29.center) to (26.center);
		\draw (30.center) to (32.center);
		\draw (33.center) to (31.center);
		\draw (3) to (34.center);
		\draw (17.center) to (6);
		\draw (4) to (5);
		\draw (10.center) to (17.center);
	\end{pgfonlayer}
\end{tikzpicture}
 }
\end{equation}
and there are two maximal decompositions:
\begin{enumerate}
\item The mesonic branch
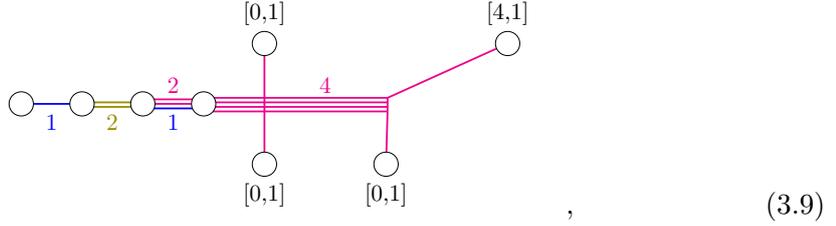
\begin{equation}
    \centering
    \scalebox{0.8}{
\begin{tikzpicture}
	\begin{pgfonlayer}{nodelayer}
		\node [style=gaugeX] (0) at (-2, 0) {};
		\node [style=gaugeX] (1) at (-3, 0) {};
		\node [style=gaugeX] (2) at (-4, 0) {};
		\node [style=gaugeX] (3) at (-5, 0) {};
		\node [style=gaugeX] (4) at (-1, 1) {};
		\node [style=gaugeX] (5) at (-1, -1) {};
		\node [style=gaugeX] (6) at (1, -1) {};
		\node [style=gaugeX] (9) at (3, 1) {};
		\node [style=none] (10) at (1.025, 0.1) {};
		\node [style=none] (11) at (1.025, 0.025) {};
		\node [style=none] (12) at (1.025, -0.05) {};
		\node [style=none] (13) at (-2, 0.1) {};
		\node [style=none] (14) at (-2, 0.025) {};
		\node [style=none] (15) at (-2, -0.05) {};
		\node [style=none] (16) at (-2, -0.125) {};
		\node [style=none] (17) at (1.025, -0.125) {};
		\node [style=none] (18) at (0, 0.3) {\color{magenta}{4}};
		\node [style=none] (20) at (-1, 1.5) {[0,1]};
		\node [style=none] (21) at (-1, -1.5) {[0,1]};
		\node [style=none] (22) at (1, -1.5) {[0,1]};
		\node [style=none] (23) at (3, 1.5) {[4,1]};
		\node [style=none] (24) at (-2, 0.075) {};
		\node [style=none] (25) at (-2, 0) {};
		\node [style=none] (26) at (-2, -0.075) {};
		\node [style=none] (27) at (-3, 0.075) {};
		\node [style=none] (28) at (-3, 0) {};
		\node [style=none] (29) at (-3, -0.075) {};
		\node [style=none] (30) at (-4, 0.025) {};
		\node [style=none] (31) at (-4, -0.05) {};
		\node [style=none] (32) at (-3, 0.025) {};
		\node [style=none] (33) at (-3, -0.05) {};
		\node [style=none] (34) at (-4, 0) {};
		\node [style=none] (35) at (-2.5, 0.3) {\color{magenta}{2}};
		\node [style=none] (36) at (-3.5, -0.3) {\color{olive}{2}};
		\node [style=none] (37) at (-4.5, -0.3) {\color{blue}{1}};
		\node [style=none] (38) at (-2.5, -0.3) {\color{blue}{1}};
	\end{pgfonlayer}
	\begin{pgfonlayer}{edgelayer}
		\draw [style=magentae] (16.center) to (17.center);
		\draw [style=magentae] (15.center) to (12.center);
		\draw [style=magentae] (14.center) to (11.center);
		\draw [style=magentae] (13.center) to (10.center);
		\draw [style=magentae] (24.center) to (27.center);
		\draw [style=magentae] (25.center) to (28.center);
		\draw [style=olivee] (31.center) to (33.center);
		\draw [style=olivee] (30.center) to (32.center);
		\draw [style=magentae] (9) to (10.center);
		\draw [style=bluee] (29.center) to (26.center);
		\draw [style=bluee] (3) to (34.center);
		\draw [style=magentae] (4) to (5);
		\draw [style=magentae] (6) to (17.center);
		\draw [style=magentae] (17.center) to (10.center);
	\end{pgfonlayer}
\end{tikzpicture}}\quad,
\end{equation}
from which we can read the magnetic quiver
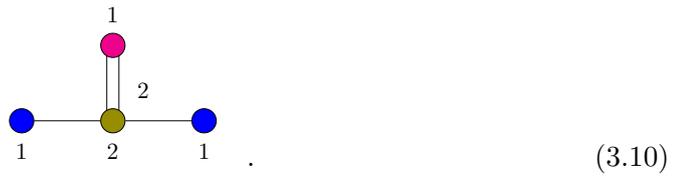
\begin{equation}
  \centering
\scalebox{0.8}{\begin{tikzpicture}
	\begin{pgfonlayer}{nodelayer}
		\node [style=none] (4) at (-0.1, 1.25) {};
		\node [style=none] (5) at (0.1, 1.25) {};
		\node [style=none] (6) at (-0.1, 0) {};
		\node [style=none] (7) at (0.1, 0) {};
		\node [style=none] (8) at (0, 1.75) {$1$};
		\node [style=none] (9) at (1.5, -0.5) {$1$};
		\node [style=none] (10) at (0, -0.5) {$2$};
		\node [style=none] (11) at (-1.5, -0.5) {$1$};
		\node [style=none] (12) at (0.5, 0.5) {$2$};
		\node [style=olivegauge] (13) at (0, 0) {};
		\node [style=bluegauge] (14) at (1.5, 0) {};
		\node [style=bluegauge] (15) at (-1.5, 0) {};
		\node [style=magentagauge] (16) at (0, 1.25) {};
	\end{pgfonlayer}
	\begin{pgfonlayer}{edgelayer}
		\draw (5.center) to (7.center);
		\draw [in=90, out=-90] (4.center) to (6.center);
		\draw (13) to (14);
		\draw (13) to (15);
	\end{pgfonlayer}
\end{tikzpicture}}\quad.
\end{equation}
The mesonic branch is therefore conjectured to be the same as in the previous example, the closure of the next-to-minimal nilpotent orbit of $\mathfrak{sl}(4,\mathbb{C})$, with HWG 
\begin{equation}
\mathrm{HWG}_\textrm{M}=\mathrm{PE}[\mu_1\mu_3t^2+\mu_2^2t^4] \,. 
\end{equation}
The unrefined Hilbert series is 
\begin{equation}
\label{HS34mes}
   \mathrm{HS}_\textrm{M} = \frac{\left(1+t^2\right)^2 \left(1+5 t^2+t^4\right)}{(1-t)^8   (1+t)^8} \,. 
\end{equation}
\item The baryonic branch
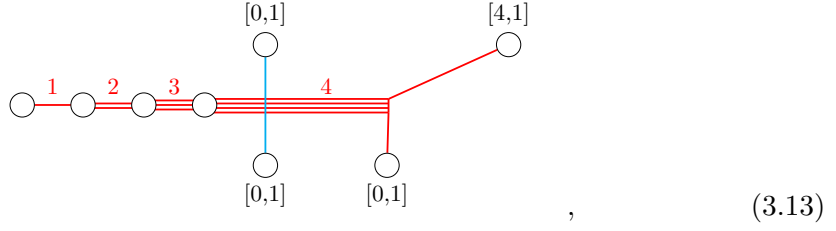
\begin{equation}
    \
    \scalebox{0.8}{\begin{tikzpicture}
	\begin{pgfonlayer}{nodelayer}
		\node [style=gaugeX] (0) at (-2, 0) {};
		\node [style=gaugeX] (1) at (-3, 0) {};
		\node [style=gaugeX] (2) at (-4, 0) {};
		\node [style=gaugeX] (3) at (-5, 0) {};
		\node [style=gaugeX] (4) at (-1, 1) {};
		\node [style=gaugeX] (5) at (-1, -1) {};
		\node [style=gaugeX] (6) at (1, -1) {};
		\node [style=gaugeX] (7) at (3, 1) {};
		\node [style=none] (8) at (1.025, 0.1) {};
		\node [style=none] (9) at (1.025, 0.025) {};
		\node [style=none] (10) at (1.025, -0.05) {};
		\node [style=none] (11) at (-2, 0.1) {};
		\node [style=none] (12) at (-2, 0.025) {};
		\node [style=none] (13) at (-2, -0.05) {};
		\node [style=none] (14) at (-2, -0.125) {};
		\node [style=none] (15) at (1.025, -0.125) {};
		\node [style=none] (16) at (0, 0.3) {\color{red}{4}};
		\node [style=none] (17) at (-1, 1.5) {[0,1]};
		\node [style=none] (18) at (-1, -1.5) {[0,1]};
		\node [style=none] (19) at (1, -1.5) {[0,1]};
		\node [style=none] (20) at (3, 1.5) {[4,1]};
		\node [style=none] (21) at (-2, 0.075) {};
		\node [style=none] (22) at (-2, 0) {};
		\node [style=none] (23) at (-2, -0.075) {};
		\node [style=none] (24) at (-3, 0.075) {};
		\node [style=none] (25) at (-3, 0) {};
		\node [style=none] (26) at (-3, -0.075) {};
		\node [style=none] (27) at (-4, 0.025) {};
		\node [style=none] (28) at (-4, -0.05) {};
		\node [style=none] (29) at (-3, 0.025) {};
		\node [style=none] (30) at (-3, -0.05) {};
		\node [style=none] (31) at (-4, 0) {};
		\node [style=none] (32) at (-2.5, 0.3) {\color{red}{3}};
		\node [style=none] (33) at (-3.5, 0.3) {\color{red}{2}};
		\node [style=none] (34) at (-4.5, 0.3) {\color{red}{1}};
	\end{pgfonlayer}
	\begin{pgfonlayer}{edgelayer}
		\draw [style=rede] (14.center) to (15.center);
		\draw [style=rede] (13.center) to (10.center);
		\draw [style=rede] (12.center) to (9.center);
		\draw [style=rede] (11.center) to (8.center);
		\draw [style=rede] (21.center) to (24.center);
		\draw [style=rede] (22.center) to (25.center);
		\draw [style=rede] (28.center) to (30.center);
		\draw [style=rede] (27.center) to (29.center);
		\draw [style=rede] (7) to (8.center);
		\draw [style=rede] (26.center) to (23.center);
		\draw [style=rede] (3) to (31.center);
		\draw [style=rede] (6) to (15.center);
		\draw [style=cyane] (4) to (5);
		\draw [style=rede] (8.center) to (15.center);
	\end{pgfonlayer}
\end{tikzpicture}}\quad,
\end{equation}
from which we can read the magnetic quiver
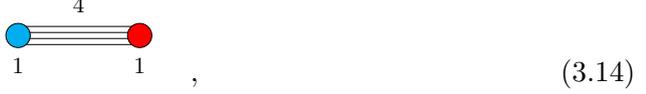
\begin{equation}
     \centering
     \scalebox{0.8}{
\begin{tikzpicture}
	\begin{pgfonlayer}{nodelayer}
		\node [style=cyangauge] (0) at (-1, 0) {};
		\node [style=redgauge] (1) at (1, 0) {};
		\node [style=none] (2) at (-1, 0.15) {};
		\node [style=none] (3) at (-1, 0.05) {};
		\node [style=none] (4) at (-1, -0.05) {};
		\node [style=none] (5) at (-1, -0.15) {};
		\node [style=none] (6) at (1, 0.15) {};
		\node [style=none] (7) at (1, 0.05) {};
		\node [style=none] (8) at (1, -0.05) {};
		\node [style=none] (9) at (1, -0.15) {};
		\node [style=none] (10) at (-1, -0.5) {1};
		\node [style=none] (11) at (1, -0.5) {1};
		\node [style=none] (12) at (0, 0.5) {4};
	\end{pgfonlayer}
	\begin{pgfonlayer}{edgelayer}
		\draw (2.center) to (6.center);
		\draw (3.center) to (7.center);
		\draw (4.center) to (8.center);
		\draw (5.center) to (9.center);
	\end{pgfonlayer}
\end{tikzpicture}  }\quad,
\end{equation}
which represents the variety $\mathbb{C}^2/\mathbb{Z}_4$ with the HWG
\begin{equation}
\mathrm{HWG}_\textrm{B}=\mathrm{PE}[t^2+(q+q^{-1})t^4 - t^8]\,,
\end{equation}
where $q$ is a fugacity for the global $U(1)$ symmetry. The unrefined Hilbert series is
\begin{equation}
    \mathrm{HS}_\textrm{B}= \frac{1-t^8}{(1-t^2)(1-t^4)^2} \,.
\end{equation}
\end{enumerate}
We can find the common part of both decompositions:
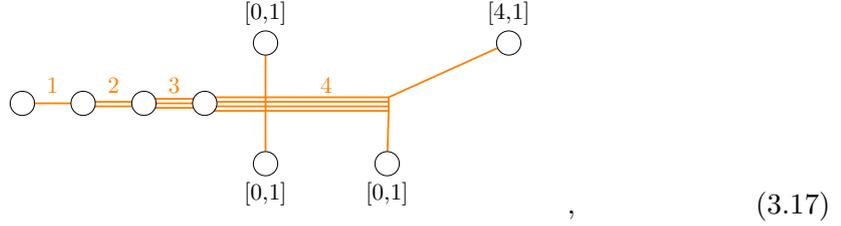
\begin{enumerate}
\setcounter{enumi}{2}
\item Intersection
\begin{equation}
       \centering
       \scalebox{0.8}{\begin{tikzpicture}
	\begin{pgfonlayer}{nodelayer}
		\node [style=gaugeX] (0) at (-2, 0) {};
		\node [style=gaugeX] (1) at (-3, 0) {};
		\node [style=gaugeX] (2) at (-4, 0) {};
		\node [style=gaugeX] (3) at (-5, 0) {};
		\node [style=gaugeX] (4) at (-1, 1) {};
		\node [style=gaugeX] (5) at (-1, -1) {};
		\node [style=gaugeX] (6) at (1, -1) {};
		\node [style=gaugeX] (7) at (3, 1) {};
		\node [style=none] (8) at (1.025, 0.1) {};
		\node [style=none] (9) at (1.025, 0.025) {};
		\node [style=none] (10) at (1.025, -0.05) {};
		\node [style=none] (11) at (-2, 0.1) {};
		\node [style=none] (12) at (-2, 0.025) {};
		\node [style=none] (13) at (-2, -0.05) {};
		\node [style=none] (14) at (-2, -0.125) {};
		\node [style=none] (15) at (1.025, -0.125) {};
		\node [style=none] (16) at (0, 0.3) {\color{orange}{4}};
		\node [style=none] (17) at (-1, 1.5) {[0,1]};
		\node [style=none] (18) at (-1, -1.5) {[0,1]};
		\node [style=none] (19) at (1, -1.5) {[0,1]};
		\node [style=none] (20) at (3, 1.5) {[4,1]};
		\node [style=none] (21) at (-2, 0.075) {};
		\node [style=none] (22) at (-2, 0) {};
		\node [style=none] (23) at (-2, -0.075) {};
		\node [style=none] (24) at (-3, 0.075) {};
		\node [style=none] (25) at (-3, 0) {};
		\node [style=none] (26) at (-3, -0.075) {};
		\node [style=none] (27) at (-4, 0.025) {};
		\node [style=none] (28) at (-4, -0.05) {};
		\node [style=none] (29) at (-3, 0.025) {};
		\node [style=none] (30) at (-3, -0.05) {};
		\node [style=none] (31) at (-4, 0) {};
		\node [style=none] (32) at (-2.5, 0.3) {\color{orange}{3}};
		\node [style=none] (33) at (-3.5, 0.3) {\color{orange}{2}};
		\node [style=none] (34) at (-4.5, 0.3) {\color{orange}{1}};
	\end{pgfonlayer}
	\begin{pgfonlayer}{edgelayer}
		\draw [style=orangee] (14.center) to (15.center);
		\draw [style=orangee] (13.center) to (10.center);
		\draw [style=orangee] (12.center) to (9.center);
		\draw [style=orangee] (11.center) to (8.center);
		\draw [style=orangee] (21.center) to (24.center);
		\draw [style=orangee] (22.center) to (25.center);
		\draw [style=orangee] (28.center) to (30.center);
		\draw [style=orangee] (27.center) to (29.center);
		\draw [style=orangee] (7) to (8.center);
		\draw [style=orangee] (26.center) to (23.center);
		\draw [style=orangee] (3) to (31.center);
		\draw [style=orangee] (6) to (15.center);
		\draw [style=orangee] (4) to (5);
		\draw [style=orangee] (8.center) to (15.center);
	\end{pgfonlayer}
\end{tikzpicture}}\quad,
\end{equation}
which is the entire brane web, therefore the intersection is trivial. We can read the magnetic quiver
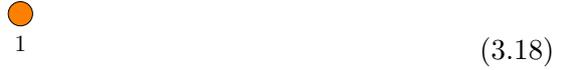
\begin{equation}
        \centering
        \scalebox{0.8}{
\begin{tikzpicture}
	\begin{pgfonlayer}{nodelayer}
		\node [style=orangegauge] (0) at (0, 0) {};
		\node [style=none] (1) at (0, -0.5) {1};
	\end{pgfonlayer}\quad,
\end{tikzpicture}}
\end{equation}
which has the HWG
\begin{equation}
    \mathrm{HWG}_I=1 \,.
\end{equation}
\end{enumerate}
We can now write the proposed HWG of the full Higgs Branch
\begin{equation}
\label{HSbraneWeb44}
\begin{split}
\mathrm{HWG}_{\textrm{Brane Web}}=&\mathrm{HWG}_\textrm{M}+\mathrm{HWG}_\textrm{B}-\mathrm{HWG}_I\\
=&\mathrm{PE}[\mu_1\mu_3t^2+\mu_2^2t^4]+\mathrm{PE}[t^2+(q+q^{-1})t^4 - t^8]-1 
\end{split}
\end{equation}
and can see that the global symmetry of the Higgs Branch is $\mathrm{SU}(4)\times \mathrm{U}(1)$, where $\mathrm{U}(1)$ is the baryon symmetry. If we unrefine this expression, we find 
\begin{equation*}
    \mathrm{HS}_{\textnormal{Brane Web}}=\frac{\left(1+t^2\right)^2 \left(1+5t^2+t^4\right)}{(1-t^2)^8 } + \frac{1-t^8}{(1-t^2)(1-t^4)^2} -1 = 
\end{equation*}
\begin{equation}
    =\frac{1+9t^2+15t^4+21t^6+24t^8-39t^{10}+44t^{12}-26t^{14}+8t^{16}-t^{18}}{(1-t^2)^8  \left(1+t^2\right)} \,. 
\end{equation}
We note that this does \emph{not} agree with the Hilbert series given by Table \ref{tabSUSQCD}. We come back to this disagreement in section \ref{sectionSU44multiplicities}. 

\subsection{\texorpdfstring{$\mathrm{SU}(3)$ with $N_f=4$ -- Two Cones with Non-Trivial Intersection}{SU(3) with 4 flavours}}

The toric diagram for $\mathrm{SU}(3)$ with $N_f=4$ massless flavours at finite coupling and CS level $\pm1$ is given by
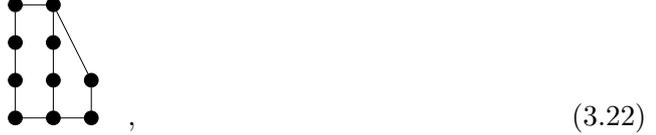
\begin{equation}
         \centering
\scalebox{0.5}{\begin{tikzpicture}
	\begin{pgfonlayer}{nodelayer}
		\node [style=smallgauge] (0) at (0, 3) {};
		\node [style=smallgauge] (1) at (0, 2) {};
		\node [style=smallgauge] (2) at (0, 1) {};
		\node [style=smallgauge] (3) at (0, 0) {};
		\node [style=smallgauge] (4) at (-1, 3) {};
		\node [style=smallgauge] (5) at (1, 1) {};
		\node [style=smallgauge] (6) at (1, 0) {};
		\node [style=smallgauge] (7) at (-1, 2) {};
		\node [style=smallgauge] (8) at (-1, 1) {};
		\node [style=smallgauge] (9) at (-1, 0) {};
	\end{pgfonlayer}
	\begin{pgfonlayer}{edgelayer}
		\draw (4) to (9);
		\draw (4) to (0);
		\draw (0) to (5);
		\draw (5) to (6);
		\draw (9) to (6);
		\draw (0) to (3);
	\end{pgfonlayer}
\end{tikzpicture}}\quad,
\end{equation}
the corresponding brane web is
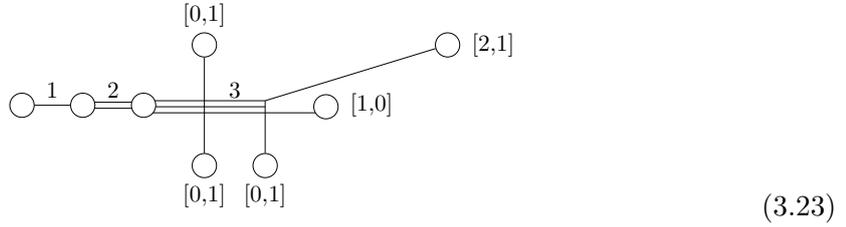
\begin{equation}
        \centering
        \scalebox{0.8}{
\begin{tikzpicture}
	\begin{pgfonlayer}{nodelayer}
		\node [style=gaugeX] (0) at (-1, 0) {};
		\node [style=gaugeX] (1) at (-2, 0) {};
		\node [style=gaugeX] (2) at (-3, 0) {};
		\node [style=gaugeX] (3) at (0, 1) {};
		\node [style=gaugeX] (4) at (0, -1) {};
		\node [style=gaugeX] (5) at (4, 1) {};
		\node [style=gaugeX] (6) at (2, -0.025) {};
		\node [style=gaugeX] (7) at (1, -1) {};
		\node [style=none] (8) at (1, -0.125) {};
		\node [style=none] (9) at (1, 0.075) {};
		\node [style=none] (10) at (1, -0.025) {};
		\node [style=none] (11) at (-1, 0.075) {};
		\node [style=none] (12) at (-1, -0.025) {};
		\node [style=none] (13) at (-1, -0.125) {};
		\node [style=none] (14) at (-1, 0.05) {};
		\node [style=none] (15) at (-1, -0.05) {};
		\node [style=none] (16) at (-2, 0.05) {};
		\node [style=none] (17) at (-2, -0.05) {};
		\node [style=none] (18) at (0, 1.5) {[0,1]};
		\node [style=none] (19) at (0, -1.5) {[0,1]};
		\node [style=none] (20) at (1, -1.5) {[0,1]};
		\node [style=none] (21) at (2.75, 0) {[1,0]};
		\node [style=none] (22) at (4.75, 1) {[2,1]};
		\node [style=none] (23) at (0.5, 0.25) {3};
		\node [style=none] (24) at (-1.5, 0.25) {2};
		\node [style=none] (25) at (-2.5, 0.25) {1};
		\node [style=none] (26) at (2, -0.125) {};
	\end{pgfonlayer}
	\begin{pgfonlayer}{edgelayer}
		\draw (11.center) to (9.center);
		\draw (10.center) to (12.center);
		\draw [in=180, out=0, looseness=1.25] (13.center) to (8.center);
		\draw (8.center) to (7);
		\draw (9.center) to (5);
		\draw (3) to (4);
		\draw (16.center) to (14.center);
		\draw (15.center) to (17.center);
		\draw (2) to (1);
		\draw (26.center) to (8.center);
		\draw (9.center) to (8.center);
	\end{pgfonlayer}
\end{tikzpicture}}
\end{equation}
and there are two maximal decompositions:

\begin{enumerate}
\item The mesonic branch
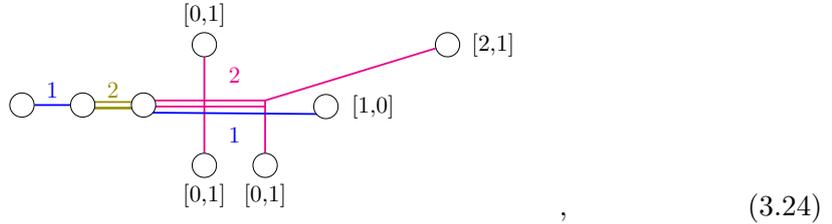
\begin{equation}
        \centering
        \scalebox{0.8}{
\begin{tikzpicture}
	\begin{pgfonlayer}{nodelayer}
		\node [style=gaugeX] (0) at (-2, 0) {};
		\node [style=gaugeX] (1) at (-3, 0) {};
		\node [style=gaugeX] (2) at (-4, 0) {};
		\node [style=gaugeX] (3) at (-1, 1) {};
		\node [style=gaugeX] (4) at (-1, -1) {};
		\node [style=gaugeX] (5) at (3, 1) {};
		\node [style=gaugeX] (6) at (1, -0.025) {};
		\node [style=gaugeX] (7) at (0, -1) {};
		\node [style=none] (9) at (0, 0.075) {};
		\node [style=none] (10) at (0, -0.025) {};
		\node [style=none] (11) at (-2, 0.075) {};
		\node [style=none] (12) at (-2, -0.025) {};
		\node [style=none] (13) at (-2, -0.125) {};
		\node [style=none] (14) at (-2, 0.05) {};
		\node [style=none] (15) at (-2, -0.05) {};
		\node [style=none] (16) at (-3, 0.05) {};
		\node [style=none] (17) at (-3, -0.05) {};
		\node [style=none] (18) at (-1, 1.5) {[0,1]};
		\node [style=none] (19) at (-1, -1.5) {[0,1]};
		\node [style=none] (20) at (0, -1.5) {[0,1]};
		\node [style=none] (21) at (1.775, -0.025) {[1,0]};
		\node [style=none] (22) at (3.75, 1) {[2,1]};
		\node [style=none] (23) at (-0.5, 0.5) {\color{magenta}{2}};
		\node [style=none] (24) at (-2.5, 0.25) {\color{olive}{2}};
		\node [style=none] (25) at (-3.5, 0.25) {\color{blue}{1}};
		\node [style=none] (26) at (1, -0.15) {};
		\node [style=none] (27) at (-0.5, -0.5) {\color{blue}{1}};
	\end{pgfonlayer}
	\begin{pgfonlayer}{edgelayer}
		\draw [style=magentae] (9.center) to (5);
		\draw [style=magentae] (3) to (4);
		\draw [style=magentae] (9.center) to (11.center);
		\draw [style=magentae] (10.center) to (12.center);
		\draw [style=magentae] (10.center) to (7);
		\draw [style=bluee] (13.center) to (26.center);
		\draw [style=olivee] (17.center) to (15.center);
		\draw [style=olivee] (17.center) to (15.center);
		\draw [style=olivee] (17.center) to (15.center);
		\draw [style=olivee] (16.center) to (14.center);
		\draw [style=bluee] (2) to (1);
		\draw [style=magentae] (9.center) to (10.center);
	\end{pgfonlayer}
\end{tikzpicture}
}\quad,
\end{equation}
from which we can read the magnetic quiver
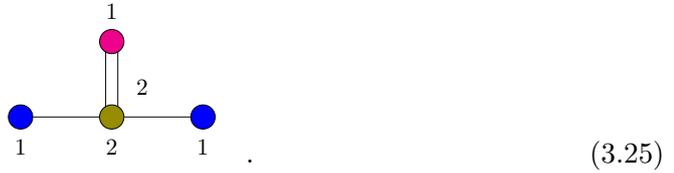
\begin{equation}
    \centering
    \scalebox{0.8}{
\begin{tikzpicture}
	\begin{pgfonlayer}{nodelayer}
		\node [style=none] (4) at (-0.1, 1.25) {};
		\node [style=none] (5) at (0.1, 1.25) {};
		\node [style=none] (6) at (-0.1, 0) {};
		\node [style=none] (7) at (0.1, 0) {};
		\node [style=none] (8) at (0, 1.75) {1};
		\node [style=none] (9) at (1.5, -0.5) {1};
		\node [style=none] (10) at (0, -0.5) {2};
		\node [style=none] (11) at (-1.5, -0.5) {1};
		\node [style=none] (12) at (0.5, 0.5) {2};
		\node [style=olivegauge] (13) at (0, 0) {};
		\node [style=bluegauge] (14) at (1.5, 0) {};
		\node [style=bluegauge] (15) at (-1.5, 0) {};
		\node [style=magentagauge] (16) at (0, 1.25) {};
	\end{pgfonlayer}
	\begin{pgfonlayer}{edgelayer}
		\draw (5.center) to (7.center);
		\draw [in=90, out=-90] (4.center) to (6.center);
		\draw (13) to (14);
		\draw (13) to (15);
	\end{pgfonlayer}
\end{tikzpicture}}\quad.
\end{equation}
The mesonic branch is therefore conjectured to be the same as in the previous two examples, the closure of the next-to-minimal nilpotent orbit of $\mathfrak{sl}(4,\mathbb{C})$, with HWG
\begin{equation}
\mathrm{HWG}_\textrm{M}=\mathrm{PE}[\mu_1\mu_3t^2+\mu_2^2t^4] \,. 
\end{equation}
\item The baryonic branch
\begin{equation}
        \centering
        \scalebox{0.8}{\begin{tikzpicture}
	\begin{pgfonlayer}{nodelayer}
		\node [style=gaugeX] (0) at (-2, 0) {};
		\node [style=gaugeX] (1) at (-3, 0) {};
		\node [style=gaugeX] (2) at (-4, 0) {};
		\node [style=gaugeX] (3) at (-1, 1) {};
		\node [style=gaugeX] (4) at (-1, -1) {};
		\node [style=gaugeX] (5) at (3, 1) {};
		\node [style=gaugeX] (6) at (1, -0.025) {};
		\node [style=gaugeX] (7) at (0, -1) {};
		\node [style=none] (8) at (0, 0.075) {};
		\node [style=none] (9) at (0, -0.025) {};
		\node [style=none] (10) at (-2, 0.075) {};
		\node [style=none] (11) at (-2, -0.025) {};
		\node [style=none] (12) at (-2, -0.125) {};
		\node [style=none] (13) at (-2, 0.05) {};
		\node [style=none] (14) at (-2, -0.05) {};
		\node [style=none] (15) at (-3, 0.05) {};
		\node [style=none] (16) at (-3, -0.05) {};
		\node [style=none] (17) at (-1, 1.5) {[0,1]};
		\node [style=none] (18) at (-1, -1.5) {[0,1]};
		\node [style=none] (19) at (0, -1.5) {[0,1]};
		\node [style=none] (20) at (1.775, -0.025) {[1,0]};
		\node [style=none] (21) at (3.75, 1) {[2,1]};
		\node [style=none] (22) at (-0.5, 0.3) {\color{red}{2}};
		\node [style=none] (23) at (-2.5, 0.3) {\color{red}{1}};
		\node [style=none] (24) at (-3.5, 0.3) {\color{blue}{1}};
		\node [style=none] (25) at (1, -0.15) {};
		\node [style=none] (26) at (-0.5, -0.3) {\color{blue}{1}};
		\node [style=none] (27) at (-2.5, -0.3) {\color{olive}{1}};
	\end{pgfonlayer}
	\begin{pgfonlayer}{edgelayer}
		\draw [style=rede] (8.center) to (5);
		\draw [style=rede] (8.center) to (10.center);
		\draw [style=rede] (9.center) to (11.center);
		\draw [style=rede] (9.center) to (7);
		\draw [style=bluee] (12.center) to (25.center);
		\draw [style=olivee] (16.center) to (14.center);
		\draw [style=rede] (15.center) to (13.center);
		\draw [style=bluee] (2) to (1);
		\draw [style=cyane] (3) to (4);
		\draw [style=rede] (8.center) to (9.center);
	\end{pgfonlayer}
\end{tikzpicture}
}\quad,
\end{equation}
from which we can read the magnetic quiver
\begin{equation}
  \centering
\scalebox{0.8}{\begin{tikzpicture}
	\begin{pgfonlayer}{nodelayer}
		\node [style=redgauge] (0) at (-0.75, 0.875) {};
		\node [style=cyangauge] (1) at (0.75, 0.875) {};
		\node [style=bluegauge] (2) at (-1.75, -0.5) {};
		\node [style=bluegauge] (4) at (1.75, -0.5) {};
		\node [style=olivegauge] (6) at (0, -0.5) {};
		\node [style=none] (7) at (-0.75, 1.375) {1};
		\node [style=none] (8) at (0.75, 1.375) {1};
		\node [style=none] (9) at (1.75, -1) {1};
		\node [style=none] (10) at (0, -1) {1};
		\node [style=none] (11) at (-1.75, -1) {};
		\node [style=none] (12) at (-1.75, -1) {1};
		\node [style=none] (13) at (-0.625, 0.95) {};
		\node [style=none] (14) at (-0.625, 0.8) {};
		\node [style=none] (15) at (0.65, 0.95) {};
		\node [style=none] (16) at (0.65, 0.8) {};
		\node [style=none] (17) at (0, 1.125) {2};
	\end{pgfonlayer}
	\begin{pgfonlayer}{edgelayer}
		\draw (2) to (6);
		\draw (6) to (4);
		\draw (0) to (2);
		\draw (1) to (4);
		\draw (13.center) to (15.center);
		\draw (14.center) to (16.center);
	\end{pgfonlayer}
\end{tikzpicture}
}\quad,
\end{equation}
which is a baryonic extension (see Appendix \ref{appendixNO}) of the closure of the minimal nilpotent orbit of $\mathfrak{sl}(4,\mathbb{C})$, with the HWG
\begin{equation}
\mathrm{HWG}_\textrm{B}=\mathrm{PE}[t^2+\mu_1\mu_3t^2+(q\mu_1+q^{-1}\mu_3)t^3-\mu_1\mu_3t^{6}] \,. 
\end{equation}
The unrefined Hilbert series is 
\begin{equation}
\label{HS34bar}
  \mathrm{HS}_\textrm{B} =  \frac{\left( \begin{array}{c}
      1 +2 t+13 t^2+28 t^3 +62 t^4+88 t^5+128 t^6+132 t^7 \\
     +128 t^8+ 88 t^9 +62 t^{10}+28 t^{11}+13 t^{12}+2 t^{13}+t^{14}
  \end{array} \right)  }{(1-t)^8 (1+t)^6
   \left(1+t+t^2\right)^4} \,. 
\end{equation}
\end{enumerate}
We can find the common part of both decompositions
\begin{enumerate}
\setcounter{enumi}{2}
\item Intersection
\begin{equation}
        \centering
        \scalebox{0.8}{
\begin{tikzpicture}
	\begin{pgfonlayer}{nodelayer}
		\node [style=gaugeX] (0) at (-2, 0) {};
		\node [style=gaugeX] (1) at (-3, 0) {};
		\node [style=gaugeX] (2) at (-4, 0) {};
		\node [style=gaugeX] (3) at (-1, 1) {};
		\node [style=gaugeX] (4) at (-1, -1) {};
		\node [style=gaugeX] (5) at (3, 1) {};
		\node [style=gaugeX] (6) at (1, -0.025) {};
		\node [style=gaugeX] (7) at (0, -1) {};
		\node [style=none] (8) at (0, 0.075) {};
		\node [style=none] (9) at (0, -0.025) {};
		\node [style=none] (10) at (-2, 0.075) {};
		\node [style=none] (11) at (-2, -0.025) {};
		\node [style=none] (12) at (-2, -0.125) {};
		\node [style=none] (13) at (-2, 0.05) {};
		\node [style=none] (14) at (-2, -0.05) {};
		\node [style=none] (15) at (-3, 0.05) {};
		\node [style=none] (16) at (-3, -0.05) {};
		\node [style=none] (17) at (-1, 1.5) {[0,1]};
		\node [style=none] (18) at (-1, -1.5) {[0,1]};
		\node [style=none] (19) at (0, -1.5) {[0,1]};
		\node [style=none] (20) at (1.775, -0.025) {[1,0]};
		\node [style=none] (21) at (3.75, 1) {(2,1)};
		\node [style=none] (22) at (-0.5, 0.3) {\color{orange}{2}};
		\node [style=none] (23) at (-2.5, 0.3) {\color{orange}{1}};
		\node [style=none] (24) at (-3.5, 0.3) {\color{blue}{1}};
		\node [style=none] (25) at (1, -0.15) {};
		\node [style=none] (26) at (-0.5, -0.3) {\color{blue}{1}};
		\node [style=none] (27) at (-2.5, -0.3) {\color{olive}{1}};
	\end{pgfonlayer}
	\begin{pgfonlayer}{edgelayer}
		\draw [style=orangee] (8.center) to (5);
		\draw [style=orangee] (8.center) to (10.center);
		\draw [style=orangee] (9.center) to (11.center);
		\draw [style=orangee] (9.center) to (7);
		\draw [style=bluee] (12.center) to (25.center);
		\draw [style=olivee] (16.center) to (14.center);
		\draw [style=orangee] (15.center) to (13.center);
		\draw [style=bluee] (2) to (1);
		\draw [style=orangee] (3) to (4);
		\draw [style=orangee] (8.center) to (9.center);
	\end{pgfonlayer}
\end{tikzpicture}
}\quad,
\end{equation}
from which we can read the magnetic quiver
\begin{equation}
        \centering
        \scalebox{0.8}{
\begin{tikzpicture}
	\begin{pgfonlayer}{nodelayer}
		\node [style=orangegauge] (0) at (0, 1) {};
		\node [style=bluegauge] (1) at (-1, 0) {};
		\node [style=olivegauge] (2) at (0, 0) {};
		\node [style=bluegauge] (3) at (1, 0) {};
		\node [style=none] (4) at (-1, -0.5) {1};
		\node [style=none] (5) at (0, -0.5) {1};
		\node [style=none] (6) at (1, -0.5) {1};
		\node [style=none] (7) at (0, 1.5) {1};
	\end{pgfonlayer}
	\begin{pgfonlayer}{edgelayer}
		\draw (0) to (1);
		\draw (1) to (2);
		\draw (2) to (3);
		\draw (0) to (3);
	\end{pgfonlayer}
\end{tikzpicture}}\quad.
\end{equation}
Hence, the intersection of the mesonic and baryonic branch is the closure of the minimal nilpotent orbit of $\mathfrak{sl}(4,\mathbb{C})$ with the HWG
\begin{equation}
\mathrm{HWG}_I=\mathrm{PE}[\mu_1\mu_3t^2].
\end{equation}
The unrefined Hilbert series is 
\begin{equation}
\label{HS34inter}
  \mathrm{HS}_I=  \frac{\left(1+t^2\right) \left(1+8 t^2+t^4\right)}{(1-t)^6 (1+t)^6} \,. 
\end{equation}
\end{enumerate}
Now we can now write the proposed HWG of the full Higgs Branch
\begin{equation}
\label{HWGdecomposition}
\mathrm{HWG}_{\textrm{Brane Web}}=\mathrm{HWG}_\textrm{M}+\mathrm{HWG}_\textrm{B}-\mathrm{HWG}_I \,. 
\end{equation}
Correspondingly, we find the unrefined Hilbert series by combining (\ref{HS34mes}), (\ref{HS34bar}) and (\ref{HS34inter}). This time the result precisely agrees with the entry of Table \ref{tabSUSQCD}. 

\subsection{\texorpdfstring{$\mathrm{SU}(2)$ with $N_f=4$ -- Single Baryonic Cone}{SU(2) with 4 flavours}}

The toric diagram for $SU(2)$ with $N_f=4$ massless flavours at finite gauge coupling and CS level $0$ is given by
\begin{equation}
         \centering
\scalebox{0.5}{\begin{tikzpicture}
	\begin{pgfonlayer}{nodelayer}
		\node [style=smallgauge] (1) at (0, 2) {};
		\node [style=smallgauge] (2) at (0, 1) {};
		\node [style=smallgauge] (3) at (0, 0) {};
		\node [style=smallgauge] (5) at (1, 1) {};
		\node [style=smallgauge] (6) at (1, 0) {};
		\node [style=smallgauge] (7) at (-1, 2) {};
		\node [style=smallgauge] (8) at (-1, 1) {};
		\node [style=smallgauge] (9) at (-1, 0) {};
		\node [style=smallgauge] (10) at (1, 2) {};
	\end{pgfonlayer}
	\begin{pgfonlayer}{edgelayer}
		\draw (5) to (6);
		\draw (9) to (6);
		\draw (7) to (9);
		\draw (7) to (10);
		\draw (10) to (5);
		\draw (1) to (3);
	\end{pgfonlayer}
\end{tikzpicture}
}\quad,
\end{equation}
the corresponding brane web is
\begin{equation}
    \scalebox{0.8}{\begin{tikzpicture}
	\begin{pgfonlayer}{nodelayer}
		\node [style=gaugeX] (0) at (-1, 0) {};
		\node [style=gaugeX] (1) at (-2, 0) {};
		\node [style=gaugeX] (3) at (0, 1) {};
		\node [style=gaugeX] (4) at (0, -1) {};
		\node [style=gaugeX] (6) at (2, 0) {};
		\node [style=gaugeX] (7) at (1, -1) {};
		\node [style=none] (18) at (0, 1.5) {[0,1]};
		\node [style=none] (19) at (0, -1.5) {[0,1]};
		\node [style=none] (20) at (1, -1.5) {[0,1]};
		\node [style=gaugeX] (22) at (1, 1) {};
		\node [style=none] (23) at (-1, 0.125) {};
		\node [style=none] (24) at (-1, -0.125) {};
		\node [style=none] (25) at (2, 0.125) {};
		\node [style=none] (26) at (2, -0.125) {};
		\node [style=gaugeX] (27) at (3, 0) {};
		\node [style=none] (28) at (1, 1.5) {$[0,1]$};
		\node [style=none] (29) at (3.7, 0) {$[1,0]$};
		\node [style=none] (30) at (0.5, 0.4) {2};
		\node [style=none] (31) at (2.5, 0.275) {1};
		\node [style=none] (32) at (-1.5, 0.275) {1};
	\end{pgfonlayer}
	\begin{pgfonlayer}{edgelayer}
		\draw (3) to (4);
		\draw (22) to (7);
		\draw (24.center) to (26.center);
		\draw (25.center) to (23.center);
		\draw (1) to (0);
		\draw (27) to (6);
	\end{pgfonlayer}
\end{tikzpicture}
}
\end{equation}
and there is only one maximal decomposition, corresponding to the baryonic branch,
\begin{equation}
\scalebox{0.8}{
\begin{tikzpicture}
	\begin{pgfonlayer}{nodelayer}
		\node [style=gaugeX] (0) at (-1, 0) {};
		\node [style=gaugeX] (1) at (-2, 0) {};
		\node [style=gaugeX] (3) at (0, 1) {};
		\node [style=gaugeX] (4) at (0, -1) {};
		\node [style=gaugeX] (6) at (2, 0) {};
		\node [style=gaugeX] (7) at (1, -1) {};
		\node [style=none] (18) at (0, 1.5) {[0,1]};
		\node [style=none] (19) at (0, -1.5) {[0,1]};
		\node [style=none] (20) at (1, -1.5) {[0,1]};
		\node [style=gaugeX] (22) at (1, 1) {};
		\node [style=none] (23) at (-1, 0.125) {};
		\node [style=none] (24) at (-1, -0.125) {};
		\node [style=none] (25) at (2, 0.125) {};
		\node [style=none] (26) at (2, -0.125) {};
		\node [style=gaugeX] (27) at (3, 0) {};
		\node [style=none] (28) at (1, 1.5) {$[0,1]$};
		\node [style=none] (29) at (3.7, 0) {$[1,0]$};
		\node [style=none] (30) at (0.5, 0.4) {{\color{olive}{2}}};
		\node [style=none] (31) at (2.5, 0.275) {{\color{blue}{1}}};
		\node [style=none] (32) at (-1.5, 0.275) {{\color{blue}{1}}};
	\end{pgfonlayer}
	\begin{pgfonlayer}{edgelayer}
		\draw [style=rede] (22) to (7);
		\draw [style=cyane] (3) to (4);
		\draw [style=olivee] (25.center) to (23.center);
		\draw [style=olivee] (24.center) to (26.center);
		\draw [style=bluee] (1) to (0);
		\draw [style=bluee] (27) to (6);
	\end{pgfonlayer}
\end{tikzpicture}}\quad ,
\end{equation}
which leads to the magnetic quiver
\begin{equation}
        \centering
        \scalebox{0.8}{
\begin{tikzpicture}
	\begin{pgfonlayer}{nodelayer}
		\node [style=bluegauge] (1) at (-1, 0) {};
		\node [style=olivegauge] (2) at (0, 0) {};
		\node [style=bluegauge] (3) at (1, 0) {};
		\node [style=none] (4) at (-1, -0.5) {1};
		\node [style=none] (5) at (0, -0.5) {2};
		\node [style=none] (6) at (1, -0.5) {1};
		\node [style=cyangauge] (7) at (-0.75, 1) {};
		\node [style=redgauge] (8) at (0.75, 1) {};
		\node [style=none] (9) at (-0.75, 1.5) {1};
		\node [style=none] (10) at (0.75, 1.5) {1};
	\end{pgfonlayer}
	\begin{pgfonlayer}{edgelayer}
		\draw (1) to (2);
		\draw (2) to (3);
		\draw (8) to (2);
		\draw (2) to (7);
	\end{pgfonlayer}
\end{tikzpicture}
}\quad.
\end{equation}
The Higgs branch is therefore conjectured to be the closure of the minimal nilpotent orbit of $\mathfrak{so}(8)$, whith the corresponding HWG
\begin{equation}
    \mathrm{HWG}_{\textnormal{Brane Web}}=\mathrm{PE}[\mu_2t^2]\, .
    \label{HWGSU2,4}
\end{equation}
The global symmetry of the Higgs Branch is $SO(8)$. Upon unrefining all the $SO(8)$ fugacities and expanding the HWG (\ref{HWGSU2,4}) into a standard Hilbert series, one obtains
\begin{equation}
    \mathrm{HS}_{\textnormal{Brane Web}}=\frac{1 + 18 t^2 + 65 t^4 + 65 t^6 + 18 t^8 + t^{10}}{(1-t)^{10} (1+t)^{10}}\, .
\end{equation}
This is result precisely agrees with the entry of Table \ref{tabSUSQCD}.

\subsection{General results}

The moduli spaces obtained with the brane web method all consist of unions of nilpotent orbit closures or baryonic extensions of nilpotent orbit closures (see table \ref{finitebranches}). The corresponding Hasse diagrams \cite{Bourget:2019aer} are presented in Table \ref{tableHasse}. As an illustration, we draw explicitly the Hasse diagrams for $N_c=6$ and low values of $N_f$ in Figure \ref{figForest}. 

\begin{table}[t]
\centering
\renewcommand{\arraystretch}{2}
\begin{tabular}{|c|c|c|c|}
\hline
Range & Baryonic Branch & Intersection & Mesonic Branch \\
\hline
$N_f \geq 2N_c$ &\raisebox{-0.25\height}{\scalebox{0.8}{\begin{tikzpicture}
\draw (0,0)--(3,0) (1,0)--(1,1) (2,0)--(2,1);
\end{tikzpicture} }}& \multicolumn{2}{c|}{ Intersection = Mesonic} \\
\hline
$N_f=2N_c-1$ &\raisebox{-0.25\height}{\scalebox{0.8}{ \begin{tikzpicture}
\draw (0,0)--(3,0) (1,0)--(4/3,1) (4/3,1)--(5/3,1) (2,0)--(5/3,1);
\end{tikzpicture} }}& \multicolumn{2}{c|}{ Intersection = Mesonic}\\
\hline
$N_c\leq N_f<2N_c-1$ &\raisebox{-0.25\height}{\scalebox{0.8}{ \begin{tikzpicture}
\draw (0,0)--(3,0) (1,0)--(4/3,1) (2,0)--(5/3,1);
\draw[thick] (4/3,1)--(5/3,1);
\end{tikzpicture} }}& \raisebox{-0.25\height}{\scalebox{0.8}{
\begin{tikzpicture}
\draw (0,0)--(3,0) (1,0)--(3/2,1) (2,0)--(3/2,1);
\end{tikzpicture}}} & \raisebox{-0.25\height}{\scalebox{0.8}{
\begin{tikzpicture} 
\draw (0,0)--(3,0) (14/10,0)--(3/2,1) (16/10,0)--(3/2,1);
\end{tikzpicture}}} \\
\hline
$N_f<N_c$ & no baryons! &  \multicolumn{2}{c|}{\raisebox{-0.25\height}{\scalebox{0.8}{
\begin{tikzpicture}
\draw (0,0)--(3,0) (14/10,0)--(3/2,1) (16/10,0)--(3/2,1);
\end{tikzpicture}}}
}\\
\hline
\end{tabular}
\caption{Schematic representation of the magnetic quivers for $\mathrm{SU}(N_c)$ SQCD with 8 supercharges obtained from five-brane webs. The drawing in a given square suggests the shape of the magnetic quiver. The actual magnetic quivers can be found in Figures \ref{compfirst}-\ref{complast}. }
\label{finitebranches}
\end{table}

\begin{table}[]
    \centering
    \begin{tabular}{|c|c|}
    \hline 
        $N_f \leq 2N_c -2$ & $N_f \geq 2 N_c-2$  \\ \hline 
\begin{tikzpicture}[scale=1]
\tikzstyle{hasse} = [circle, fill,inner sep=2pt]
\node[hasse] (0) at (0,0) {};
\node[hasse] (1) at (0,1) {};
\node[hasse] (2) at (0,2) {};
\node at (0,2.6) {$\vdots$};
\node[hasse] (3) at (0,3) {};
\node[hasse] (4) at (0,4) {};
\node[hasse] (5) at (0,5) {};
\node at (0,5.6) {$\vdots$};
\node[hasse] (6) at (0,6) {};
\node[hasse] (7) at (0,7) {};
\node[hasse] (8) at (0,8) {};
\node at (0,-0.5) {$ $};
\node at (0,9) {$ $};
\node at (0,8.5) {Meson};
\node[hasse] (9) at (-1,5) {};
\node at (-1,5.5) {Baryon};
\draw (0)--(1)--(2);
\draw (3)--(4)--(5);
\draw (6)--(7)--(8);
\draw (4)--(9);
\node at (.7,7.5) {$a_{1+X}$};
\node at (.7,6.5) {$a_{3+X}$};
\node at (1,4.5) {$a_{2N_c-N_f-1}$};
\node at (1,3.5) {$a_{2N_c-N_f+1}$};
\node at (.7,1.5) {$a_{N_f-3}$};
\node at (.7,0.5) {$a_{N_f-1}$};
\node at (-1.2,4.2) {$A_{2N_c-N_f-1}$};
\end{tikzpicture} 
&
\begin{tikzpicture}[scale=1]
\tikzstyle{hasse} = [circle, fill,inner sep=2pt]
\node[hasse] (0) at (0,0) {};
\node[hasse] (1) at (0,1) {};
\node[hasse] (2) at (0,2) {};
\node (3) at (0,3.7) {$\vdots$};
\node[hasse] (4) at (0,5) {};
\node[hasse] (5) at (0,6) {};
\node[hasse] (6) at (0,7) {};
\node[hasse] (7) at (0,8) {};
\node at (-1,-0.5) {$ $};
\node at (2.5,9) {$ $};
\node at (0,8.5) {Baryon};
\draw (0)--(1)--(2);
\draw (4)--(5)--(6)--(7);
\node at (1,7.5) {$d_{N_f-2N_c+4}$};
\node at (1,6.5) {$a_{N_f-2N_c+5}$};
\node at (1,5.5) {$a_{N_f-2N_c+7}$};
\node at (.7,1.5) {$a_{N_f-3}$};
\node at (.7,0.5) {$a_{N_f-1}$};
\end{tikzpicture} 
\\
\hline 
    \end{tabular}
    \caption{Hasse diagrams for the classical Higgs branch of $SU(N_c)$ SQCD with $N_f$ flavors. Multiplicities of the various leaves are not included, see Section \ref{sectionDiscreteFactors}. \\ For the diagram on the left, the bifurcation with the $A_{2N_c-N_f-1}$ branch occurs only if the bottom of the diagram $a_{N_f-1}$ is not reached before, i.e. when $N_c\leq N_f$. Moreover, $X=0$ for $N_f$ even and $X=1$ for $N_f$ odd. \\ In the case $N_f=2N_c-2$, both diagrams can be used: the diagram on the left gives at the top two $a_1$ transitions, while the diagram on the right gives a $d_2$ transitions -- these are the same objects under two different names.  }
    \label{tableHasse}
\end{table}

The magnetic Quivers for the individual Branches of the Higgs Branch are obtained in Figures \ref{compfirst} to \ref{complast} at CS levels chosen for our convenience. However it can be shown that changing the CS level does not have an impact on the magnetic quivers obtained, as long as we stay at the finite coupling regime. In each case, the HWG is known, and we refer to \cite{zhenghao} for the general expressions. 

\begin{landscape}

\begin{figure}
    \centering
\begin{tikzpicture}[scale=.8]
\tikzstyle{hasse} = [circle, fill,inner sep=2pt]
\node[hasse] (10) at (0,0) {};
\node at (0,-1) {$N_f=1$};
\node at (2,-1) {$N_f=2$};
\node at (4,-1) {$N_f=3$};
\node at (6,-1) {$N_f=4$};
\node at (8,-1) {$N_f=5$};
\node at (10,-1) {$N_f=6$};
\node at (12,-1) {$N_f=7$};
\node at (14,-1) {$N_f=8$};
\node at (16,-1) {$N_f=9$};
\node at (18,-1) {$N_f=10$};
\node at (20,-1) {$N_f=11$};
\node at (22,-1) {$N_f=12$};
\node at (24,-1) {$N_f=13$};
\node[hasse] (20) at (2,0) {};
\node[hasse] (21) at (2,1) {};
\node at (2.5,0.5) {$a_1$};
\draw (20)--(21);
\node[hasse] (30) at (4,0) {};
\node[hasse] (31) at (4,1) {};
\node at (4.5,0.5) {$a_2$};
\draw (30)--(31);
\node[hasse] (40) at (6,0) {};
\node[hasse] (41) at (6,1) {};
\node[hasse] (42) at (6,2) {};
\node at (6.5,1.5) {$a_1$};
\node at (6.5,0.5) {$a_3$};
\draw (40)--(41)--(42);
\node[hasse] (50) at (8,0) {};
\node[hasse] (51) at (8,1) {};
\node[hasse] (52) at (8,2) {};
\node at (8.5,1.5) {$a_2$};
\node at (8.5,0.5) {$a_4$};
\draw (50)--(51)--(52);
\node[hasse] (60) at (10,0) {};
\node[hasse] (61) at (10,1) {};
\node[hasse] (62) at (10,2) {};
\node[hasse] (63) at (10,3) {};
\node[hasse] (64) at (9,1) {};
\node at (9.2,0.2) {$A_5$};
\node at (10.5,2.5) {$a_1$};
\node at (10.5,1.5) {$a_3$};
\node at (10.5,0.5) {$a_5$};
\draw (64)--(60)--(61)--(62)--(63);
\node[hasse] (70) at (12,0) {};
\node[hasse] (71) at (12,1) {};
\node[hasse] (72) at (12,2) {};
\node[hasse] (73) at (12,3) {};
\node[hasse] (74) at (11,2) {};
\node at (11.2,1.2) {$A_4$};
\node at (12.5,2.5) {$a_2$};
\node at (12.5,1.5) {$a_4$};
\node at (12.5,0.5) {$a_6$};
\draw (74)--(71);
\draw (70)--(71)--(72)--(73);
\node[hasse] (80) at (14,0) {};
\node[hasse] (81) at (14,1) {};
\node[hasse] (82) at (14,2) {};
\node[hasse] (83) at (14,3) {};
\node[hasse] (84) at (14,4) {};
\node[hasse] (85) at (13,3) {};
\node at (13.2,2.2) {$A_3$};
\node at (14.5,3.5) {$a_1$};
\node at (14.5,2.5) {$a_3$};
\node at (14.5,1.5) {$a_5$};
\node at (14.5,0.5) {$a_7$};
\draw (85)--(82);
\draw (80)--(81)--(82)--(83)--(84);
\node[hasse] (90) at (16,0) {};
\node[hasse] (91) at (16,1) {};
\node[hasse] (92) at (16,2) {};
\node[hasse] (93) at (16,3) {};
\node[hasse] (94) at (16,4) {};
\node[hasse] (95) at (15,4) {};
\node at (15.2,3.2) {$A_2$};
\node at (16.5,3.5) {$a_2$};
\node at (16.5,2.5) {$a_4$};
\node at (16.5,1.5) {$a_6$};
\node at (16.5,0.5) {$a_8$};
\draw (95)--(93);
\draw (90)--(91)--(92)--(93)--(94);
\node[hasse] (100) at (18,0) {};
\node[hasse] (101) at (18,1) {};
\node[hasse] (102) at (18,2) {};
\node[hasse] (103) at (18,3) {};
\node[hasse] (104) at (18,4) {};
\node[hasse] (105) at (18,5) {};
\node[hasse] (106) at (17,5) {};
\node at (17.2,4.2) {$A_1$};
\node at (18.5,4.5) {$a_1$};
\node at (18.5,3.5) {$a_3$};
\node at (18.5,2.5) {$a_5$};
\node at (18.5,1.5) {$a_7$};
\node at (18.5,.5) {$a_9$};
\draw (106)--(104);
\draw (100)--(101)--(102)--(103)--(104)--(105);
\node[hasse] (110) at (20,0) {};
\node[hasse] (111) at (20,1) {};
\node[hasse] (112) at (20,2) {};
\node[hasse] (113) at (20,3) {};
\node[hasse] (114) at (20,4) {};
\node[hasse] (115) at (20,5) {};
\node at (20.5,4.5) {$d_3$};
\node at (20.5,3.5) {$a_4$};
\node at (20.5,2.5) {$a_6$};
\node at (20.5,1.5) {$a_8$};
\node at (20.5,.5) {$a_{10}$};
\draw (110)--(111)--(112)--(113)--(114)--(115);
\node[hasse] (120) at (22,0) {};
\node[hasse] (121) at (22,1) {};
\node[hasse] (122) at (22,2) {};
\node[hasse] (123) at (22,3) {};
\node[hasse] (124) at (22,4) {};
\node[hasse] (125) at (22,5) {};
\node at (22.5,4.5) {$d_4$};
\node at (22.5,3.5) {$a_5$};
\node at (22.5,2.5) {$a_7$};
\node at (22.5,1.5) {$a_9$};
\node at (22.5,.5) {$a_{11}$};
\draw (120)--(121)--(122)--(123)--(124)--(125);
\node[hasse] (130) at (24,0) {};
\node[hasse] (131) at (24,1) {};
\node[hasse] (132) at (24,2) {};
\node[hasse] (133) at (24,3) {};
\node[hasse] (134) at (24,4) {};
\node[hasse] (135) at (24,5) {};
\node at (24.5,4.5) {$d_5$};
\node at (24.5,3.5) {$a_6$};
\node at (24.5,2.5) {$a_8$};
\node at (24.5,1.5) {$a_{10}$};
\node at (24.5,.5) {$a_{12}$};
\draw (130)--(131)--(132)--(133)--(134)--(135);
\end{tikzpicture}     
    \caption{This figure displays the explicit Hasse diagrams as given in generic form in Table \ref{tableHasse}, in the particular case $N_c=6$, and for $1 \leq N_f \leq 13$. \\ The case $N_f=10$ could also be represented as a $d_2-a_3-a_5-a_7-a_9$ diagram. \\ Note that the maximal height of the diagram is 5, which is the rank of the gauge group. It is reached only when $N_f \geq 2N_c-2=10$. \\ One can see that for $N_f < N_c=6$ there is only the mesonic branch. The baryonic branch appears for $N_f=6$, and then grows in dimension from 1 (for $N_f=N_c=6$) to 25 (for $N_f=2N_c-2=10$) where it equals the dimension of the mesonic branch. Then the baryonic branch takes over, and contains the mesonic branch as a sub-cone for $N_f \geq 2N_c-1=11$. }
    \label{figForest}
\end{figure}

\begin{figure}[t]
\centering
 
\caption{(1) Toric Diagram representing SQCD with $N_f\leq N_c$ and $N_f$ even at finite gauge coupling where masses are set to zero. (2) Brane Web indicating the mesonic branch of the Higgs Branch (3) magnetic Quiver whose 3d Coulomb Branch is the mesonic branch of the Higgs Branch of our SQCD. For $N_f<N_c$ the Higgs branch consists only of the mesonic branch.}
\label{compfirst}
\end{figure}

\begin{figure}[t]
\centering
%
 
\caption{(1) Toric Diagram representing SQCD with $N_f\leq N_c$ and $N_f$ odd at finite gauge coupling where masses are set to zero. (2) Brane Web indicating the mesonic branch of the Higgs Branch (3) magnetic Quiver whose 3d Coulomb Branch is the mesonic branch of the Higgs Branch of our SQCD. For $N_f<N_c$ the Higgs branch consists only of the mesonic branch.}
\label{NfsmallerNcOdd}
\end{figure}

\begin{figure}[t]
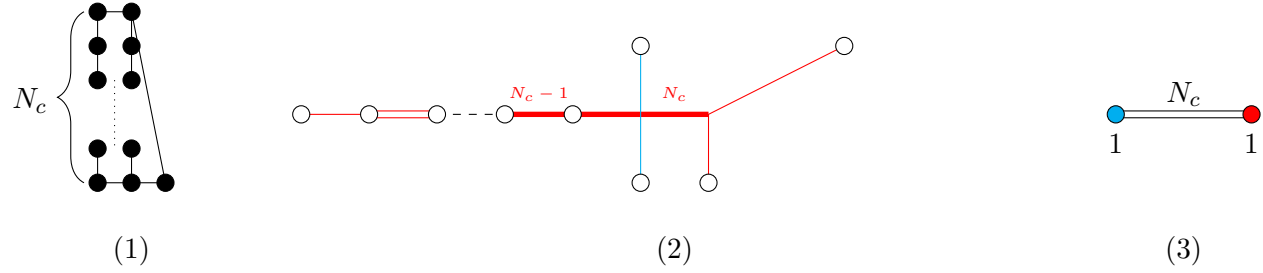

\centering
%
 
\caption{(1) Toric Diagram representing SQCD with $N_f=N_c$ at finite gauge coupling where masses are set to zero. (2) Brane Web indicating the baryonic branch of the Higgs Branch (3) magnetic Quiver whose 3d Coulomb Branch is the baryonic branch of the Higgs Branch of our SQCD. For $N_f<N_c$ there is no baryonic branch.}
\label{nfequalncB}
\end{figure}

\begin{figure}[t]
\centering
%
 
\caption{(1) Toric Diagram representing SQCD with $N_f=N_c$ at finite gauge coupling where masses are set to zero. (2) Brane Web indicating the intersection of the baryonic branch and the mesonic branch of the Higgs Branch (3) magnetic Quiver whose 3d Coulomb Branch is the intersection of the baryonic branch and the mesonic branch of the Higgs Branch of our SQCD.}
\label{intersectionNfeqNc}
\end{figure}

\begin{figure}[t]
\centering
%
\caption{(1) Toric Diagram representing SQCD with $N_c<N_f<2N_c-1$, $N_f$ even, at finite gauge coupling where masses are set to zero. (2) Brane Web indicating the Mesonic cone of the Higgs Branch (3) magnetic Quiver whose 3d Coulomb Branch is the Mesonic cone of the Higgs Branch of our SQCD. 
}
\label{NfinbetweenME}
\end{figure}

\begin{figure}[t]
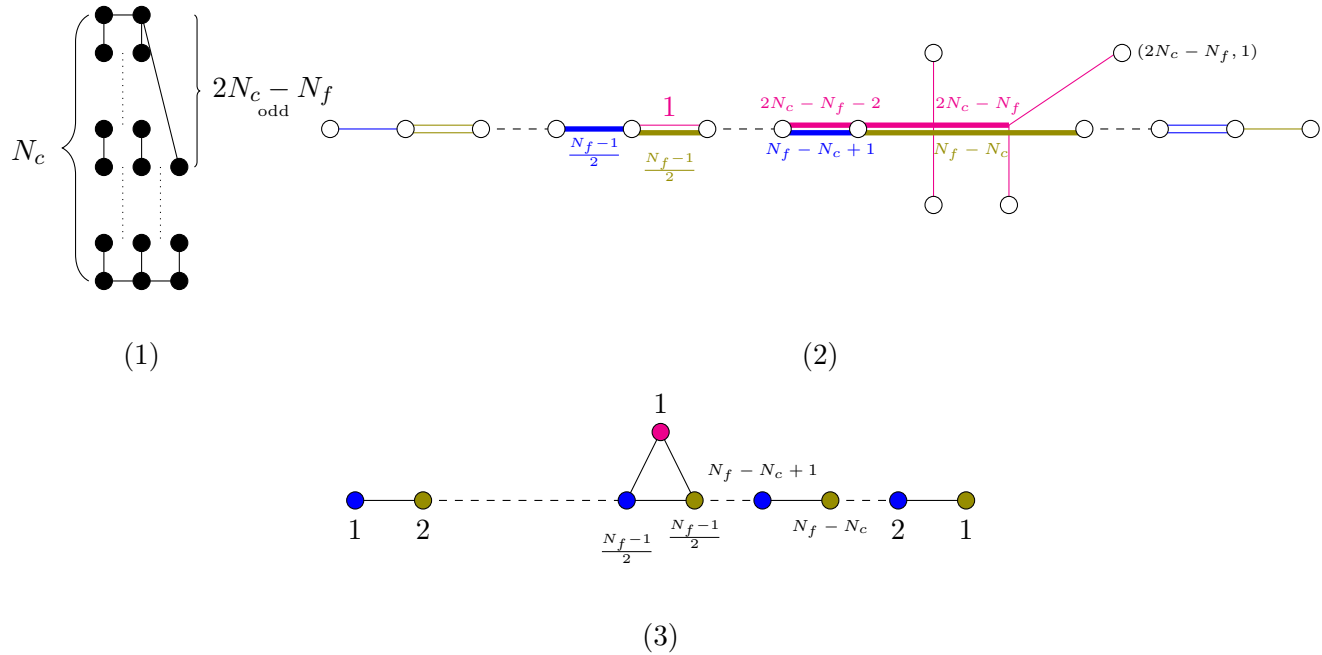

\centering
%
 
\caption{(1) Toric Diagram representing SQCD with $N_c<N_f<2N_c-1$, $N_f$ odd, at finite gauge coupling where masses are set to zero. (2) Brane Web indicating the Mesonic cone of the Higgs Branch (3) magnetic Quiver whose 3d Coulomb Branch is the Mesonic cone of the Higgs Branch of our SQCD. It should be noted that the quiver is symmetric, however only the nodes visible in the brane web are drawn.}
\label{NfinbetweenO}
\end{figure}

\begin{figure}[t]
\centering
%
 
\caption{(1) Toric Diagram representing SQCD with $N_c<N_f<2N_c-1$ at finite gauge coupling where masses are set to zero. (2) Brane Web indicating the Baryonic cone of the Higgs Branch (3) magnetic Quiver whose 3d Coulomb Branch is the Baryonic cone of the Higgs Branch of our SQCD.}
\label{NfinbetweenBary}
\end{figure}

\begin{figure}[t]
\centering
%
 
\caption{(1) Toric Diagram representing SQCD with $N_c<N_f<2N_c-1$ at finite gauge coupling where masses are set to zero. (2) Brane Web indicating the intersection of the Mesonic and Baryonic cone of the Higgs Branch (3) magnetic Quiver whose 3d Coulomb Branch is the intersection of the Mesonic and Baryonic cone of the Higgs Branch of our SQCD.}
\label{NfinbetweenIntersection}
\end{figure}

\begin{figure}[t]
\centering
%
 
\caption{(1) Toric Diagram representing SQCD with $N_f=2N_c-1$ at finite gauge coupling where masses are set to zero. (2) Brane Web indicating the Higgs Branch (3) magnetic Quiver whose 3d Coulomb Branch is the Higgs Branch of our SQCD.}
\label{Nf2Ncminus1}
\end{figure}

\begin{figure}[t]
\centering
%
 
\caption{(1) Toric Diagram representing SQCD with $N_f=2N_c$ at finite gauge coupling where masses are set to zero. (2) Brane Web indicating the Higgs Branch (3) magnetic Quiver whose 3d Coulomb Branch is the Higgs Branch of our SQCD.}
\label{Nf2Nc}
\end{figure}

\begin{figure}[t]
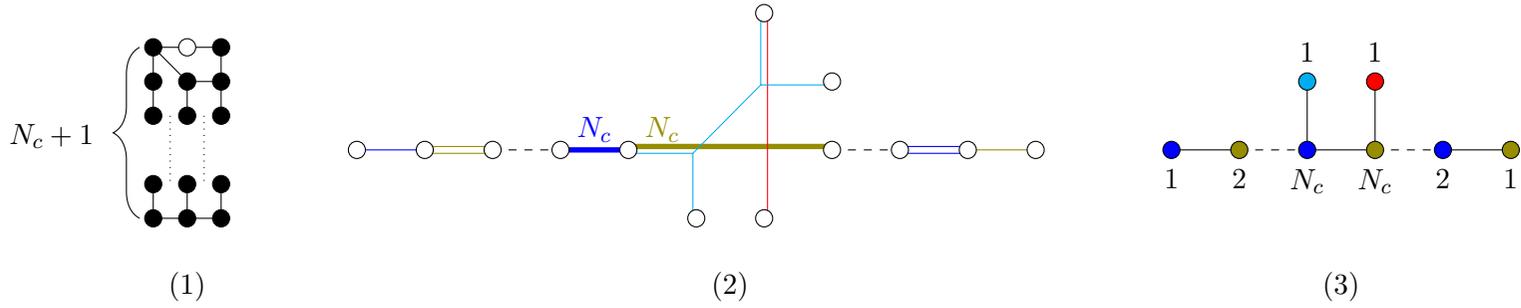

\centering
%
 
\caption{(1) Toric Diagram representing SQCD with $N_f=2N_c+1$ at finite gauge coupling where masses are set to zero. This is an abuse of notation since the diagram corresponds to a non-toric geometry; the white dots stand for shrunken faces following \cite{Benini:2009gi}. This comment also applies to Figure \ref{Nf2Ncpl2}.  (2) Brane Web indicating the Higgs Branch (3) magnetic Quiver whose 3d Coulomb Branch is the Higgs Branch of our SQCD.}
\label{Nf2Ncpl1}
\end{figure}

\begin{figure}[t]
\centering
%
 
\caption{(1) Toric Diagram representing SQCD with $N_f=2N_c+2$ at finite gauge coupling where masses are set to zero. (2) Brane Web indicating the Higgs Branch (3) magnetic Quiver whose 3d Coulomb Branch is the Higgs Branch of our SQCD.}
\label{Nf2Ncpl2}
\end{figure}

\begin{figure}[t]
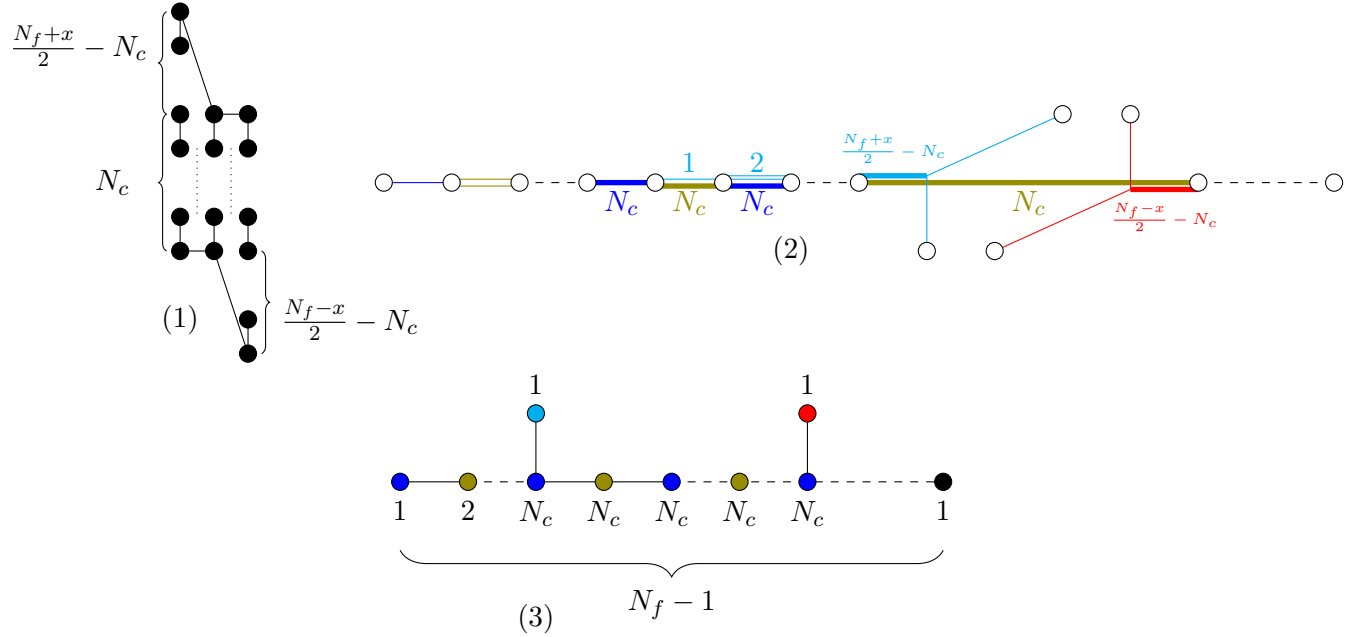

    \centering
    %
    \caption{(1) Toric Diagram representing SQCD with $N_f>2N_c+2$ at finite gauge coupling where masses are set to zero. (2) Brane Web indicating the Higgs Branch (3) magnetic Quiver whose 3d Coulomb Branch is the Higgs Branch of our SQCD. In this figure, $x=0$ if $N_f$ even, and $x=1$ if $N_f$ odd. The black node in the quiver is not displayed in the brane web. The quiver is symmetric. Note that the grid diagram is not convex and we cannot obtain the brane web construction for the infinite coupling regime by contracting it in the horizontal direction even if a UV fixed point exists.}
    \label{complast}
\end{figure}
\end{landscape}

\section{Brane Webs and Radical Ideals}
\label{sectionRadical}

In the previous section, we saw how the brane webs give a prediction for the Hilbert series of $\mathrm{SU}(N_c)$ SQCD. However, a quick comparison with the results of section \ref{sectionSUgroup} shows that there is a mismatch. For instance, consider the case $N_c=5$ with $N_f=4$. The Hilbert series in this case appears in Table \ref{tabSUSQCD}, and it does not agree with the brane web ansatz (\ref{HSBW45}). 

In retrospect, this disagreement is to be expected. As we see in section \ref{sectionDecompositionPrimary}, the \hyperlink{HiggsRing}{Higgs ring} of $\mathrm{SU}(N_c)$ SQCD can contain nilpotent elements. As such, they are not the coordinate ring of algebraic varieties, which do not contain nilpotent elements as stated by the \emph{Nullstellensatz}\footnote{See appendix \ref{AppendixAlgebra} for an explanation of the Nullstellensatz}. On the other hand, the Higgs branch as seen by the brane webs of section \ref{section3} is an algebraic variety, and hence the corresponding ring is defined by a \emph{radical} ideal. It turns out that the ring of Higgs Branch operators we obtain using the brane web method corresponds to the reduced part of the \hyperlink{HiggsRing}{Higgs ring} obtained using the F-terms. This means, we obtain the ring predicted by the brane web method by adding the nilpotent operators in the \hyperlink{HiggsRing}{Higgs ring} to the ideal generated by the equations (\ref{rel1})-(\ref{rel5}), i.e. taking the radical of the ideal $\langle\textnormal{Relations 1. 2. 3. 4. 5.}\rangle$. Hence, the brane web method provides us with the \hyperlink{HiggsVar}{Higgs variety}. On the level of the Hilbert series this translates to
\begin{eqnarray}
    \mathrm{HS}\left(\textnormal{\hyperlink{HiggsVar}{Higgs variety}}\right)&:=& \mathrm{HS}\left(\frac{\mathbb{C}[M,B,\tilde{B}]}{\sqrt{\langle\textnormal{Relations 1. 2. 3. 4. 5.}\rangle}}\right) \nonumber \\ &=& \mathrm{HS}\left(\frac{\mathbb{C}[M,B,\tilde{B}]}{\langle\textnormal{Relations 1. 2. 3. 4. 5., nilpotent operators}\rangle}\right) \\ &=& \mathrm{HS}_{\textnormal{Brane Web}} \nonumber \,,
\end{eqnarray}
where $HS_{\textnormal{Brane Web}}$ is the Hilbert series one obtains by using the brane web method.\\

In this section, we check that indeed the brane webs of Section \ref{section3} correctly reproduce the algebraic \emph{varieties} which correspond to the equations of Section \ref{sectionSUgroup}. 

\subsection{\texorpdfstring{The case $N_f<N_c$}{The case Nf<Nc}}

Let us begin with a simple situation. In the case of $N_f<N_c$ there is only a mesonic branch, so all the equations containing baryons disappear, and the Higgs Ring is 
\begin{equation}
\mathcal{C}=\frac{\mathbb{C}[M]}{\langle M M' \rangle} \, ,  
\end{equation}
using the notations of section \ref{sectionSUgroup}. The equation $M M' =0$ can be written $M^2 = \alpha \mathrm{Tr}(M) M$, with $\alpha = \frac{1}{N_c}$. We have observed that the structure of the quotient ring does not change if $\alpha$ is varied continuously from $\alpha = \frac{1}{N_c}$ to $\alpha = 0$ (and it would be interesting to gain a deeper understanding of why this is the case).\footnote{In particular, we have computed the Hilbert series both for $\alpha = \frac{1}{N_c}$ and $\alpha = 0$ using \texttt{Maclaulay2} \cite{Macaulay2} and found the same result. In addition, the fact that $N_c$ can be as large as wanted (as long as $N_c > N_f$) without changing the geometry suggests that taking $\alpha \rightarrow 0$ does not change the geometry.} As a consequence, we can write the simpler equation $M^2=0$ instead: 
\begin{equation}
\label{RingM^2=0}
\mathcal{C} \equiv \frac{\mathbb{C}[M]}{\langle M^2 \rangle},\qquad M\in \mathbb{C}^{N_f\times N_f} \, . 
\end{equation}
For the discussion below, we use the presentation (\ref{RingM^2=0}). Now the crucial fact is that $M^2=0$ implies that $\mathrm{Tr}(M)^{N_f+1}=0$ but it does not imply that $\mathrm{Tr}(M)^{N_f}=0$. In other words, in the ring (\ref{RingM^2=0}), the element $\mathrm{Tr}(M)$ is nilpotent, with nilpotency degree $N_f+1$. We prove this claim below in the case $N_f=2$; the general situation can be obtained following the same strategy, using a recursion argument. 

\paragraph{The case $N_f=2$}
The characteristic polynomial of the $2 \times 2$ matrix $M$ reads
\begin{equation}
    p(\lambda) = \lambda^2 - \mathrm{Tr}(M) \lambda + \mathrm{det} (M) \, . 
\end{equation}
By the Cayley-Hamilton theorem, we have the algebraic identity $p(M)=0$. Using $M^2=0$, we get 
\begin{equation}
\label{eqTrDet}
    \mathrm{Tr}(M) M = \mathrm{det} (M) \, . 
\end{equation}
Multiplying this equation by $M$ and using once more $M^2=0$, we get $\mathrm{det} (M) M = 0$, and taking the trace, $\mathrm{det} (M) \mathrm{Tr}(M) = 0$. We can now multiply (\ref{eqTrDet}) by $\mathrm{Tr}(M)$ to get $\mathrm{Tr}(M)^2 M =0$. Finally, the trace of this last equality gives $\mathrm{Tr}(M)^3 =0$ as claimed. 

\vspace{1em}

Having identified the nilpotent element $\mathrm{Tr}(M)$, we know that this element is a generator of the ideal radical. It is natural to propose that the radical of the ideal $I= \langle M^2 \rangle$ is precisely $\sqrt{I}= \langle M^2,\mathrm{Tr}(M)\rangle$, and indeed this can be checked. Thus taking the radical we recover the coordinate ring of the closure of the biggest height 2 nilpotent orbit of $A_{N_f-1}$, as a variety
\begin{equation}
\label{resultNO}
\overline{\mathcal{O}}_{(2^{(N_f-\varepsilon)/2},\varepsilon)}=\{M\in \mathfrak{sl}(N_f,\mathbb{C})|M^2=0\}=\{M\in\mathbb{C}^{N_f \times N_f}|M^2=0, \mathrm{Tr}(M)=0\}
\end{equation}
\begin{equation}
    \varepsilon=\left\{\begin{array}{ll}
        1, &  N_f \textnormal{ odd}\\
        0, & N_f \textnormal{ even}
    \end{array}\right.
\end{equation}
which is the prediction of the brane web method for the purely mesonic Higgs branch. Hence, we have shown that in the case where there are no baryons the brane web computation provides us with the \hyperlink{HiggsVar}{Higgs variety}.

In general however, it is much more difficult to compute the radical of the ideal generated by the equations of section \ref{sectionSUgroup}. We illustrate this on some examples with $N_f=4$ in the following subsections. Since $N_c>N_f$ is treated in the present subsection, we focus on $N_c=4,3,2$. 

\subsection{\texorpdfstring{$\mathrm{SU}(4)$ with $N_f=4$ flavours}{SU(4) with 4 flavours}}

Here we look at $\mathrm{SU}(4)$ with $N_f=4$ flavours. In this case, $\mathrm{Tr}(M)$ is no longer nilpotent. However, there are still nilpotent operators. One can show that ${(\mathrm{Tr}(M){M'}_i^j)}^3=0$ for every $1 \leq i,j \leq N_f$, and as a consequence $\mathrm{Tr}(M){M'}_i^j$ is a generator of the radical ideal. Another computation shows that this generator, along with the other generators from section \ref{sectionSUgroup}, generate the full radical. The Hilbert series for the corresponding ring is
\begin{equation}
\begin{split}
    &\mathrm{HS}\left(\textnormal{\hyperlink{HiggsVar}{Higgs variety}}\right)\\
    &=\mathrm{HS}\left(\frac{\mathbb{C}[B,B,\tilde{B}]}{\langle \textrm{Relations 1. 2. 3. 4. 5.}, Tr(M){M'}_i^j\rangle}\right)\\
    &=\frac{1+9 t^2+15 t^4+21 t^6+24 t^8-39 t^{10}+44 t^{12}-26 t^{14}+8 t^{16}-t^{18}}{(1-t)^8 (1+t)^8 \left(1+t^2\right)} \\
    &=\frac{\left(1+5 t^2+t^4\right) \left(1+t^2\right)^2}{\left(1-t^2\right)^8}+\frac{1-t^8}{\left(1-t^2\right)
   \left(1-t^4\right)^2}-1\\
   &=\mathrm{HS}_{\textnormal{Brane Web}}\,.
\end{split}
\end{equation}
We conjecture that for $N_c=N_f$ arbitrary, the only nilpotent operator is $\mathrm{Tr}(M) {M'}_i^j$, which satisfies ${(\mathrm{Tr}(M){M'}_{i}^{j})}^{N_f-1}=0$. However we do not know how to prove this in general. It should be noted that the algebraic computations become very time and memory consuming when $N_c$ and $N_f$ become large. 

\subsection{\texorpdfstring{$\mathrm{SU}(2)$ and $\mathrm{SU}(3)$ with $N_f=4$ flavours}{SU(2) and SU(3) with 4 flavours}}

When the gauge group is $\mathrm{SU}(2)$ or $\mathrm{SU}(3)$, the ideal is radical to begin with. As a consequence, the Hilbert series of the \hyperlink{HiggsScheme}{Higgs scheme} agrees with the brane web calculation.

\section{Finite Factors and Multiplicities}
\label{sectionDiscreteFactors}

In section \ref{sectionRadical} we see, that we can obtain the reduced part of the \hyperlink{HiggsRing}{Higgs ring} by obtaining magnetic quivers from brane web decompositions. However, we hope to reconstruct the full ring, or at least its Hilbert series, combining the hyper-K\"ahler quotient with the brane web method. Let us start with examples where there is only a mesonic branch and try to analyse the ideals in more detail.

\subsection{\texorpdfstring{The Purely Mesonic Branch, $N_f <N_c$}{The Mesonic Branch, NfleqNc}}

In this subsection, we assume $N_f <N_c$. Under that hypothesis, the Higgs scheme is entirely independent of $N_c$, so it is parametrized by $N_f$ only. 

\paragraph{The case $\mathbf{N_f=1}$}
If we take $\mathrm{SU}(N_c)$ with 1 flavour (with $N_c > 1$), the \hyperlink{HiggsRing}{Higgs ring} is generated by a scalar meson. The F-terms imply that it squares to zero. Thus the \hyperlink{HiggsRing}{Higgs ring} is isomorphic to
\begin{equation}
    \frac{\mathbb{C}[x]}{\langle x^2\rangle}=\mathrm{Vect}(1,x)\, ,
\end{equation}
the ring of polynomials of order 1,  with the finite Hilbert series
\begin{equation}
    \mathrm{HS}\left(\frac{\mathbb{C}[x]}{\langle x^2\rangle}\right)=1+t^2 \, .
\end{equation}
The notation $\mathrm{Vect}(\dots)$ stands for the vector space spanned by the elements inside the bracket. 
The \hyperlink{HiggsVar}{Higgs variety}
\begin{equation}
    \{x\in\mathbb{C}|x^2=0\}=\{x\in\mathbb{C}|x=0\}
\end{equation}
is the origin, which has the coordinate ring (the \hyperlink{HiggsVar}{reduced Higgs ring})
\begin{equation}
    \frac{\mathbb{C}[x]}{\langle x\rangle}=\{1\}
\end{equation}
with Hilbert series
\begin{equation}
    \mathrm{HS}\left(\frac{\mathbb{C}[x]}{\langle x\rangle}\right)=1\, .
\end{equation}
This is different from our \hyperlink{HiggsRing}{Higgs ring}, which contains an additional nilpotent scalar meson. The Higgs branch as a geometric space is the origin with multiplicity two. The notion of this multiplicity is reflected explicitly in the Hilbert series as a multiplication with a finite factor \footnote{Note that a finite factor in the Hilbert series does not necessarily imply that the corresponding ring contains nilpotent elements. For example the Higgs branch of $U(1)$ with 2 flavours has the unrefined Hilbert series $(1-t^4)/(1-t^2)^3$ which may be written as $(1+t^2)\times 1/(1-t^2)^2$, which in turn may indicate that there is a nilpotent element, but this moduli space does not contain nilpotent elements.}
\begin{equation}
    \mathrm{HS}\left(\frac{\mathbb{C}[x]}{\langle x^2\rangle}\right)=\mathrm{HS}\left(\frac{\mathbb{C}[x]}{\langle x\rangle}\right)\parunderbrace{\times (1+t^2)}{Multiplicity of the Origin}\,.
\end{equation}
This characterises the Higgs branch as the scheme \footnote{See appendix \ref{AppendixAlgebra} for the notion of an affine scheme.} $\mathrm{Spec}\frac{\mathbb{C}[x]}{\langle x^2\rangle}$, a `fat point', which is schematically depicted as
\begin{equation}
    \scalebox{0.7}{\begin{tikzpicture}
	\begin{pgfonlayer}{nodelayer}
	\tikzstyle{disc} = [circle, fill,inner sep=2.5pt]
        \node [disc] (4) at (0,0) {};
        \node [disc] (5) at (0,1) {};
                \node [] (8) at (-0.5,0) {1};
        \node [] (9) at (-0.5,1) {$x$};
	\end{pgfonlayer}
    \end{tikzpicture}}   \quad ,
\end{equation}
with labels $1$ and $x$ indicating the points in the scheme. From the point of view of quantum field theory, there is nothing wrong with having nilpotent scalar operators in the \hyperlink{HiggsRing}{Higgs ring}. This simply means, they correspond to operators which square to zero, or vanish at some power, inside the path integral, up to operators that are trivial in the \hyperlink{HiggsRing}{Higgs ring}, like derivatives etc.

\paragraph{The case $\mathbf{N_f=2}$}
Let us try to study in detail the case $N_f=2$ ($N_c>2$). In this case, the ideal $\langle M^2 \rangle$ admits the following primary decomposition:
\begin{equation}
    \langle M^2 \rangle = \underbrace{\langle M^2 , \mathrm{Tr} M \rangle}_{I_1} \cap \underbrace{\langle b,c,a^2,d^2 \rangle}_{I_2}
\end{equation}
using the notation $M = \begin{pmatrix} a&b\\c&d\end{pmatrix}$. Let us look at each ideal, the associated coordinate ring and the associated scheme in some detail. 
\begin{itemize}
    \item The first ideal is $I_1 = \langle a^2+bc , a+d \rangle $. Note that $ \mathrm{Tr} M =a+d$ belongs to this ideal. The coordinate ring is an infinite dimensional vector space, it has Hilbert series equal to 
    \begin{equation}
        \mathrm{HS}\left(\frac{\mathbb{C}[a,b,c,d]}{I_1}\right)=\frac{1-t^4}{(1-t^2)^3} \,. 
    \end{equation}
    The algebraic variety is the non-trivial nilpotent orbit closure of $\mathfrak{sl}(2)$, that we depict as a cone
    \begin{equation}
    \label{coneNO}
         \scalebox{0.7}
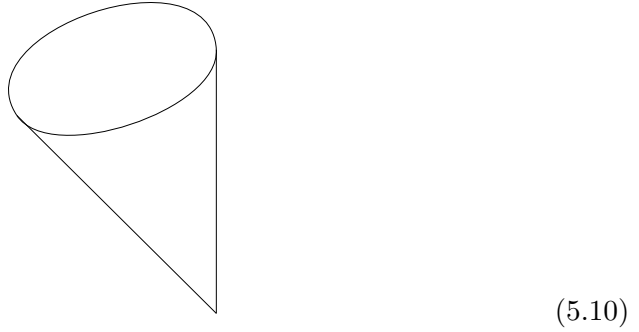
{\begin{tikzpicture}
	\begin{pgfonlayer}{nodelayer}
		\node [style=none] (1) at (0, 0) {};
		\node [style=none] (2) at (0, 5) {};
		\node [style=none] (3) at (-3.75, 3.75) {};
	\end{pgfonlayer}
	\begin{pgfonlayer}{edgelayer}
		\draw (1.center) to (2.center);
		\draw (1.center) to (3.center);
		\draw [bend left=-105, looseness=1.25] (2.center) to (3.center);
		\draw [bend right=-75, looseness=0.75] (2.center) to (3.center);
	\end{pgfonlayer}
\end{tikzpicture}}   
    \end{equation}
    \item The second ideal is $I_2 = \langle b,c,a^2,d^2 \rangle$. We note that $ \mathrm{Tr} M \notin I_2$  and $ (\mathrm{Tr} M )^2 \notin I_2$, but $(\mathrm{Tr} M )^3 = a^2(a+3d)+d^2(3a+d) \in I_2$. It is useful to write down explicitly a graded basis of this ring:
    \begin{eqnarray}
        \mathbb{C}[a,b,c,d]/I_2 &=& \mathrm{Vect} \left( 1, a, d, ad \right) \label{changebasis1}\\ 
        &=&  \mathrm{Vect} \left( 1, a-d , a+d , ad \right)\label{changebasis2}\\ &=&  \mathrm{Vect} \left( 1, a-d , \mathrm{Tr} M , (\mathrm{Tr} M)^2 \right) \label{changebasis3}\,.
    \end{eqnarray}
    Note that from (\ref{changebasis1}) to (\ref{changebasis2}) a change of basis in the subspace of weight $t^2$ is performed, and from (\ref{changebasis2}) to (\ref{changebasis3}) the ideal relations are used to rewrite
    \begin{equation}
        ad = \frac{1}{2}(a^2+d^2-(\mathrm{Tr} M)^2) \equiv - \frac{1}{2}(\mathrm{Tr} M)^2 \,. 
    \end{equation}
    Correspondingly, the associated ring is finite-dimensional, with Hilbert series
    \begin{equation}
        \mathrm{HS}\left(\frac{\mathbb{C}[a,b,c,d]}{I_2}\right)=(1+t^2)^2=1+2t^2+t^4\,.
    \end{equation}
    The corresponding scheme is the origin with multiplicity 4. It is depicted as 
        \begin{equation}
    \label{scheme4}
         \scalebox{0.7}{\begin{tikzpicture}
	\begin{pgfonlayer}{nodelayer}
	\tikzstyle{disc} = [circle, fill,inner sep=2.5pt]
        \node [disc] (4) at (0,0) {};
        \node [disc] (5) at (0,1) {};
        \node [disc] (6) at (1,0) {};
        \node [disc] (7) at (1,1) {};
        \node [] (8) at (-.5,0) {1};
        \node [] (9) at (-1,1) {$a-d$};
        \node [] (10) at (2,0) {$\mathrm{Tr} M$};
        \node [] (11) at (2,1) {$(\mathrm{Tr} M)^2$};
	\end{pgfonlayer}
\end{tikzpicture}}   
    \end{equation}
    \item Finally the sum of the two ideals is $I_1 + I_2 = \langle b,c,a+d,(a-d)^2 \rangle$. It contains $ \mathrm{Tr} M$ but does not contain $a-d$. The ring is 
      \begin{equation}
        \mathbb{C}[a,b,c,d]/(I_1+I_2) =\mathrm{Vect} \left( 1, a-d  \right) \, , 
    \end{equation}
    and the scheme is the origin with multiplicity 2. Geometrically, this is the intersection between the nilpotent orbit (\ref{coneNO}) and the scheme (\ref{scheme4}), depicted as
            \begin{equation}
         \scalebox{0.7}{\begin{tikzpicture}
	\begin{pgfonlayer}{nodelayer}
	\tikzstyle{disc} = [circle, fill,inner sep=2.5pt]
        \node [disc] (4) at (0,0) {};
        \node [disc] (5) at (0,1) {};
                \node [] (8) at (-.5,0) {1};
        \node [] (9) at (-1,1) {$a-d$};
	\end{pgfonlayer}
\end{tikzpicture}}   
    \end{equation}
\end{itemize}
The full picture is
        \begin{equation}
        \label{cone2}
         \scalebox{0.7}{\begin{tikzpicture}
	\begin{pgfonlayer}{nodelayer}
	\tikzstyle{disc} = [circle, fill,inner sep=2.5pt]
		\node [style=none] (1) at (0, 0) {};
		\node [style=none] (2) at (0, 5) {};
		\node [style=none] (3) at (-3.75, 3.75) {};
        \node [disc] (4) at (0,0) {};
        \node [disc] (5) at (0,1) {};
        \node [disc] (6) at (1,0) {};
        \node [disc] (7) at (1,1) {};
                \node [] (8) at (-.5,0) {1};
        \node [] (9) at (-.5,1.5) {$a-d$};
        \node [] (10) at (2,0) {$\mathrm{Tr} M$};
        \node [] (11) at (2,1) {$(\mathrm{Tr} M)^2$};
	\end{pgfonlayer}
	\begin{pgfonlayer}{edgelayer}
		\draw (1.center) to (2.center);
		\draw (1.center) to (3.center);
		\draw [bend left=-105, looseness=1.25] (2.center) to (3.center);
		\draw [bend right=-75, looseness=0.75] (2.center) to (3.center);
	\end{pgfonlayer}
\end{tikzpicture}}  \,,
    \end{equation}
where all dots are part of $\mathrm{Spec}\left(\frac{\mathbb{C}[M]}{I_2}\right)$ and the left dots are part of $\mathrm{Spec}\left(\frac{\mathbb{C}[M]}{I_1+I_2}\right)$. At the level of the Hilbert series, the picture translates into 
\begin{equation}
\begin{split}
    \mathrm{HS}\left(\frac{\mathbb{C}[M]}{\langle M^2\rangle}\right)&=\frac{(1+t^2) (1 + t^2 - 2 t^4 + t^6)}{(1-t^2 )^2}= \frac{1-t^4}{(1-t^2)^3} + (1+2t^2 + t^4)-(1+t^2)\\
    &= \parunderbrace{\frac{1-t^4}{(1-t^2)^3}}{Cone in (\ref{cone2})} +\parunderbrace{t^2+t^4}{Two right dots in (\ref{cone2})}\,. 
\end{split}
\end{equation}
We can rewrite the Hilbert series as
\begin{equation}
    \mathrm{HS}\left(\frac{\mathbb{C}[M]}{\langle M^2\rangle}\right)= \frac{1-t^4}{(1-t^2)^3} + 1\parunderbrace{\times (1+t^2 + t^4)}{Multiplicity of the origin}\underbrace{-1}_{\mathrm{Intersection}}\,,
\end{equation}
suggesting that the \hyperlink{HiggsScheme}{Higgs scheme} is the union of $\overline{\mathcal{O}}_{(2)}$ and (the origin with multiplicity 3). It turns out that this is the right form to generalise for higher number of flavours.

\paragraph{The case $\mathbf{N_f=3}$}

For the case $N_f=3$ ($N_c>3$) the primary decomposition $\langle M^2 \rangle =I_1\cap I_2$ with \texttt{Maclaulay2} \cite{Macaulay2} yields 
\begin{itemize}
\item $I_1=\langle M^2,(\mathrm{Tr}(M))^2\rangle$, with the Hilbert series
\begin{equation}
    HS\left(\frac{\mathbb{C}[M]}{I_1}\right)=\underbrace{\frac{1+4t^2+t^4}{(1-t^2)^4}}_{\overline{\mathcal{O}}_{(2,1)}}\parunderbrace{\times(1+t^2)}{Multiplicity of $\overline{\mathcal{O}}_{(2,1)}$}.
\end{equation}
Telling us, that the scheme is the closure of the minimal nilpotent orbit of $\mathfrak{sl}(3,\mathbb{C})$ with multiplicity 2. Here the multiplicity comes from the operator $\mathrm{Tr}M$ contained in the ring $\frac{\mathbb{C}[M^2]}{\langle M^2,(\mathrm{Tr}M)^2\rangle}$.
\item $I_2$: Different strategies lead to different ideals $I_2$ \footnote{Demonstrating the non-uniqueness of the primary decomposition.}, all of which are quite complicated, therefore we only reproduce the associated Hilbert series:
\begin{enumerate}
    \item Shimoyama-Yokoyama \cite{shimoyama1996localization} strategy:
    \begin{equation}
        HS\left(\frac{\mathbb{C}[M]}{I_2}\right)=1+8t^2+25t^4+43t^6+54t^8+59t^{10}+50t^{12}+14t^{14}
    \end{equation}
    and
    \begin{equation}
        HS\left(\frac{\mathbb{C}[M]}{I_1+I_2}\right)=1+8t^2+24t^4+43t^6+54t^8+59t^{10}+50t^{12}+14t^{14}.
    \end{equation}
    \item Eisenbud-Huneke-Vasconcelos \cite{eisenbud1992direct} strategy:
    \begin{equation}
        HS\left(\frac{\mathbb{C}[M]}{I_2}\right)=1+9t^2+36t^4+92t^6
    \end{equation}
    and
    \begin{equation}
        HS\left(\frac{\mathbb{C}[M]}{I_1+I_2}\right)=1+9t^2+35t^4+91t^6.
    \end{equation}
\end{enumerate}
We see that in both cases
\begin{equation}
    HS\left(\frac{\mathbb{C}[M]}{I_2}\right)-HS\left(\frac{\mathbb{C}[M]}{I_1+I_2}\right)=t^4+t^6.
\end{equation}
\end{itemize}
The two monomials contained in $I_1+I_2$ but not in $I_2$ are $(\mathrm{Tr}M)^2$ and $(\mathrm{Tr}M)^3$. Hence we can again find a schematic graphical depiction of the scheme,
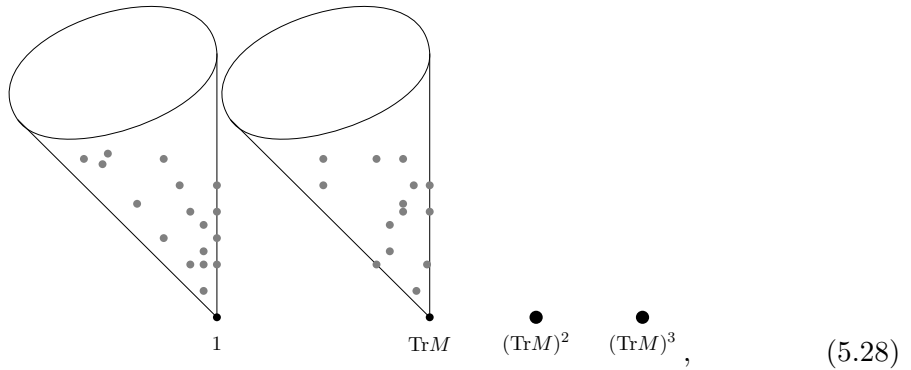
\begin{equation}
    \label{cone3}
         \scalebox{0.7}{\begin{tikzpicture}
	\begin{pgfonlayer}{nodelayer}
	\tikzstyle{disc} = [circle, fill,inner sep=2.5pt]
	\tikzstyle{discinn} = [circle, fill,inner sep=1.5pt]
		\node [style=none] (1) at (0, 0) {};
		\node [style=none] (2) at (0, 5) {};
		\node [style=none] (3) at (-3.75, 3.75) {};
        \node [discinn] (4) at (0,0) {};
        \node [discinn] (22) at (4,0) {};
        \node [discinn, gray] (5) at (0,1) {};
        \node [discinn, gray] (6) at (-2.15,2.9) {};
        \node [discinn, gray] (7) at (-2.05,3.1) {};
        \node [discinn, gray] (8) at (0,1.5) {};
        \node [discinn, gray] (9) at (0,2) {};
        \node [discinn, gray] (10) at (0,2.5) {};
        \node [discinn, gray] (11) at (-0.25,1.25) {};
        \node [discinn, gray] (12) at (-0.25,1.75) {};
        \node [discinn, gray] (13) at (-0.5,2) {};
        \node [discinn, gray] (14) at (-0.7,2.5) {};
        \node [discinn, gray] (15) at (-0.25,0.5) {};
        \node [discinn, gray] (16) at (-0.25,1) {};
        \node [discinn, gray] (17) at (-0.5,1) {};
        \node [discinn, gray] (18) at (-1,1.5) {};
        \node [discinn, gray] (19) at (-1,3) {};
        \node [discinn, gray] (20) at (-1.5,2.15) {};
        \node [discinn, gray] (21) at (-2.5,3) {};
		\node [style=none] (23) at (4, 0) {};
		\node [style=none] (24) at (4, 5) {};
		\node [style=none] (25) at (0.25, 3.75) {};
        
        \node [discinn, gray] (26) at (4,2) {};
        \node [discinn, gray] (27) at (4,2.5) {};
        \node [discinn, gray] (28) at (3.25,1.25) {};
        \node [discinn, gray] (29) at (3.25,1.75) {};
        \node [discinn, gray] (30) at (3.5,2) {};
        \node [discinn, gray] (31) at (3.7,2.5) {};
        \node [discinn, gray] (32) at (3.75,0.5) {};
        \node [discinn, gray] (33) at (3.95,1) {};
        \node [discinn, gray] (34) at (3,1) {};
        \node [discinn, gray] (35) at (2,2.5) {};
        \node [discinn, gray] (36) at (3,3) {};
        \node [discinn, gray] (37) at (3.5,2.15) {};
        \node [discinn, gray] (38) at (3.5,3) {};
        \node [discinn, gray] (39) at (2,3) {};

        \node [disc] (48) at (6,0) {};
        \node [disc] (49) at (8,0) {};
        \node [] (52) at (0,-0.5) {$1$};
        \node [] (53) at (4,-0.5) {$\mathrm{Tr} M$};
        \node [] (52) at (6,-0.5) {$(\mathrm{Tr} M)^2$};
        \node [] (53) at (8,-0.5) {$(\mathrm{Tr} M)^3$};
	\end{pgfonlayer}
	\begin{pgfonlayer}{edgelayer}
		\draw (1.center) to (2.center);
		\draw (1.center) to (3.center);
		\draw [bend left=-105, looseness=1.25] (2.center) to (3.center);
		\draw [bend right=-75, looseness=0.75] (2.center) to (3.center);
		\draw (23.center) to (24.center);
		\draw (23.center) to (25.center);
		\draw [bend left=-105, looseness=1.25] (24.center) to (25.center);
		\draw [bend right=-75, looseness=0.75] (24.center) to (25.center);
	\end{pgfonlayer}
\end{tikzpicture}},
\end{equation}
where all dots are part of $\mathrm{Spec}[\frac{\mathbb{C}[M]}{I_2}]$ and the smaller dots are also part of $\mathrm{Spec}[\frac{\mathbb{C}[M]}{I_1+I_2}]$. The grey dots are arbitrary remnants of primary decomposition (the points on the picture are purely illustrative). The two cones represent $\mathrm{Spec}[\frac{\mathbb{C}[M]}{I_1}]$. On the level of the Hilbert series this is expressed as
\begin{equation}
    \begin{split}
    \mathrm{HS}\left(\frac{\mathbb{C}[M]}{\langle M^2\rangle}\right)&=\dfrac{\left(1+t^2\right) \left(1 + 4 t^2 + 2 t^4 - 4 t^6 + 6 t^8 - 4 t^{10} + t^{12}\right)}{(1-t)^4 (1+t)^4}\\
    &=\parunderbrace{\frac{1+4t^2+t^4}{(1-t^2)^4}(1+t^2)}{Two cones in (\ref{cone3})}+\parunderbrace{t^4+t^6}{Two big dots (\ref{cone3})}\\
    &=\mathrm{HS}\left(\overline{\mathcal{O}}_{(2,1)}\right)\parunderbrace{\times (1+t^2)}{Multiplicity of $\overline{\mathcal{O}}_{(2,1)}$} + \parunderbrace{(1+t^2+t^4+t^6)}{Multiplicity of the origin} - \underbrace{(1+t^2)}_{\textnormal{Intersection}}
    \end{split}\,,
\end{equation}
where the $-(1+t^2)$ term represents the intersection of (the origin with multiplicity four) and  ($\overline{\mathcal{O}}_{(2,1)}$ with multiplicity two), i.e. (the origin with multiplicity two).
\\

\paragraph{Interlude} 
For $N_f >3$, it turns out the computation of the primary decomposition becomes untractable on a standard machine. However we can compare the Hilbert series obtained through the brane web method ($\mathrm{HS}_{\textnormal{Brane Web}}$) with the Hilbert series obtained through the hyper-K\"ahler quotient ($\mathrm{HS}_{\textnormal{HK Quotient}}$). Their difference leaves us with the terms in the Hilbert series corresponding to the nilpotent elements in the \hyperlink{HiggsRing}{Higgs ring},
\begin{equation}
\label{formulaDifference}
    \mathrm{HS}_{\textnormal{HK Quotient}}-\mathrm{HS}_{\textrm{Brane Web}}=\textnormal{nilpotent part}\, .
\end{equation}
In a way, the brane web methods offers an alternative to the usual primary decomposition algorithms to obtain the wanted decomposition, through the formula (\ref{formulaDifference}). Using it, we can go one step further, as illustrated in the next paragraph.

\paragraph{The case $\mathbf{N_f=4}$}
The difference in Hilbert series is
\begin{equation}
\mathrm{HS}_{\textnormal{HK Quotient}}-\mathrm{HS}_{\textrm{Brane Web}}=\mathrm{HS}(\overline{\mathcal{O}}_{(2,1^2)})(t^2+t^4)+t^6+t^8.
\end{equation}
This leads us to an expression of $\mathrm{HS}_{\textnormal{HK Quotient}}$ including multiplicities,
\begin{equation}
    \mathrm{HS}_{\textnormal{HK Quotient}}=\mathrm{HS}(\overline{\mathcal{O}}_{(2^2)})+\mathrm{HS}(\overline{\mathcal{O}}_{(2,1^2)})(t^2+t^4) + (t^6+t^8) \, , 
\end{equation}
using the result (\ref{resultNO}). This can be interpreted as the computation of the Hilbert series for a union of cones (or schemes), with intersections which can themselves have intersections, etc. In such a situation, the Hilbert series is evaluated as a sum with alternating signs, using the inclusion–exclusion principle. In the present case, we get
\begin{equation}
    \begin{split}
    \mathrm{HS}_{\textnormal{HK Quotient}}=&\mathrm{HS}(\overline{\mathcal{O}}_{(2^2)})+\mathrm{HS}(\overline{\mathcal{O}}_{(2,1^2)})\underbrace{\times(1+t^2+t^4)}_{\textnormal{Multiplicity of $\overline{\mathcal{O}}_{(2,1^2)}$}} + \underbrace{(1+t^2+t^4+t^6+t^8)}_{\textnormal{Multiplicity of $\overline{\mathcal{O}}_{(1^4)}$}}\\
    &- \left(\mathrm{HS}(\overline{\mathcal{O}}_{(2,1^2)}) + (1+t^2+t^4) + 1\right)\\
    &+ 1 ,
    \end{split}
\end{equation}
Here the first line has three terms, corresponding to the three components of the \hyperlink{HiggsScheme}{Higgs scheme}; the second line removes the $\binom{3}{2}=3$ pairwise intersections, and the last line adds back the full intersection, which is supported at the origin. To compute the multiplicities of the intersections, we use the conjectured rule that the multiplicity of the intersection of mesonic schemes is the minimum of the multiplicities of the intersecting parts. 
The multiplicity of the origin is simply the degree of nilpotence of the operator $\mathrm{Tr}(M)$, the multiplicity of the higher orbits shows a decrease by two.

\paragraph{Summary}
We summarize the results obtained in this subsection: 
\begin{itemize}
    \item for $N_f=1<N_c$ we get
\begin{equation}
    \mathrm{HS}_{\textnormal{HK Quotient}}-\mathrm{HS}_{\textrm{Brane Web}}=t^2\, ,
\end{equation}
\item for $N_f=2<N_c$ we get
\begin{equation}
    \mathrm{HS}_{\textnormal{HK Quotient}}-\mathrm{HS}_{\textrm{Brane Web}}=t^2+t^4\, ,
\end{equation}
\item for $N_f=3<N_c$ we get
\begin{equation}
    \mathrm{HS}_{\textnormal{HK Quotient}}-\mathrm{HS}_{\textrm{Brane Web}}=\mathrm{HS}\left(\overline{\mathcal{O}}_{(2,1)}\right)(t^2)+(t^4+t^6)\, .
\end{equation}
\item for $N_f=4<N_c$ we get
\begin{equation}
\mathrm{HS}_{\textnormal{HK Quotient}}-\mathrm{HS}_{\textrm{Brane Web}}=\mathrm{HS}(\overline{\mathcal{O}}_{(2,1^2)})(t^2+t^4)+t^6+t^8.
\end{equation}
\end{itemize}

\subsection{Enter the Baryonic Branch, $N_f \geq N_c$}

When $N_f \geq N_c$, the analysis depends on both $N_f$ and $N_c$. In order to study some examples, let's fix $N_f=4$ and vary $N_c$. 

\subsubsection{\texorpdfstring{$\mathrm{SU}(4)$ with $N_f=4$ flavours}{SU(4) with 4 flavours}}
\label{sectionSU44multiplicities}

The difference in Hilbert series is
\begin{equation}
\mathrm{HS}_{\textnormal{HK Quotient}}-\mathrm{HS}_{\textrm{Brane Web}}=\mathrm{HS}(\overline{\mathcal{O}}_{(2,1^2)})(t^2+t^4)-(t^2+t^4).
\end{equation}
Therefore, using (\ref{HSbraneWeb44}) we can write the Hilbert series as
\begin{equation}
    \mathrm{HS}_{\textnormal{HK Quotient}} = \mathrm{HS}(\overline{\mathcal{O}}_{(2^2)}) + \mathrm{HS}(\overline{\mathcal{O}}_{(2,1^2)})(t^2+t^4) + \mathrm{HS}(\textrm{Baryon}) - (1+t^2+t^4)
\end{equation}
We can see that the Baryon branch is not affected by the multiplicities induced by $\mathrm{Tr}(M)$. The intersection of the baryonic and mesonic branch is the origin, which has multiplicity 1.

\subsubsection{\texorpdfstring{$\mathrm{SU}(3)$ with $N_f=4$ flavours}{SU(3) with 4 flavours}}

The difference in Hilbert series in this case is
\begin{equation}
    \mathrm{HS}_{\textnormal{HK Quotient}}-\mathrm{HS}_{\textrm{Brane Web}}=0
\end{equation}
Thus we can see, that there are no multiplicities of orbits. and the Hilbert series is
\begin{equation}
    \mathrm{HS}_{\textnormal{HK Quotient}}=\mathrm{HS}(\overline{\mathcal{O}}_{(2^2)})+\mathrm{HS}(\textrm{Baryon})-\mathrm{HS}(\overline{\mathcal{O}}_{(2,1^2)})
\end{equation}
The intersection of the baryonic branch and the mesonic branch is $\overline{\mathcal{O}}_{(2,1^2)}$ which has multiplicity 1. 

\subsubsection{\texorpdfstring{$\mathrm{SU}(2)$ with $N_f=4$ flavours}{SU(2) with 4 flavours}}

The difference in Hilbert series is again zero and the Higgs branch is a single hyper-Kähler cone without multiplicities, a baryon branch containing the meson.

\subsection{General results}

In the previous examples we saw that the mesonic branch contains multiplicities, while the baryonic branch does not. Let us first consider Higgs Branches which consist of only a mesonic branch.

\subsubsection{Mesonic Branch}

\begin{figure}[t]
\centering
\begin{tikzpicture}[scale=1]
\tikzstyle{hasse} = [circle, fill,inner sep=2pt]
\tikzstyle{hasseg} = [circle, fill=gray,inner sep=2pt]
\node at (0,-0.5) {$1$};
\node[hasse] at (0,0) {}; \node[hasse] at (0,1) {}; \node[hasse] at (0,3) {}; \node[hasse] at (0,4) {}; \node[hasse] at (0,5) {}; \draw (0,0)--(0,1.5) (0,2.5)--(0,5); \node at (0,2) {$\vdots$};

\node[gray] at (1.5,-1) {$Tr(M)^{\varepsilon}$};
\node[hasseg] at (1.5,0) {}; \node[hasseg] at (1.5,1) {}; \node[hasseg] at (1.5,3) {}; \node[hasseg] at (1.5,4) {}; \node[hasseg] at (1.5,5) {}; \draw[gray] (1.5,0)--(1.5,1.5) (1.5,2.5)--(1.5,5); \node[gray] at (1.5,2) {$\vdots$};

\node at (3,-0.5) {$Tr(M)^{1+\varepsilon}$};
\node[hasse] at (3,0) {}; \node[hasse] at (3,1) {}; \node[hasse] at (3,3) {}; \node[hasse] at (3,4) {}; \draw (3,0)--(3,1.5) (3,2.5)--(3,4); \node at (3,2) {$\vdots$};

\node at (4.5,-1) {$Tr(M)^{2+\varepsilon}$};
\node[hasse] at (4.5,0) {}; \node[hasse] at (4.5,1) {}; \node[hasse] at (4.5,3) {}; \node[hasse] at (4.5,4) {}; \draw (4.5,0)--(4.5,1.5) (4.5,2.5)--(4.5,4); \node at (4.5,2) {$\vdots$};

\node at (8,-0.5) {$Tr(M)^{N_f-3}$};
\node[hasse] at (8,0) {}; \node[hasse] at (8,1) {}; \draw (8,0)--(8,1);

\node at (9.5,-1) {$Tr(M)^{N_f-2}$};
\node[hasse] at (9.5,0) {}; \node[hasse] at (9.5,1) {}; \draw (9.5,0)--(9.5,1);

\node at (11,-0.5) {$Tr(M)^{N_f-1}$};
\node[hasse] at (11,0) {};

\node at (12.5,-1) {$Tr(M)^{N_f}$};
\node[hasse] at (12.5,0) {};

\node at (6,2) {$\ddots$}; \node at (6,0) {$\dots$};

\end{tikzpicture}
\caption{Conjectured stratification of the Hilbert series of pure mesonic Higgs branches, the grey is present if $N_f$ is odd. Again $\varepsilon=0$ if $N_f$ is even and $\varepsilon=1$ if $N_f$ is odd.}
\label{stratification}
\end{figure}
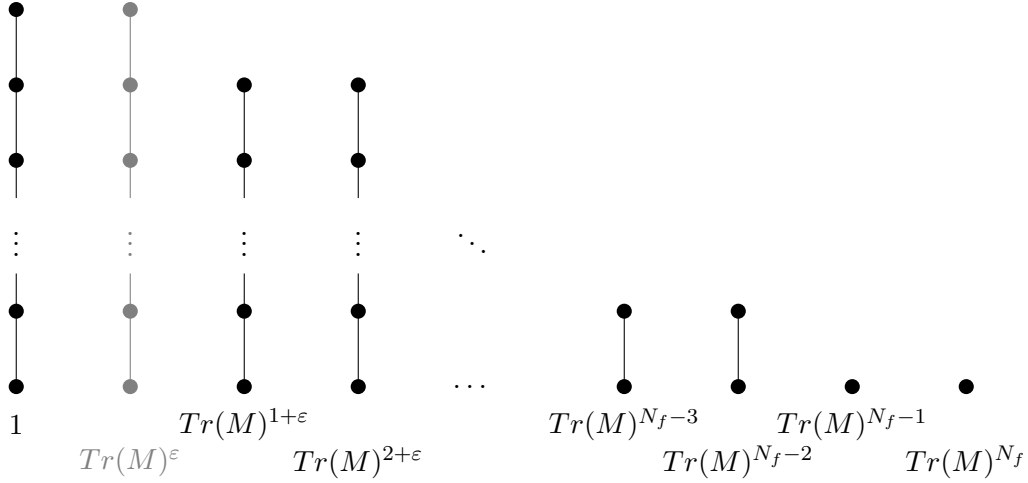

As we saw in the examples of $N_f=1,\dots,4$ with $N_f<N_c$, there is a multiplicity of orbits which decreases by two, where the multiplicity of the origin is fixed by the nilpotence of $\mathrm{Tr}(M)$. The computations using computer codes quickly become untractable and only the Hilbert series in Table \ref{tabSUSQCD} were checked explicitly. However, based on this inspection of low rank examples, we conjecture the following Hilbert series for $N_f<N_c$:
\begin{enumerate}
    \item $N_f$ even \begin{equation}
    \mathrm{HS}=\mathrm{HS}\left(\overline{\mathcal{O}}_{(2^{N_f/2})}\right)+\sum_{i=1}^{\frac{N_f}{2}}\left((t^{4i-2}+t^{4i})\mathrm{HS}\left(\overline{\mathcal{O}}_{(2^{N_f/2-i},1^{2i})}\right)\right)
\end{equation}
    \item $N_f$ odd \begin{equation}
    \mathrm{HS}=\sum_{i=0}^{\frac{N_f-1}{2}}\left((t^{4i}+t^{4i+2})\mathrm{HS}\left(\overline{\mathcal{O}}_{(2^{i-(N_f-1)/2},1^{2i+1})}\right)\right).
\end{equation}
\end{enumerate}
We can explicitly write the highest weight generating functions as
\begin{enumerate}
    \item $N_f$ even
\begin{equation}
    \mathrm{HWG}=\mathrm{PE}[\sum_{k=1}^{\frac{N_f}{2}}\mu_k\mu_{N_f-k}t^{2k}]+\sum_{i=1}^{\frac{N_f}{2}-1}\left((t^{4i-2}+t^{4i})\mathrm{PE}[\sum_{k=1}^{\frac{N_f}{2}-i}\mu_k\mu_{N_f-k}t^{2k}]\right)+(t^{2N_f-2}+t^{2N_f})
\end{equation}
    \item $N_f$ odd
\begin{equation}
    \mathrm{HWG}=\sum_{i=0}^{\frac{N_f-1}{2}-1}\left((t^{4i}+t^{4i+2})\mathrm{PE}[\sum_{k=1}^{\frac{N_f-1}{2}-i}\mu_k\mu_{N_f-k}t^{2k}]\right)+(t^{2N_f-2}+t^{2N_f}).
\end{equation}
\end{enumerate}
Where we summed over all orbits of height 2 with a multiplicity prefactor. The structure of the HS/HWG as a sum over orbits with multiplicities is indicated in the Hasse diagram Figure \ref{HasseMultMes}. Instead of talking about the multiplicity of an orbit, we may also think of a stratification with respect to $\mathrm{Tr}(M)$, see Figure \ref{stratification}, with (\ref{coneNO}) and (\ref{cone3}) in mind.

\begin{figure}[t]
\centering
\begin{tikzpicture}[scale=1]
\tikzstyle{hasse} = [circle, fill,inner sep=2pt]
\node at (-2.5,0) {$(1^{N_f})$}; \node[hasse] at (0,0) {}; \node at (2.5,0) {$N_f+1$};
\node at (-2.5,1.5) {$(2,1^{N_f-2})$}; \node[hasse] at (0,1.5) {}; \node at (2.5,1.5) {$N_f-1$};
\node at (-2.5,2.5) {$\vdots$}; \node at (0,2.5) {$\vdots$}; \node at (2.5,2.5) {$\vdots$};
\node at (-2.5,3.5) {$(2^{(N_f-\varepsilon)/2-k-1},1^{\varepsilon+2k+2})$}; \node[hasse] at (0,3.5) {}; \node at (2.5,3.5) {$\varepsilon+2k+3$};
\node at (-2.5,5) {$(2^{(N_f-\varepsilon)/2-k},1^{\varepsilon+2k})$}; \node[hasse] at (0,5) {}; \node at (2.5,5) {$\varepsilon+2k+1$};
\node at (-2.5,6) {$\vdots$}; \node at (0,6) {$\vdots$}; \node at (2.5,6) {$\vdots$};
\node at (-2.5,7) {$(2^{(N_f-\varepsilon)/2-1},1^{\varepsilon+2})$}; \node[hasse] at (0,7) {}; \node at (2.5,7) {$\varepsilon+3$};
\node at (-2.5,8.5) {$(2^{(N_f-\varepsilon)/2},1^{\varepsilon})$}; \node[hasse] at (0,8.5) {}; \node at (2.5,8.5) {$\varepsilon+1$};

\draw (0,0)--(0,2) (0,3)--(0,5.5) (0,6.5)--(0,8.5);

\node at (-2.5,9.5) {Partition}; \node at (2.5,9.5) {Multiplicity};
\end{tikzpicture}
\caption{The height two part of the Hasse diagram of the $\mathfrak{sl}(N_f,\mathbb{C})$ nilpotent orbits, with associated conjectured orbit multiplicities in the Hilbert series for pure mesonic Higgs branches. Here $\varepsilon=0$ if $N_f$ is even and $\varepsilon=1$ if $N_f$ is odd.}
\label{HasseMultMes}
\end{figure}

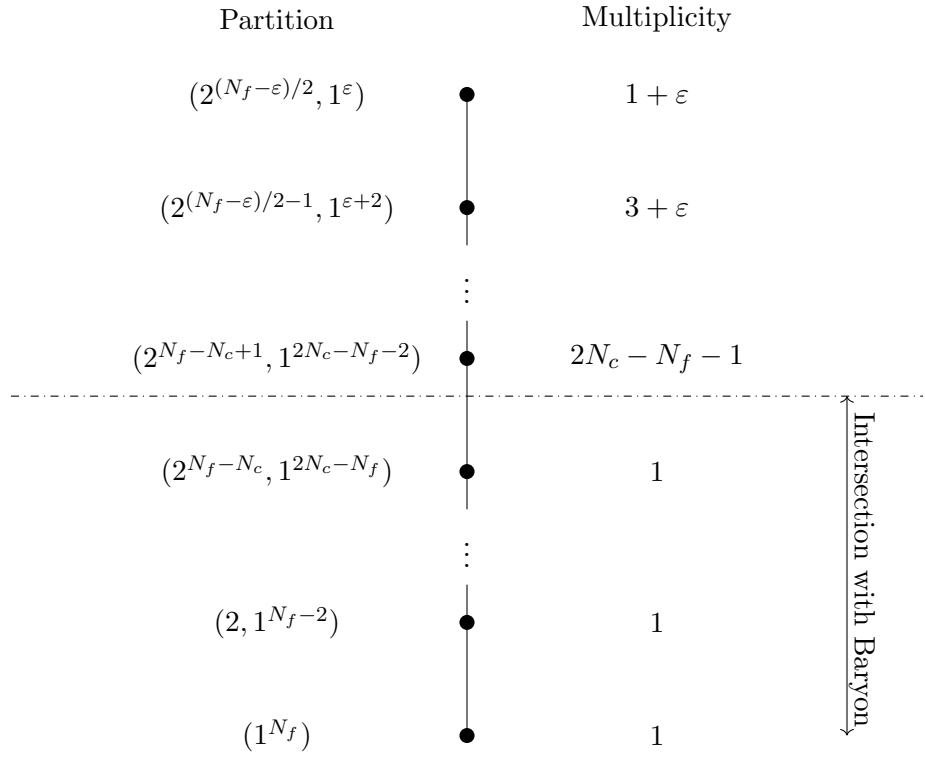
\begin{figure}[t]
\centering
\begin{tikzpicture}[scale=1]
\tikzstyle{hasse} = [circle, fill,inner sep=2pt]
\node at (-2.5,0) {$(1^{N_f})$}; \node[hasse] at (0,0) {}; \node at (2.5,0) {$1$};
\node at (-2.5,1.5) {$(2,1^{N_f-2})$}; \node[hasse] at (0,1.5) {}; \node at (2.5,1.5) {$1$};
\node at (-2.5,3.5) {$(2^{N_f-N_c},1^{2N_c-N_f})$}; \node[hasse] at (0,3.5) {}; \node at (2.5,3.5) {$1$};
\node at (-2.5,5) {$(2^{N_f-N_c+1},1^{2N_c-N_f-2})$}; \node[hasse] at (0,5) {}; \node at (2.5,5) {$2N_c-N_f-1$};ö
\node at (-2.5,7) {$(2^{(N_f-\varepsilon)/2-1},1^{\varepsilon+2})$}; \node[hasse] at (0,7) {}; \node at (2.5,7) {$3+\varepsilon$};
\node at (-2.5,8.5) {$(2^{(N_f-\varepsilon)/2},1^{\varepsilon})$}; \node[hasse] at (0,8.5) {}; \node at (2.5,8.5) {$1+\varepsilon$};

\draw (0,0)--(0,2) (0,3)--(0,5.5) (0,6.5)--(0,8.5);
\node at (0,2.5) {$\vdots$}; \node at (0,6) {$\vdots$};

\draw[dash dot] (-6,4.5)--(6,4.5);
\draw[<->] (5,4.5)--(5,0);
\node[rotate=-90] at (5.2,2.2) {Intersection with Baryon};

\node at (-2.5,9.5) {Partition}; \node at (2.5,9.5) {Multiplicity};
\end{tikzpicture}
\caption{The height two part of the Hasse diagram of $\mathfrak{sl}(N_f,\mathbb{C})$, with the conjectured associated orbit multiplicities of the mesonic branch in the presence of a baryonic branch. $\varepsilon=0$ if $N_f$ is even and $\varepsilon=1$ if $N_f$ is odd.
}
\label{HasseMultBar}
\end{figure}

\subsubsection{Mesons and Baryons}

In the cases involving baryons the situation becomes more complicated. From the previous examples it seems that the multiplicities in the mesonic part stay the same, with the exception of the multiplicity of the intersection always being 1, this is indicated in Figure \ref{HasseMultBar}. There is no stratification picture like the one in the purely mesonic case. However the computation of the Hilbert series using the hyper-K\"ahler quotient becomes too complex to compute many cases and the detailed study of baryonic branches is postponed to future work. The conjectured HWGs for $N_c\leq N_f\leq 2N_c-1$ are
\begin{enumerate}
    \item $N_f$ even
    \begin{equation}
        \begin{split}
        \mathrm{HS}=&\mathrm{HS}\left(\overline{\mathcal{O}}_{(2^{N_f/2})}\right)+\left(\sum_{i=1}^{N_c-\frac{N_f}{2}-1}(t^{4i-2}+t^{4i})\mathrm{HS}\left(\overline{\mathcal{O}}_{(2^{N_f/2-i},1^{2i})}\right)\right)\\
        &+\mathrm{HS}\left(\mathrm{Baryon}\right)\\
        &- \left( \sum_{i=0}^{2N_c-N_f-2}(t^{2i}) \right)\mathrm{HS}\left(\overline{\mathcal{O}}_{(2^{N_f-N_c},1^{2N_c-N_f})}\right)
        \end{split}
    \end{equation}
\begin{equation}
\begin{split}
    \mathrm{HWG}=&\mathrm{PE}\left[\sum_{k=1}^{\frac{N_f}{2}}\mu_k\mu_{N_f-k}t^{2k}\right] + \left(\sum_{i=1}^{N_c-\frac{N_f}{2}-1}(t^{4i-2}+t^{4i})\mathrm{PE}\left[\sum_{k=1}^{\frac{N_f}{2}-i}\mu_k\mu_{N_f-k}t^{2k}\right]\right)\\
    &+ \mathrm{HWG(Baryon)}\\
    &- \left( \sum_{i=0}^{2N_c-N_f-2}(t^{2i}) \right) \mathrm{PE}\left[\sum_{k=1}^{N_f-N_c}\mu_k\mu_{N_f-k}t^{2k}\right]
\end{split}
\end{equation}
    \item $N_f$ odd
    \begin{equation}
    \begin{split}
    \mathrm{HS}=&\left(\sum_{i=0}^{N_c-\frac{N_f-1}{2}-2}(t^{4i}+t^{4i+2})\mathrm{HS}\left(\overline{\mathcal{O}}_{(2^{(N_f-1)/2-i},1^{2i+1})}\right)\right)\\
    &+ \mathrm{HS(Baryon)}\\
    &- \left(\sum_{i=0}^{2N_c-N_f-2}(t^{2i})\right)\mathrm{HS}\left(\overline{\mathcal{O}}_{(2^{N_f-N_c},1^{2N_c-N_f})}\right)
    \end{split}
\end{equation}
\begin{equation}
    \begin{split}
    \mathrm{HWG}=&\left(\sum_{i=0}^{N_c-\frac{N_f-1}{2}-2}(t^{4i}+t^{4i+2})\mathrm{PE}\left[\sum_{k=1}^{\frac{N_f-1}{2}-i}\mu_k\mu_{N_f-k}t^{2k}\right]\right)\\
    &+ \mathrm{HWG(Baryon)}\\
    &-\left(\sum_{i=0}^{2N_c-N_f-2}(t^{2i})\right)\mathrm{PE}\left[\sum_{k=1}^{N_f-N_c}\mu_k\mu_{N_f-k}t^{2k}\right]
    \end{split}
\end{equation}
\end{enumerate}
where 
\begin{equation}
    \mathrm{HWG(Baryon)}=\mathrm{PE}\left[t^2+(\mu_{N_c}q+\mu_{N_f-N_c}q^{-1})t^{N_c}+\sum_{k=1}^{N_f-N_c}\mu_k\mu_{N_f-k}t^{2k}-\mu_{N_c}\mu_{N_f-N_c}t^{2N_c}\right]
\end{equation}
and 
\begin{equation}
    \sum_{k=1}^{0}\equiv 0 \quad\textnormal{for the limiting case $N_f=N_c$.}
\end{equation}
It should be noted, that while the Hilbert series proposed here agree with those in Table \ref{tabSUSQCD}, checks of cases where the nilpotent operators in the \hyperlink{HiggsRing}{Higgs ring} and the intersection of the baryonic and mesonic branch are non-trivial were not possible. The simplest example to check would be $SU(5)$ with 6 flavours, but this is already out of reach with standard computers.

\subsubsection{Baryonic branch}

For $N_f\geq 2N_c-1$ the Higgs branch consists of only the baryonic branch and contains no nilpotent elements. The brane web method yields the same answer as the hyper-Kähler quotient,
\begin{equation}
    \mathrm{HWG}=\mathrm{PE}[t^2+(\mu_{N_c}q+\mu_{N_f-N_c}q^{-1})t^{N_c}+\sum_{k=1}^{N_c}\mu_k\mu_{N_f-k}t^{2k}-\mu_{N_c}\mu_{N_f-N_c}t^{2N_c}] .
\end{equation}

\section{Future Directions}

In this work, using Hilbert series techniques, we have shown that the Higgs branch of supersymmetric theories with 8 supercharges has a coordinate ring containing nilpotent operators. We have shown how the \hyperlink{HiggsVar}{Higgs variety} can be obtained by restricting to the radical of the equations relating the gauge invariant operators, and we have found perfect agreement with the geometry predicted by the techniques of brane webs and magnetic quivers. This demonstrates the usefulness of magnetic quivers allowing to explore a lot of the structure of the full Higgs branch, such as global symmetry and cone structure. This is done with little effort, since the technique of magnetic quivers allows us to make use of symmetries of the problem, while many computations on the level of rings involve the computation of a Gröbner basis, ignoring useful symmetries entirely. However, two crucial points clearly deserve further study. Firstly, what is the physical meaning of the nilpotent operators on the Higgs branch, or equivalently, what distinguishes the various copies of an irreducible component of a Higgs branch having non-trivial multiplicity? It would be interesting, for instance, to describe a tunnelling mechanism from one vacuum to another, the two vacua corresponding to various distinct powers of a nilpotent operator. Secondly, how can the multiplicity of the cones in the Higgs branch be seen and predicted by brane methods? It would be interesting to derive a way to compute the multiplicity polynomials from purely geometric methods. 

\section*{Acknowledgements}

We are grateful to Andrés Collinucci, Simone Giacomelli, Rudolph Kalveks, Victor Lekeu, Dominik Miketa, Diego Rodríguez-Gómez and Anton Zajac for useful discussions. This work was supported by STFC grant ST/P000762/1, STFC Consolidated Grant ST/J0003533/1, and EPSRC Programme Grant EP/K034456/1.  
The work of S.C. is supported by an EPSRC DTP studentship EP/M507878/1.  

\appendix

\section{Tropical Brane Webs and Magnetic Quivers}
\label{Appendixmag}
The aim of this appendix is to review \textit{Conjecture 1} of \cite{Cabrera:2018jxt}. This conjecture states that we can obtain cones in the Higgs branch of a 5d $\mathcal{N}=1$ gauge theory living on a brane web by constructing so called magnetic quivers. The number of inequivalent maximal decompositions is equal to the number of cones in the Higgs branch. For every decomposition of the brane web into subwebs there is a corresponding magnetic quiver. Using the monopole formula on the magnetic quiver, i.e. formally computing the 3d $\mathcal{N}=4$ Coulomb branch Hilbert series, we obtain the Hilbert series of a cone in the Higgs branch. The positions of the subwebs along the 7-branes parameterize the moduli space of dressed monopole operators of the magnetic quiver. In the following we focus on techniques to obtain the magnetic quiver corresponding to a brane web decomposition. These quivers only consist of $\mathrm{U}(n)$ gauge nodes and edges with possible multiplicity.

\begin{enumerate}
    \item \underline{Gauge Nodes:}\\
    For every subweb in the decomposition we naively get a $\mathrm{U}(1)$ gauge node in the magnetic quiver. For a number of $n$ identical subwebs on top of each other the $n$ $\mathrm{U}(1)$ groups get enhanced to a $\mathrm{U}(n)$ gauge node.
    \item \underline{Edges:}\\
    The number of edges between two nodes $A$ and $B$ corresponding to subwebs labelled $A$ and $B$ is given by
    \begin{equation}
        \textrm{E}(A,B)=\textrm{SI}(A,B) + \sum_{k}\textrm{X}(A,B,k) - \sum_{k}\textrm{Y}(A,B,k),
    \end{equation}
    where the set $\{k\}$ is the set of all 7-branes in the brane web, $\textrm{SI}(A,B)$ is the \textit{stable intersection} \cite{fulton1997intersection,richter2005first,mikhalkin2006tropical} of subwebs $A$ and $B$ (defined below), $\textrm{X}(A,B,k)$ labels number of combinations of two 5-branes from brane webs $A$ and $B$ attached to 7-brane $k$ from opposite sides and $\textrm{Y}(A,B,k)$ labels number of combinations of two 5-branes from brane webs $A$ and $B$ attached to 7-brane $k$ from the same side\footnote{Note that a $(p,q)5$-brane can only end on a $[P,Q]7$ brane if $p=P$ and $q=Q$. Hence two $(P,Q)5$-branes can end on a $[P,Q]7$-brane only from the same side or from opposite sides.}. For every node of rank $>1$ we only consider one of the identical subwebs.
\end{enumerate}
The stable intersection (SI) is a notion borrowed from tropical geometry\footnote{Here we view the brane web as a collection of tropical curves.}. For two intersecting 5-branes, $(p,q)$ and $(p',q')$ the intersection number I is given by
\begin{equation}
    \textrm{I}\{(p,q),(p',q')\}=\textrm{abs}\left(\det\begin{pmatrix}
    p & q\\
    p' & q'
    \end{pmatrix}\right)=|pq'-qp'|.
\end{equation}
The SI of two subwebs is computed by moving them relative to each other, until all intersections are intersections of two straight 5-branes. Now the SI can be computed as the sum of all of its individual intersection numbers of 5-brane intersections.
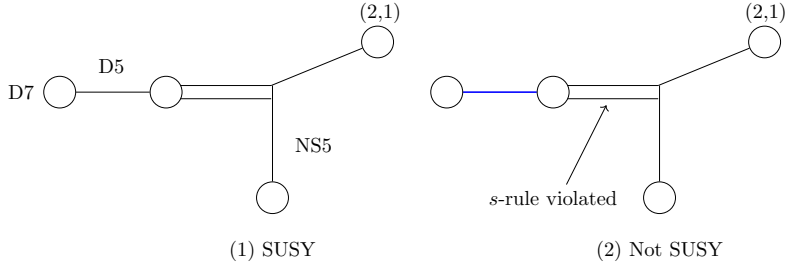
\begin{figure}[t]
\centering
\scalebox{0.7}{\begin{tikzpicture}
	\begin{pgfonlayer}{nodelayer}
		\node [style=gauge1] (0) at (3.5, 0) {};
		\node [style=gauge1] (1) at (5.5, -2) {};
		\node [style=none] (2) at (5.5, 0.125) {};
		\node [style=none] (3) at (5.475, 0.125) {};
		\node [style=none] (4) at (5.475, -0.125) {};
		\node [style=none] (5) at (3.55, 0.125) {};
		\node [style=none] (6) at (3.55, -0.125) {};
		\node [style=gauge1] (7) at (7.475, 0.95) {};
		\node [style=none] (8) at (5.475, 0.125) {};
		\node [style=none] (9) at (5.5, -0.125) {};
		\node [style=none] (12) at (5.5, -3) {(2) Not SUSY};
		\node [style=gauge1] (30) at (1.5, 0) {};
		\node [style=gauge1] (31) at (-3.775, 0) {};
		\node [style=gauge1] (32) at (-1.775, -2) {};
		\node [style=none] (33) at (-1.775, 0.125) {};
		\node [style=none] (34) at (-1.8, 0.125) {};
		\node [style=none] (35) at (-1.8, -0.125) {};
		\node [style=none] (36) at (-3.725, 0.125) {};
		\node [style=none] (37) at (-3.725, -0.125) {};
		\node [style=gauge1] (38) at (0.2, 0.95) {};
		\node [style=none] (39) at (-1.8, 0.125) {};
		\node [style=none] (40) at (-1.775, -0.125) {};
		\node [style=none] (41) at (-1.775, -3) {(1) SUSY};
		\node [style=gauge1] (42) at (-5.775, 0) {};
		\node [style=none] (43) at (7.5, 1.5) {(2,1)};
		\node [style=none] (44) at (0.25, 1.5) {(2,1)};
		\node [style=none] (45) at (4.5, -0.25) {};
		\node [style=none] (46) at (3.75, -1.75) {};
		\node [style=none] (47) at (3.5, -2) {$s$-rule violated};
	\end{pgfonlayer}
	\begin{pgfonlayer}{edgelayer}
		\draw (2.center) to (1);
		\draw (5.center) to (3.center);
		\draw (6.center) to (4.center);
		\draw (8.center) to (7);
		\draw [style=bluee] (30) to (0);
		\draw (33.center) to (32);
		\draw (36.center) to (34.center);
		\draw (37.center) to (35.center);
		\draw (39.center) to (38);
		\draw (42) to (31);
		\draw [style=->] (46.center) to (45.center);
	\end{pgfonlayer}
	\node at (-6.5,0) {D7};
	\node at (-4.8,0.5) {D5};
	\node at (-1,-1) {NS5};
\end{tikzpicture}
}
\caption{The brane web (1) shows two D5 branes stretched between two different D7 branes and ending on the same NS5 brane. This configuration does not violate the $s$-rule \cite{Hanany:1996ie} and is supersymmetric. (2) If we attempt to decompose the brane web into two subwebs, black and blue, the resulting configuration is not supersymmetric. This is because there are two D5 branes stretched between the same D7 and NS5 brane which violates the $s$-rule.}
\label{srulediff}
\end{figure}

\begin{figure}[t]
    \centering
    \begin{tikzpicture}
        \tikzstyle{gauge1} = [inner sep=0.8mm,draw=none,fill=white,minimum size=0.35mm,circle, draw];
        \draw (2,2)--(0,1) (0,1)--(-1,0) (-1,0)--(-1,-1);
        \draw[dashed] (2,1)--(-2,1) (2,0)--(-2,0);
        \node at (-3,1) {\tiny monodromy cut};
        \node at (-3,0) {\tiny monodromy cut};
        \node[gauge1] at (2,2) {};
        \node at (2.6,2) {$[2,1]7$};
        \node[gauge1] at (2,1) {};
        \node at (2.6,1) {$[1,0]7$};
        \node at (0.3,0.5) {$(1,1)5$};
        \node[gauge1] at (2,0) {};
        \node at (2.6,0) {$[1,0]7$};
        \node[gauge1] at (-1,-1) {};
        \node at (-0.4,-1) {$[0,1]7$};
        \node at (-1,2) {(1)};
        
        \draw (5,0)--(6,0) (5.5,1)--(7,1) (6,-1)--(6,0) (6,0)--(7,1) (7,1)--(9,2);
        \draw[dashed] (4.5,0)--(5,0) (4.5,1)--(5.5,1);
        
        \node[gauge1] at (5,0) {};
        \node[gauge1] at (5.5,1) {};
        \node[gauge1] at (6,-1) {};
        \node[gauge1] at (9,2) {};
        
        \node at (6,2) {(2)};
    \end{tikzpicture}
    \caption{Shown is an example similar to the one in Figure \ref{srulediff}. (1) The monodromy cut of the D7 branes turns the $(2,1)5$-brane into a $(1,1)5$-brane and $(0,1)5$-brane respectively. (2) The same system after Hanany-Witten transition. If the D7 branes are brought together on the same horizontal axis one gets back the example in Figure \ref{srulediff}.}
    \label{monod}
\end{figure}
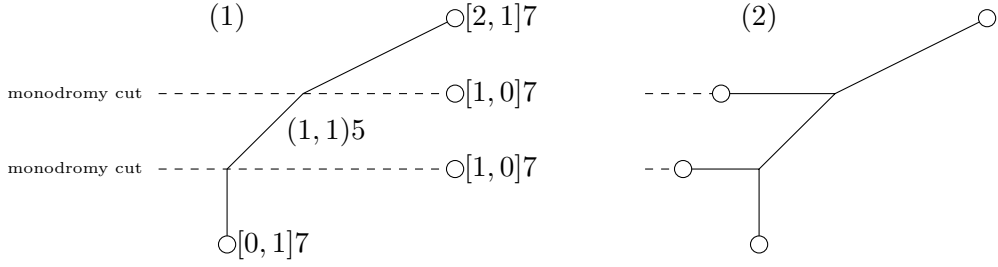

The remaining piece is how to obtain a valid decomposition. There are two conditions every decomposition into subwebs has to obey
\begin{enumerate}
    \item \underline{Charge conservation:}\\
    For every subweb obtained $(p,q)$-charge has to be conserved at the vertices.
    \item \underline{Supersymmetry conservation ($s$-rule):}\\
    Every web has to obey the \textit{s-rule} first introduced in \cite{Hanany:1996ie} and presented in a generalised form applicable to $(p,q)$ webs in \cite{Mikhailov:1998bx,DeWolfe:1998bi,Bergman:1998ej,Bachas:1997kn,Benini:2009gi}. In the set up of \cite{Hanany:1996ie} using NS5, D5 and D3-branes the s-rule states, that any configuration with two or more identical D3-branes between the same NS5 and D5-brane breaks Supersymmetry. In our setting this leads to the rule, that any configuration with two or more identical D5-branes between the same NS5 and D7 breaks Supersymmetry, as illustrated in Figure \ref{srulediff}. More generally there can either be $0$ or $|pQ-Pq|$ $(P,Q)5$-branes between a $[P,Q]7$-brane and a $(p,q)5$-brane, this can be seen from the brane creation effect, when a $(p,q)5$-brane crosses a $[P,Q]7$-brane $|pQ-Pq|$ $(P,Q)5$-branes are created in order to conserve charge at the vertex, which previously crossed the $SL(2,\mathbb{Z})$ monodromy cut of the $[P,Q]7$-brane. An example is given in Figure \ref{monod}.
\end{enumerate}
The monodromy matrix $M_{[P,Q]}$ for a $[P,Q]7$-brane can be written as 
\begin{equation}
    M_{[P,Q]}=\begin{pmatrix}
    1+PQ & - P^2\\
    Q^2 & 1 - PQ
    \end{pmatrix}
\end{equation}
when a $(p,q)5$-brane crosses the monodromy cut of a $[P,Q]7$-brane (counterclockwise in the pictures) then it gets tilted and becomes a $(r,s)5$-brane, where
\begin{equation}
    \begin{pmatrix}
    r\\    s
    \end{pmatrix}=M_{[P,Q]}\begin{pmatrix}
    p\\    q
    \end{pmatrix} \, .
\end{equation}

\section{\texorpdfstring{Height 2 Nilpotent Orbits of $\mathfrak{sl}(n,\mathbb{C})$ and their Extensions}{Nilpotent orbits and their extensions}}
\label{appendixNO}

An $\mathfrak{sl}(n,\mathbb{C})$ nilpotent adjoint orbit is characterised by a partition of $n$, $\lambda\in\mathcal{P}(n)$, where $\mathcal{P}(n)$ is the set of all tuples of positive integers $\lambda_i$, such that $\lambda_i>\lambda_j\forall i<j$ and $\sum_i \lambda_i=n$. As an example
\begin{equation}
\begin{split}
    \mathcal{P}(5)=&\{(1,1,1,1,1),(2,1,1,1),(2,2,1),(3,1,1),(3,2),(4,1),(5)\}\\
    =&\{(1^5),(2,1^3),(2^2,1),(3,1^2),(4,1),(5)\}\qquad \textnormal{(exponent notation)}
\end{split}
\end{equation}
An elementary Jordan block of order $d$, $J_d$, is a $d\times d$ matrix with all entries $0$ except for superdiagonal entries, which are $1$
\begin{equation}
    J_d=\begin{pmatrix}
    0 & 1 & 0 & \dots & 0 & 0\\
    0 & 0 & 1 & \dots & 0 & 0\\
    \vdots & \vdots & \vdots & \ddots & \vdots & \vdots\\
    0 & 0 & 0 & \dots & 0 & 1\\
    0 & 0 & 0 & \dots & 0 & 0
    \end{pmatrix}\in\mathbb{R}^{d\times d}.
\end{equation}
for every partition $\lambda=(\lambda_1,\lambda_2,\dots,\lambda_k)$ we can build the nilpotent matrix
\begin{equation}
    X_{\lambda}=\begin{pmatrix}
    J_{\lambda_1} & 0 & \dots & 0\\
    0 & J_{\lambda_2} & \dots & 0\\
    \vdots & \vdots & \ddots & \vdots\\
    0 & 0 & \dots & J_{\lambda_k}
    \end{pmatrix}\in \mathfrak{sl}(n,\mathbb{C}).
\end{equation}
A nilpotent adjoint orbit of $\mathfrak{sl}(n,\mathbb{C})$ is now given as
\begin{equation}
    \mathcal{O}_{\lambda}=\{M\in \mathfrak{sl}(n,\mathbb{C})|M= \mathrm{Ad}_g(X_{\lambda}), g\in \mathrm{PSL}(n,\mathbb{C})\}.
\end{equation}
Two nilpotent orbits of corresponding to different partitions are disjoint sets in $\mathfrak{sl}(n,\mathbb{C})$.
However, the Zariski closures of nilpotent orbits are partially ordered by inclusion. A graphical representation of this partial order is given by a Hasse diagram. The closure of an orbit is its union with all of its lower orbits in the Hasse diagram. 

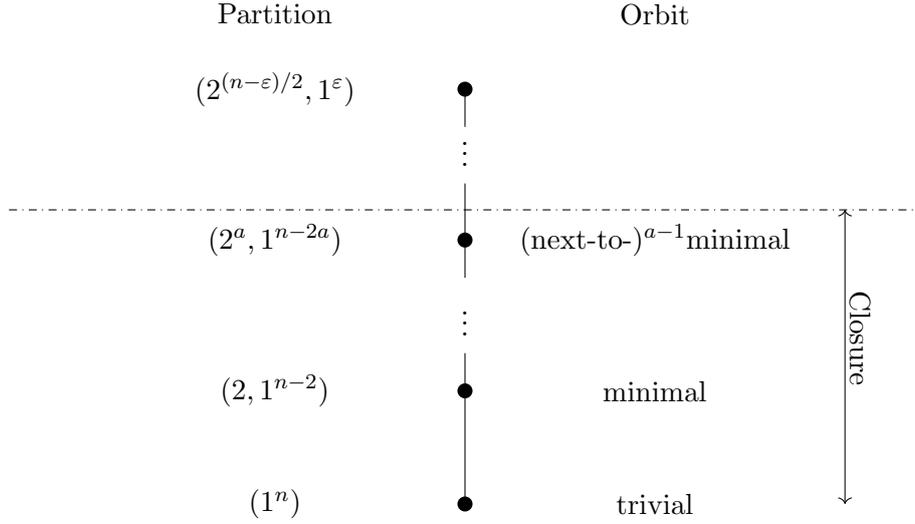
\begin{figure}[t]
\centering
\begin{tikzpicture}[scale=1]
\tikzstyle{hasse} = [circle, fill,inner sep=2pt]
\node at (-2.5,0) {$(1^{n})$}; \node[hasse] at (0,0) {}; \node at (2.5,0) {trivial};
\node at (-2.5,1.5) {$(2,1^{n-2})$}; \node[hasse] at (0,1.5) {}; \node at (2.5,1.5) {minimal};
\node at (-2.5,3.5) {$(2^{a},1^{n-2a})$}; \node[hasse] at (0,3.5) {}; \node at (2.5,3.5) {(next-to-)$^{a-1}$minimal};
\node at (-2.5,5.5) {$(2^{(n-\varepsilon)/2},1^\varepsilon)$}; \node[hasse] at (0,5.5) {}; \node at (2.5,5.5) {};

\draw (0,0)--(0,2) (0,3)--(0,4.25) (0,5)--(0,5.5);
\node at (0,2.5) {$\vdots$}; \node at (0,4.75) {$\vdots$};

\node at (-2.5,6.5) {Partition}; \node at (2.5,6.5) {Orbit};

\draw[dash dot] (-6,3.9)--(6,3.9);
\draw[<->] (5,3.9)--(5,0);
\node[rotate=-90] at (5.2,2.2) {Closure};
\end{tikzpicture}
\caption{The height two part of the Hasse diagram of $\mathfrak{sl}(n,\mathbb{C})$, along with the common names for the orbits. Here $\varepsilon=0$ if $n$ is even and $\varepsilon=1$ if $n$ is odd.}
\label{orbitpartitionHasse}
\end{figure}

In the following we are only concerned with orbits of height two, i.e. the corresponding partition only contains factors of $1$ and $2$, which arrange in a subset of the Hasse diagram, Figure \ref{orbitpartitionHasse}. This subset is defined as the full subset of orbits which are lower than $\mathcal{O}_{(2^{(n-\varepsilon)/2},1^\varepsilon)}$ with 
\begin{equation}
\label{defx}
    \varepsilon= \begin{cases}
        1  &  \textnormal{for }n \textnormal{ odd}\\
        0  & \textnormal{for } n \textnormal{ even.}
    \end{cases} 
\end{equation}
In this subset, the order is total. As a consequence, for orbits of height two there is a simple expression
\begin{equation}
	\overline{\mathcal{O}}_{(2^a,1^{n-2a})}=\mathcal{O}_{(2^a,1^{n-2a})}\cup \mathcal{O}_{(2^{a-1},1^{n-2a+2})} \cup \dots \cup \mathcal{O}_{(1^n)}
\end{equation}
indicated in Figure \ref{orbitpartitionHasse}.

The closure of the largest nilpotent orbit of height two has a simple expression as an algebraic variety, 
\begin{equation}
\overline{\mathcal{O}}_{(2^{(n-\varepsilon)/2},1^\varepsilon)}=\{M\in \mathfrak{sl}(n,\mathbb{C})|M^2=0\}=\{M\in\mathbb{C}^{n \times n}|M^2=0, \, \mathrm{Tr}(M)=0\} \, , 
\end{equation}
where $\varepsilon$ is given by (\ref{defx}). Similarly, the other orbits at height 2 are described as algebraic varieties involving a rank condition 
\begin{equation}
\overline{\mathcal{O}}_{(2^{a},1^{n-2a})}=\{M\in\mathbb{C}^{n \times n}|M^2=0, \,  \mathrm{Tr}(M)=0 , \, \mathrm{rank}(M) \leq a \} \, , 
\end{equation}

There is a simple class of quivers whose 3d Coulomb branch is a closure of nilpotent orbits of height two of $\mathfrak{sl}(n,\mathbb{C})$. $\overline{\mathcal{O}}_{(1^n)}$ is the 3d Coulomb branch of
\begin{equation}
    \begin{tikzpicture}
        \tikzstyle{gauge1} = [inner sep=0.8mm,draw=none,fill=white,minimum size=0.35mm,circle, draw];
        \node at (0,0) [gauge1] {};
        \node at (0,-0.5) {1};
        \end{tikzpicture} 
\end{equation}
$\overline{\mathcal{O}}_{(2^a,1^{n-2a})}$ is the 3d Coulomb branch of
\begin{equation}
\label{NOquiver}
    \begin{tikzpicture}
        \tikzstyle{gauge1} = [inner sep=0.8mm,draw=none,fill=white,minimum size=0.35mm,circle, draw];
        
        \draw (0,0)--(1,0) (2,0)--(3,0) (2,0)--(3.5,1) (3.5,1)--(5,0) (4,0)--(5,0) (6,0)--(7,0);
        \draw[dashed] (1,0)--(2,0) (3,0)--(4,0) (5,0)--(6,0);

        \node at (0,0) [gauge1, label=below:{1}] {};
        \node at (1,0) [gauge1, label=below:{2}] {};
        \node at (2,0) [gauge1, label=below:{$a$}] {};
        \node at (3,0) [gauge1, label=below:{$a$}] {};
        \node at (3.5,1) [gauge1, label=above:{$1$}] {};
        \node at (4,0) [gauge1, label=below:{$a$}] {};
        \node at (5,0) [gauge1, label=below:{$a$}] {};
        \node at (6,0) [gauge1, label=below:{2}] {};
        \node at (7,0) [gauge1, label=below:{1}] {};
    
        \draw [decorate,decoration={brace,amplitude=10pt}]
        (7.3,-0.5) -- (-0.3,-0.5);
        \node at (3.5,-1) {$n-1$};
        \end{tikzpicture} 
\end{equation}
The Highest Weight Generating function of a closure of an orbit of height two is given by
\begin{equation}
	\mathrm{HWG}(\overline{\mathcal{O}}_{(2^a,1^{n-2a})})=\mathrm{PE} \left[\sum_{k=1}^{a}\mu_k\mu_{n-k}t^{2k} \right].
\end{equation}

We also encounter a second class of spaces, which we refer to as \emph{baryonic extensions} of nilpotent orbits. They are characterized by an integer $b \geq 0$. They have a 3d Coulomb branch quiver representation as
\begin{equation}
    \begin{tikzpicture}\tikzstyle{gauge1} = [inner sep=0.8mm,draw=none,fill=white,minimum size=0.35mm,circle, draw];
    
    \draw (0,0)--(1,0) (2,0)--(3,0) (2,0)--(2,1) (5,1)--(5,0) (4,0)--(5,0) (6,0)--(7,0);
    \draw (2,1.05)--(5,1.05) (2,0.95)--(5,0.95); \node at (3.5,1.3) {$b$};
    \draw[dashed] (1,0)--(2,0) (3,0)--(4,0) (5,0)--(6,0);
    
    \node at (0,0) [gauge1, label=below:{1}] {};
    \node at (1,0) [gauge1, label=below:{2}] {};
    \node at (2,0) [gauge1, label=below:{$a$}] {};
    \node at (3,0) [gauge1, label=below:{$a$}] {};
    \node at (2,1) [gauge1, label=above:{$1$}] {};
    \node at (5,1) [gauge1, label=above:{$1$}] {};
    \node at (4,0) [gauge1, label=below:{$a$}] {};
    \node at (5,0) [gauge1, label=below:{$a$}] {};
    \node at (6,0) [gauge1, label=below:{$2$}] {};
    \node at (7,0) [gauge1, label=below:{$1$}] {};
    
    \draw [decorate,decoration={brace,amplitude=10pt}]
    (7.3,-0.5) -- (-0.3,-0.5);
    \node at (3.5,-1) {$n-1$};
    \end{tikzpicture}.
    \label{baryonHWGquiv}
\end{equation}
The HWG for the 3d Coulomb branch of (\ref{baryonHWGquiv}) is \cite{zhenghao}
\begin{equation}
    \mathrm{HWG}=\mathrm{PE} \left[t^2+(\mu_{a}q+\mu_{n-a}q^{-1})t^{a+b}+\sum_{k=1}^{a}\mu_k\mu_{n-k}t^{2k}-\mu_{a}\mu_{n-a}t^{2(a+b)} \right].
\end{equation}
All the magnetic quivers obtained from brane webs in this paper are either of the type (\ref{NOquiver}) or (\ref{baryonHWGquiv}).

\section{Some Commutative Algebra}
\label{AppendixAlgebra}

In this Appendix, we gather a few notions from commutative algebra that we use in the text. This is based on \cite{eisenbud2013commutative,cox2006using} in which the reader will find more details and proofs. 

\paragraph{Ideals and Varieties}
We are mainly interested in two classes of objects: 
\begin{itemize}
    \item Polynomial rings of the form $\mathbb{C}[X_1 , \dots , X_n]/I$ where $I$ is an ideal. For instance, the ring $\mathbb{C}[Q,\tilde{Q}]/\langle \textrm{F-terms} \rangle$, or the ring $\mathbb{C}[M,B,\tilde{B}]$ modulo the equations of section \ref{sectionSUgroup}. 
    \item Algebraic varieties, i.e. the subset of $\mathbb{C}^n$ of zeroes of a finite family of polynomials. 
\end{itemize}
At the heart of algebraic geometry is the correspondence between these two classes of objects. An ideal in $\mathbb{C}[X_1 , \dots , X_n]$ is always generated by a finite number of polynomials $P_1 , \dots , P_r$. In this case, we denote the ideal by $I = \langle P_1 , \dots , P_r\rangle$. Therefore to each ideal one can associate an algebraic variety. Conversely, to every algebraic variety one can associate the ideal of polynomials which vanish on this variety. 
However the first class contains more objects, because certain polynomials in the rings can be nilpotent, and as a consequence two ideals can correspond to the same variety. For instance the rings $\mathbb{C}[X]/\langle X\rangle$ and $\mathbb{C}[X]/\langle X^2\rangle$ both correspond to the algebraic variety $\{0\}$, but they are not isomorphic rings. The Hilbert series is sensitive to such a difference: if $X$ is given weight $1$, then the Hilbert series of $\mathbb{C}[X]/\langle  X\rangle$ is $1$ while the Hilbert series of $\mathbb{C}[X]/\langle X^2\rangle$ is $1+t$.

\paragraph{Radical}
To remedy this, one needs to introduce the concept of radical of an ideal. The radical of $I$ is the ideal defined by 
\begin{equation}
    \sqrt{I} = \left\{ f \mid f^m \in I \textrm{ for some integer } m>0 \right\} \,. 
\end{equation}
If an algebraic variety is defined by a set of polynomial equations $P_i = 0$ for $i=1,\dots,r$ in some variables $X_1 , \dots , X_n$ then the coordinate ring of this variety is 
\begin{equation}
    \mathbb{C}[X_1 , \dots , X_n] / \sqrt{\langle P_1 , \dots , P_r \rangle } \,.
\end{equation}
For instance, we can check that $\sqrt{\langle X^2\rangle} = \langle X\rangle$. There is a one-to-one correspondence between the algebraic varieties and the radical ideals (this is the Nullstellensatz). In particular, the Hilbert series of an algebraic variety coincides with the Hilbert series of the ring defined by the radical ideal. This is the property we use in section \ref{sectionRadical}. 

A ring without nonzero nilpotent elements is called a \emph{reduced} ring. It follows directly from the definition that a quotient ring $R/I$ is reduced if and only if $I$ is a radical ideal. 

\paragraph{Schemes}
We have just seen that to a given ideal is associated an algebraic variety which is sensitive only on the radical part of the ideal. Does this mean that the nilpotent elements have no geometrical counterpart? This is not the case: the geometric object that corresponds to a non necessarily radical ideal is an \emph{affine scheme} (see for instance \cite{eisenbud2006geometry} for a gentle introduction to these objects). As a set, the scheme associated to a ring $R$ is the spectrum (set of prime ideals) of that ring denoted $\mathrm{Spec}(R)$. We do not use any properties of schemes in this paper beyond the fact that almost by definition, the ring of global sections of the structure sheaf of the affine scheme corresponding to the ring $R = \mathbb{C}[X_1 , \dots , X_n] / \sqrt{\langle P_1 , \dots , P_r \rangle }$ is exactly $R$. In particular, it can contain nilpotent elements. Therefore, the Hilbert series $1+t$ can not be the Hilbert series of an algebraic variety, but it is the Hilbert series of an affine scheme. 

\paragraph{Intersections and Unions of varieties} 
Given two algebraic varieties $V_1$ and $V_2$, their intersection $V_1 \cap V_2$ is again an algebraic variety. At the level of ideals, this translates into a sum. Namely, let $I_1$ and $I_2$ be the (radical) ideals associated to $V_1$ and $V_2$. The sum $I_1+I_2$ is simply the set of all polynomials $P + Q$ for $P \in I_1$, $Q \in I_2$. We note the useful property that if $I_1 = \langle P_1 , \dots , P_r \rangle$ and $I_2 = \langle Q_1 , \dots , Q_s \rangle$, then 
\begin{equation}
    I_1 + I_2 = \langle P_1 , \dots , P_r , Q_1 , \dots , Q_s  \rangle \,. 
\end{equation}
This makes it clear that $I_1+I_2$ is associated with the intersection $V_1 \cap V_2$. 

Similarly, a the \emph{union} $V_1 \cup V_2$ is associated to the \emph{intersection} of ideals $I_1 \cap I_2$. In the next paragraph, we explain how a given ideal can be decomposed in a canonical way as an intersection. Correspondingly, the variety is written as a union of irreducible varieties (which are cones in this paper). 

\paragraph{Primary Decomposition}
Ideal theory generalizes the theory of arithmetics in $\mathbb{Z}$. As such, there is an analogous of the fundamental theorem of existence and unicity of the decomposition of integers into a product of primes. 

An ideal $I$ is said to be \emph{prime} if $fg \in I$ implies $f \in I$ or $g \in I$. However, because of the problem of nilpotent elements evoked above, we need a slightly more general definition: we say that the ideal $I$ is \emph{primary} if $fg \in I$ implies $f \in I$ or $g^m \in I$ for some $m>0$. The Lasker-Noether theorem then states that any ideal $I \subset \mathbb{C}[X_1 , \dots , X_n]$ admits a decomposition as an intersection of primary ideals. 

The unicity of this decomposition is more tricky: in general, as stated above, there is no unicity of the primary decomposition. However, notice that the radical of a primary ideal is always prime. Consider a primary decomposition $I = \bigcap_{i=1}^r J_i$, and assume that it is \emph{minimal}, which means that the $\sqrt{J_i}$ are all distinct, and $J_j$ does not contain the intersection $\bigcap_{i\neq j} J_i$. Then the $\sqrt{J_i}$ are uniquely determined by $I$, but the $J_i$ are not in general.

\bibliographystyle{JHEP}
\providecommand{\href}[2]{#2}\begingroup\raggedright\endgroup
\end{document}